\documentclass{article}
\newif\ifamse
\amsefalse
\newif\ifclean
\cleantrue  % hide extra markup 
\usepackage{graphicx}	
\usepackage{amssymb}
\usepackage[letterpaper]{geometry}
\usepackage{subcaption} % not compatible with asmejour.sty
\usepackage{amsmath,amssymb,amsthm}
\usepackage{bm}
\usepackage[usenames,dvipsnames]{xcolor}
\usepackage[normalem]{ulem}
\newcommand{\COMMENT}[1]{\textcolor{cyan}{{[ \sc{#1} ]}}} % comments

\newcommand{\QUESTION}[1]{\textcolor{ForestGreen}{QUESTION: {#1}}}
\newcommand{\red}[1]{\textcolor{red}{{#1}}}

\newlength{\figwidth}
\newlength{\figwidthtwo}
\newlength{\figwidththree}
\newcommand{\aref}[1]{App.\,\ref{#1}}

\newcommand{\Fref}[1]{Figure\,\ref{#1}}
\newcommand{\tref}[1]{Table\,\ref{#1}}
\newcommand{\eref}[1]{Eq.\,(\ref{#1})}

\newcommand{\sref}[1]{Sec.\!~\ref{#1}}

\newcommand{\cref}[1]{Ref.\,\cite{#1}}
\newcommand{\crefs}[1]{Refs.\,\cite{#1}}

\newcommand{\ie}{{\it i.e.}\! }

\newcommand{\etc}{{\it etc.}\! }
\newcommand{\etal}{{\it et al.}\! }
\newcommand{\apriori}{{\it a priori} }

\newcommand{\Bc}{\mathcal{B}}
\newcommand{\Ic}{\mathcal{I}}

\newcommand{\Vs}{\mathsf{V}}
\newcommand{\cs}{\mathsf{c}}

\newcommand{\Dbb}{\mathbb{D}}
\newcommand{\Ibb}{\mathbb{I}}
\newcommand{\Pbb}{\mathbb{P}}

\newcommand{\ab}{\mathbf{a}}

\newcommand{\eb}{\mathbf{e}}

\newcommand{\xb}{\mathbf{x}}

\newcommand{\Ab}{\mathbf{A}}
\newcommand{\Bb}{\mathbf{B}}
\newcommand{\Cb}{\mathbf{C}}

\newcommand{\Fb}{\mathbf{F}}

\newcommand{\Eb}{\mathbf{E}}

\newcommand{\Sb}{\mathbf{S}}
\newcommand{\Ib}{\mathbf{I}}

\newcommand{\Xb}{\mathbf{X}}

\newcommand{\chib}{\boldsymbol{\chi}}

\newcommand{\tr}{\operatorname{tr}}

\newcommand{\dev}{\operatorname{dev}}

\newcommand{\partialb}{\boldsymbol{\partial}}

\newcommand{\NN}{N\!\!N}
\graphicspath{{./Figures/}}

\usepackage{comment}
\newcommand{\caution}{\red{\bf Draft: \today. Do not distribute.}}
\ifclean
\renewcommand{\COMMENT}[1]{{}}
\renewcommand{\QUESTION}[1]{{}}
\else
\fi

\newcommand{\TITLE}{Stress representations for tensor basis neural networks: alternative formulations to Finger-Rivlin-Ericksen
}

\usepackage{bm}
\usepackage{physics}

\usepackage[utf8]{inputenc}
\usepackage[english]{babel}
\ifamse
\hypersetup{%
        pdfauthor={Jan N. Fuhg, Nikolaos Bouklas, Reese. E. Jones},    
        pdftitle={\TITLE},  
        pdfkeywords={tensor basis, neural network, stress, hyperelasticity, orthogonal basis}
}

\JourName{JCISE}

\else 

\title{\bf\TITLE}

\author{
Jan N. Fuhg, Nikolaos Bouklas \\[0.02in]
{\small \it Cornell University, Ithaca, NY, USA} \\[0.10in]
Reese E. Jones\footnote{rjones@sandia.gov} \\[0.02in]
{\small \it Sandia National Laboratories, Livermore, CA, USA}
}
\ifclean
\date{}
\else
\date{\caution}
\fi

\fi

\begin{document}

\ifamse
\SetAuthorBlock{Jan N. Fuhg}{Department of Mechanical Engineering,\\
   Cornell University,\\
   Street address,\\
   Ithaca, NY, USA \\
   email: jf853@cornell.edu} 

\SetAuthorBlock{Nikolaos Bouklas}{Department of Mechanical Engineering,\\
   Cornell University,\\
   Street address,\\
   Ithaca, NY, USA \\
   email: nb589@cornell.edu } 

\SetAuthorBlock{Reese E. Jones}{Mechanics of Materials Department,\\
   Sandia National Laboratories,\\
   7011 East Avenue,\\
   Livermore, CA, USA \\
   email: rjones@sandia.gov} 
   
\title{\TITLE}
\fi

\maketitle

\begin{abstract}
Data-driven constitutive modeling frameworks based on neural networks and classical representation theorems have recently gained considerable attention due to their ability to easily incorporate constitutive constraints and their excellent generalization performance. 
In these models, the stress prediction follows from a linear combination of invariant-dependent coefficient functions and known tensor basis generators. 
However, thus far the formulations have been limited to stress representations based on the classical Rivlin and Ericksen form, while the performance of alternative representations has yet to be investigated.
In this work, we survey a variety of tensor basis neural network models for modeling hyperelastic materials in a finite deformation context, including a number of so far unexplored formulations which use theoretically equivalent invariants and generators to Finger-Rivlin-Ericksen. 
Furthermore, we compare potential-based and coefficient-based approaches, as well as different calibration techniques.
Nine variants are tested against both noisy and noiseless datasets for three different materials.
Theoretical and practical insights into the performance of each formulation are given.
\end{abstract}

%%%%%%%%%%%%%%%%%%%%%%%%%%%%%%%%%%%%%%%%%%%%%%%%%%%%%%%%%%%%%%%%%%%%%%%%%%%%
\section{Introduction} 
%%%%%%%%%%%%%%%%%%%%%%%%%%%%%%%%%%%%%%%%%%%%%%%%%%%%%%%%%%%%%%%%%%%%%%%%%%%%

Recently, there has been dramatically increased interest in machine learning (ML) in the computational sciences. 
This rise in popularity is due to: the ability of machine learning models to directly utilize experimental data in simulation environments, the potential speed up of ML models in comparison to traditional numerical models and methods, as well as the general utility and open-access ecosystem of ML tools. 
Nevertheless, many scientific ML (SciML) applications suffer from two interconnected bottlenecks: a lack of generalization capabilities due to poor extrapolations and a lack of trustworthiness due to the opaqueness of the trained models.
The main premise in SciML is that the underlying data often comply with physical laws (known or yet to be discovered) or otherwise connect to known mathematical structure, which can help surmount the aforementioned bottlenecks via a physics-informed paradigm.
The promise of SciML can lead to myriad benefits such as: more accurate predictions, reduction of unnecessary human involvement, speed-up of the processing-performance-product development cycle, and minimization of the computational costs of detailed simulations. 
Particular to the focus of this work, an automated data-driven approach for constitutive modeling can have significant payoffs in material discovery, industrial engineering simulations and research.
Many developments have been made in this arena for fluid closure models \cite{ling2016machine,fang2020neural,kaandorp2020data}; in this work we focus on constitutive models for solids.

A number of distinct approaches to forming constitutive models with ML have been investigated.
ML tools have been utilized in parameter estimation of known constitutive models \cite{wang2021metamodeling}.
This is a task that becomes more complex as model parameters increase and experimental observations are limited. This is especially true for traditional optimization approaches due to the non-convex nature of the optimization problem at hand.
Mixing traditional and ML approaches to representation and calibration via symbolic regression \cite{sun2019data,kabliman2021application,bomarito2021development,wang2022establish,de2023establishing,abdusalamov2023automatic} has been widely explored.
This approach selects from a library of known models that directly enforce physical and mechanistic constraints (depending on the specific model choices) to distill parsimonious data-driven constitutive models. 
Notable developments include the approach of  Wang \etal \cite{wang2019cooperative} who used reinforcement learning to turn model building into a competitive game.
Also Schmidt \etal \cite{schmidt2009distilling} used symbolic regression to distill constitutive laws from unlabeled data.
Later, De Lorentzis and co-workers \cite{thakolkaran2022nn,flaschel2022discovering} utilized sparse regression  to discover interpretable constitutive laws for a wide array of material classes.
An interesting extension to this work was the development of an unsupervised Bayesian framework for discovering hyperelasticity models which accounts for uncertainty \cite{joshi2022bayesian}.
Neural networks and Gaussian process models have been widely employed as replacements for human-selected, traditional model forms.
In fact, the use of ML black-box constitutive models has been extensively studied for over 30 years. 
Starting from the influential works of  Ghaboussi and collaborators \cite{wu1990representation,ghaboussi1990material,ghaboussi1991knowledge}, these tools have been employed for different material models with increasing complexity over the years \cite{lefik2003artificial,jung2006characterizing,huang2020machine,fuhg2021local}.

A significant current challenge is generating trustworthy models from low-data (constrained by experimental/computational cost) and limited-data (constrained by experimental design and observation).
To this end efforts have been made to train data-driven constitutive models that do not only train with raw stress-strain data but incorporate additional physics-based restrictions to the trained model \cite{liu2019exploring,heider2020so,xu2021learning,xu2021learning2,fuhg2023modular,fuhg2023enhancing}. 
These models, referred to as {\it physics-informed} or {\it physics-guided data-driven} constitutive models try to enforce a variety of physical principles and mechanics-informed assumptions.
From enforcing objectivity, to material symmetries \cite{ling2016machine}, thermodynamic constraints \cite{jones2018machine,jones2022neural} and polyconvexity \cite{klein2022polyconvex,fuhg2022learning} there are approaches that enforce these  condition weakly through the loss function \cite{masi2021thermodynamics,linka2021constitutive,vlassis2020geometric}  or strictly in the construction of the ML representation \cite{ling2016machine,frankel2020prediction,fuhg2022physics,linden2023neural}. 
A large majority of the proposed works in the literature for physics-guided constitutive models are based on neural networks \cite{liu2019exploring,frankel2019predicting,frankel2020prediction,heider2020so,xu2021learning,xu2021learning2,masi2021thermodynamics} due to the flexibility of this paradigm. 

Material frame indifference is a primary concern in developing constitutive models \cite{truesdell1965non}. 
Ling \etal \cite{ling2016machine} introduced the tensor basis neural network (TBNN) to embed objectivity through an equivariant NN formulation.
An anisotropic hyperelastic model was formed from the scalar invariants and tensor basis of the strain using atomistic crystal data, in addition to fluids applications.
Later Frankel \etal \cite{frankel2020prediction} adapted the tensor basis representations to a Gaussian process formalism to represent general tensor functions and hyperelastic data.
This was extended by Fuhg and Bouklas \cite{fuhg2022physics} to anisotropic materials, strictly enforcing known symmetries up to transverse isotropy; this work showed that this simplified learning approach led to significant generalization capabilities when the physics do not radically change outside of the training region. 
This approach was also utilized in Kalina \etal \cite{kalina2023fe} integrated in a multiscale framework with automated data-mining. 
In Fuhg \etal \cite{fuhg2022learning}, tensor basis NNs were utilized to discover the character of the anisotropy of the material through labeled data. 
Even though several works have focused on utilizing tensor basis representation theorems in learning of hyperelastic responses from labeled data pairs, there has not been an extensive study aimed at discovering the most efficient tensor basis representations for the learning tasks at hand in the context of finite deformation and hyperelasticity. 

In the context of hyperelasticity, strict enforcement of polyconvexity requirements \cite{ball1976convexity} for the strain energy density  has also proven extremely useful towards generalization, discovery, and robustness. 
Input convex neural networks have been utilized for the enforcement of polyconvexity towards learning hyperelastic responses \cite{klein2022polyconvex,fuhg2022learning}, and in some cases even interpretability can be achieved \cite{linka2023new} due to the non-parametric nature of the specific implementation. 
Alternately, neural ordinary differential equations have also been utilized towards strict enforcement of polyconvexity \cite{tac2022data}.  
More recently Linden \etal \cite{linden2023neural} presents a thorough review of techniques to enforce physical constraints and mechanistic assumptions towards learning hyperelasticity with NNs. 
Such approaches are crucial for the efficient utilization of the data and the development of robust material models that can efficiently generalize.  

This work provides a limited survey of the wide variety of tensor basis techniques and contrasts their performance on representative data in the low-data regime (100 training points).
We focus on stress representations for hyperelastic materials since they are the fundamental basis for finite deformation mechanics.
The contributions of this work are:
novel formulations based on the variety that the tensor basis framework  affords,
exploration of different methods of calibrating the models to data, and
demonstration of the effects of noise and incompatible representations on physics-constrained formulations.
To this end, we utilize well-known hyperelastic models as data generators.

In \sref{sec:formulation} we develop a multitude of equivariant {\it tensor basis} neural networks (TBNNs) \cite{ling2016machine} formulations from classical representation theory.
Then in \sref{sec:data} we give details of the data generation and training methodology.
\sref{sec:results} presents the results of testing the models in and out of distribution and without and with additive noise.
Finally in \sref{sec:conclusion} we summarize the findings and conclude with avenues for future work.

%%%%%%%%%%%%%%%%%%%%%%%%%%%%%%%%%%%%%%%%%%%%%%%%%%%%%%%%%%%%%%%%%%%%%%%%%%
\section{Stress representations} \label{sec:formulation}
%%%%%%%%%%%%%%%%%%%%%%%%%%%%%%%%%%%%%%%%%%%%%%%%%%%%%%%%%%%%%%%%%%%%%%%%%%

In this work, we  develop and compare a variety of tensor basis neural network (TBNN) formulations for stress representations. 
In this section, we introduce the fundamental differences between the representations and the neural network formulations that follow directly from the representations.

%.........................................................................
\subsection{Tensor basis models}
%.........................................................................
Hyperelasticity is the prevailing theory for the description of finite deformation solid mechanics for continua in the absence of inelastic phenomena. 
The theory posits a potential $\Psi$ from which the second Piola-Kirchhoff stress $\Sb$ can be derived:
\begin{equation}
    \Sb = 2 \partialb_\Cb \Psi(\Cb) \ ,
\end{equation}
as a function of the right Cauchy-Green stretch tensor $\Cb = \Fb^T\Fb$.
Here $\Fb = \partial_\Xb \chib$ is the deformation gradient of the spatial position $\xb = \chib(\Xb,t)$ at time $t$ with respect to the corresponding reference position $\Xb$ of the material.
This potential ensures deformations are reversible, and is also utilized in some 
incremental formulations of large strain plasticity \cite{stainier2019model,mota2013lie}.

In this work, we limit the discussion to isotropic hyperelasticity.
In this case material frame invariance of the potential leads to the reduction of the inputs of $\Psi$ to three scalar invariants $I_a$ of $\Cb$ and an equivariant stress function:
\begin{equation}
    \Sb = 2 \, \partialb_\Cb \Psi(I_1(\Cb), I_2(\Cb), I_3(\Cb))
\end{equation}
The chain rule results in the summation of material-specific, scalar derivative functions and an {\it a priori} known tensor basis:
\begin{equation} \label{eq:chain_rule}
    \Sb = 2 \partialb_\Cb \Psi(\Cb) 
    = 2 \, \sum_{a=1}^3 \partialb_{I_a}  \Psi \, \partialb_{\Cb} I_a
\end{equation}
Typically the principal invariants,
\begin{equation} \label{eq:cayley_invariants}
    I_1 = \tr(\Cb), \quad 
    I_2 = \tr(\Cb^{-1}) \det (\Cb), \quad 
    I_3 = \det(\Cb) \ ,
\end{equation}
from the Cayley-Hamilton theorem 
\begin{equation} \label{eq:cayley-hamilton}
\Cb^3 - I_1 \Cb^2 + I_2 \Cb - I_3 \Ib = \mathbf{0}
\end{equation}
are employed.
Note the second invariant is equivalently $I_2 = \frac{1}{2}(\tr(\Cb)^2 - \tr (\Cb^2))$.
A three-term formula for the stress
\begin{equation} \label{eq:normal_representation}
    \Sb = 2 \left[ \partial_{I_1} \Psi +  I_{1}  \partial_{I_2} \Psi \right] \Ib 
    - 2 \partial_{I_2} \Psi \, \Cb 
    + 2 I_{3} \, \partial_{I_3} \Psi \, \Cb^{-1} 
\end{equation}
comes from collecting terms with like powers of $\Cb$. 
This is a well-known and arguably the most widely used stress representation for isotropic materials.
It was first introduced by Finger \cite{finger1894potential} but was further popularized by Rivlin and Ericksen \cite{rivlin1955stress}. 

A generalization of this representation can be compactly written as a tensor basis expansion
\begin{equation} \label{eq:tbnn}
\Sb = \sum_{a=1}^3 c_a(\Ic) \, \Bb_a \ ,
\end{equation}
where the 3 coefficients $c_a$ 
are functions of a set of 3 independent invariants $\Ic$ and the basis $\Bb_a$ must span $\{\Cb^a, a=-1,0,1\}$. 
For instance, \eref{eq:normal_representation} can be expressed as
\begin{eqnarray} 
    \Ic &=& \{ I_a, \, a=1,2,3 \} \\
    \Bc &=& \{ \Cb^a, \, a=0,1,-1 \} \label{eq:power_basis}
\end{eqnarray}
and 
\begin{equation}\label{eq:cfunsICCinv}
    \begin{aligned}
            c_0 &= 2 \left[ \partial_{I_1} \Psi +  I_{1}  \partial_{I_2} \Psi \right] \\
    c_1 &= - 2 \, \partial_{I_2} \Psi \\
    c_{-1} &= 2 I_{3} \ \partial_{I_3} \Psi
    \end{aligned}
\end{equation} 
Note that the Cayley-Hamilton theorem \eref{eq:cayley-hamilton} allows the power basis to be shifted to higher or lower powers 
\begin{equation} \label{eq:cayley-hamilton_k}
\Cb^{3+k} =   I_1 \Cb^{2+k} - I_2 \Cb^{1+k} + I_3 \Cb^k  \  \text{for} \ k \in \{ \ldots, -2, -1, 0, 1, 2, \ldots \} \ ,
\end{equation}
for example
\begin{equation} \label{eq:rivlin_basis}
\Bc = \{ \Cb^a, \, a = 0,1,2 \} \ ,
\end{equation}
via 
\begin{equation} \label{eq:cayley-hamilton2}
I_3 \Cb^{-1} = \Cb^2 - I_1 \Cb + I_2 \Ib \ .
\end{equation}
This basis together with the principal invariants \eqref{eq:cayley_invariants} is another form of the Rivlin-Ericksen representation \cite{rivlin1955stress}.
Also, the basis that results from the chain rule:
\begin{eqnarray}  
    c_a &=& \{ \partial_{I_a} \Psi, \, a = 1,2,3 \}  \label{eq:derivative_coefficients} \\
   \Bc  &=& \{ \partialb_{\Cb} {I_a}, \, a = 1,2,3  \} \label{eq:derivative_basis}
\end{eqnarray}
is part of an equally valid representation.

To calibrate \eref{eq:tbnn}, the model output can be regressed directly to stress data, or the coefficients for a given basis, e.g. $\Bc = \{ \Ib, \Cb, \Cb^{-1} \}$, can be determined at each data point $(\Cb_i,\Sb_i)$ via:
\begin{equation} \label{eq:coef_solve}
\begin{bmatrix}
c_1 \\
c_2 \\
c_3 
\end{bmatrix}
= 
\begin{bmatrix}
1 & \epsilon_1 & \epsilon_1^{-1}  \\
1 & \epsilon_2 & \epsilon_2^{-1}  \\
1 & \epsilon_3 & \epsilon_3^{-1}  
\end{bmatrix}^{-1}
\begin{bmatrix}
\sigma_1 \\
\sigma_2 \\
\sigma_3 
\end{bmatrix}
\end{equation}
using the fact that any power basis, such as \eref{eq:power_basis}, is collinear with $\Sb$.
Here $\sigma_a$ and $\epsilon_a$ are the eigenvalues of the stress and stretch tensors, herein $\Sb$ and $\Cb$, respectively.
If the eigenvalues are distinct, \eref{eq:coef_solve} provides a unique solution for the coefficient values; however,  multiplicity of strain eigenvalues requires special treatment, see \crefs{gurtin1982introduction,frankel2020tensor,jones2022neural} and \aref{app:multiplicity}, which also outlines alternate solution procedures.
Alternatively, we can use the Gram-Schmidt procedure
\begin{equation} \label{eq:gram-schmidt}
    \Bb_a = \tilde{\Bb}_a - \sum_{b=1}^{a-1} \frac{\tilde{\Bb}_a : \Bb_b}{ \Bb_b : \Bb_b} \Bb_b 
\end{equation}
to orthogonalize the basis $\tilde{\Bb}_a \in \{ \Ib, \Cb, \Cb^{-1} \}$, which results in 
\begin{eqnarray} 
        \Bc &=& \left\{ \Ib, \,
   \dev(\Cb), \,
   \dev(\Cb^{-1}) - \left[ \Cb^{-1} : \frac{ \dev(\Cb)}{ \| \dev(\Cb) \| } \right] \frac{\dev(\Cb)}{\| \dev(\Cb) \|}  \right\} \label{eq:orthogonal_basis}
\end{eqnarray}
if we keep the same scalar invariants.
Herein $\dev(\Cb) = \Cb - 1/3 \tr(\Cb) \Ib$.
The fact that the Gram-Schmidt procedure starting with $\Bb_1= \Ib$ and $\Bb_2 = \Cb$ leads to a spherical-deviatoric split is noteworthy.
Orthogonality of the basis allows for direct determination of the coefficients:
\begin{equation} \label{eq:projection_coef}
    c_a = \Sb : \Bb_a
\end{equation}
Likewise, Gram-Schmidt applied to $\{ \Cb^a, a=0,1,2\}$ gives the unnormalized basis
\begin{eqnarray} \label{eq:unnormalized_basis}
        \Bc &=& \left\{ \Ib, \,
   \dev(\Cb), \,
   \| \dev(\Cb) \|^2 
   \dev(\Cb^2) -  ( \Cb^2 : \dev(\Cb)) \dev(\Cb) \right\} 
\end{eqnarray}

Similarly, we can use a formulation inspired by the work Criscione \etal \cite{criscione2000invariant} which effects an orthogonal spherical-deviatoric split of the basis via invariants:
\begin{eqnarray} 
    \Ic &=& \left\{ K_1 = \tr(\Cb), \,
K_2 = \| \dev(\Cb) \|, \,
K_3 =  \det\left( \frac{\dev(\Cb)}{\| \dev (\Cb) \| } \right) \right\}  \label{eq:criscione_invariants} \\
    \Bc &=& \left\{ \Ib, \Ab,  -\frac{1}{3}\Ib
- \tr(\Ab^3) \Ab
+ \Ab^2  \right\} \label{eq:criscione_basis}
\end{eqnarray}
where $\Ab = \dev(\Cb) / \| \dev (\Cb) \| $.
The resulting stress representation is 
\begin{eqnarray} \label{eq:criscione_stress}
\Sb  &=&
\partial_{K_1} \Psi \,  \Ib 
+ \partial_{K_2} \Psi \, \Ab
+ \partial_{K_3} \Psi \, \frac{1}{K_2} \left[
-\frac{1}{3}\Ib
- \tr(\Ab^3) \Ab
+ \Ab^2 
\right] \\
&=&
\underbrace{\left[ \partial_{K_1} \Psi - \frac{1}{3 K_{2}} \, \partial_{K_3} \Psi  \right]}_{c_0} \, \Ib 
    + \underbrace{\left[ \partial_{K_2} \Psi - \frac{3 K_3}{K_2} \, \partial_{K_3} \Psi \right]}_{c_1} \, \Ab 
    + \underbrace{\frac{1}{K_2}\, \partial_{K_3} \Psi}_{c_2} \,  \Ab^2 \nonumber
\end{eqnarray}
Note Criscione \etal \cite{criscione2000invariant} formulate the representation in terms of the spatial Hencky stretch, and here we apply the invariants of the same form to $\Cb$.
This formulation combines derivative connection of the potential, invariants and the basis in the sense that the basis is a result of the choice of invariants, as in \eref{eq:derivative_coefficients}, and orthogonality of the basis.
A better behaved set of related invariants 
\begin{equation} \label{eq:smoother_invariants}
\Ic = \left\{  \tr(\Cb), \, \tr \dev(\Cb)^2 , \, \det\left( \dev(\Cb) \right) \right\} 
\end{equation}
which eliminate the normalization in \eref{eq:criscione_invariants},
leads to the basis
\begin{equation} \label{eq:smoother_basis}
 \Bc = \{ \Ib,  2 \dev \Cb,  (\det (\dev \Cb)) \dev( (\dev(\Cb))^{-1} ) \}
\end{equation}
See \aref{app:orthogonal} for further details on the construction of an orthogonal basis.

With any of these representations, a densely connected neural network (NN) can be employed as a representation of the potential $\Psi(\Ic)$ itself or the coefficient functions $c_a(\Ic)$ directly.
Summation of the coefficients with the known basis $\Bc$, as in \eref{eq:tbnn}, completes the formulation of a {\it tensor basis neural network} (TBNN) \cite{ling2016machine}.
\sref{sec:training}, \aref{app:potentialNN} and \aref{app:coeffNN} provide details of the implementation of the TBNNs.

%.........................................................................
\subsection{Additional physical constraints} \label{sec:addl_constraints}
%.........................................................................
Other fundamental considerations, in addition to equivariance of the stress $\Sb(\Cb)$, constrain the form of the coefficient functions $c_a(\Ic)$. 
Of the various constraints (rank-1 convexity, strong ellipticity, Hadamard stability, \etc \cite{truesdell1965non,gurtin1981topics,steigmann2017finite}), polyconvexity was proved by Ball \cite{ball1976convexity} to ensure the existence of solutions in the context of hyperelasticity. 
For isotropic materials, polyconvexity requires that $\Psi$ is convex in the triplet $(\Fb, \operatorname{cof} \Fb, \det \Fb)$ which can be fulfilled when 
\begin{equation} \label{eq:polyconvexity}
\Psi = \Psi(I_1,I_2,I_3)
\ \text{is convex in each of its arguments \cite{linden2023neural}}.
\end{equation}
An  input convex neural network (ICNN) \cite{amos2017input} satisfies these conditions and has been utilized for modeling hyperelastic materials in various recent studies \cite{klein2022polyconvex,linden2023neural,fuhg2023modular}.
Alternatively, if we assume that $\Psi$ is polyconvex,
we know that the derivatives of $\Psi$ have to be non-decreasing, \ie $\partial_{I_a} \Psi \ge 0$ with regards to $I_{a}$.
Assuming a representation of the form of \eref{eq:chain_rule} 
\begin{equation} \label{eq:monotone_stress}
    \Sb  = 2 \, \sum_{a=1}^3 \partialb_{I_i}  \Psi \, \partialb_{\Cb} I_a 
    = \underbrace{\partialb_{I_1}  \Psi}_{c_0} \, \Ib 
    + \underbrace{\partialb_{I_2}  \Psi}_{c_1} \, (I_{1} \Ib- \Cb) 
    + \underbrace{I_3 \partialb_{I_3}  \Psi}_{c_{-1}} \, \Cb^{-1} \ ,
\end{equation}
this implies that $c_{1}(I_{1}, I_{2}^{0}, I_{3}^{0})$ is monotonically increasing in $I_{1}$ for fixed $I_{2}^{0}$ and $I_{3}^{0}$.
Note that the basis element $I_1 \Ib - \Cb = -\dev \Cb$ naturally arises from the Cayley-Hamilton/principal invariants, c.f. \eref{eq:cayley_invariants} and \eref{eq:cfunsICCinv}.
We enforce this condition via an input monotone (or in fact monotonically non-decreasing) neural network \cite{fuhg2023modular} which guarantees that the outputs of a neural network are monotonically non-decreasing in each of its inputs. 
Since to the best of our knowledge, no currently proposed neural network architecture enforces that each output individually is monotonically non-decreasing to only a subset of its inputs, and proposing a network of this kind is out of the scope of this work, we remark that this is an overconstrained way of enforcing the convexity condition.

Additional constitutive constraints resulting from mechanistic assumptions, include that the stress in the reference configuration is zero, 
\begin{equation} \label{eq:zero_stress}
\Sb(\Cb=\Ib) = \mathbf{0} \ \text{implies} \ \sum_a c_{a}(I^0_{1},I^0_{2},I^0_{3})= 0 \ .
\end{equation}
with $I^0_{1}=3,I^0_{2}=3,I^0_{3}=1$.
One possible solution to enforcing this is to refactor the basis to form a Saint-Venant-like expansion:
\begin{equation} \label{eq:StV-E}
\Sb = \sum_{a=1}^3 c_a \Eb^a 
\end{equation}
where $\Eb = 1/2 ( \Cb - \Ib) $.
A $\Cb$ based version is likewise:
\begin{equation} \label{eq:StV-C}
\Sb = \sum_{a=1}^3 c_a \Cb^a .
\end{equation}
Note the coefficient functions $c_a$ for these two representations are distinct but related, as are all the other representations introduced in this section.
The requirements at the reference state $\Cb = \Ib$ can be seen as a special case of the more general condition of symmetric loading where 2 or 3 of the eigenvalues are equal, examples include equibiaxial and hydrostatic/volumetric loadings. 

The set of points where the eigenvalues are unique is dense  in the invariant input space \cite{serrin1959derivation,man1995smoothness}, whereas highly symmetric cases are often used in testing and experiments since they are more easily understood and yet are sparse in the invariant input space.
Since the unique case is dense there are continuous extensions for the coefficient functions to the case of eigenvalue multiplicity; however, the formula for the solution of the coefficients \eref{eq:coef_solve} does not provide them since the determinant of the system goes to zero.
Although not well-cited, the important body of theoretical work starting with Serrin  \cite{serrin1959derivation,man1995smoothness,scheidler1996smoothness,xiao2002basic} relates the smoothness of $\Sb(\Cb)$ or $\Sb(\Eb)$ to the smoothness of the coefficient functions with respect to the scalar invariants.
Since most classical work treated only polynomial functions of the invariants, these developments have not been fully utilized; however, in the present context, we are forming general coefficient functions with neural networks.
Man \cite{man1995smoothness} proved that $\Sb$ needs to be two degrees more continuous than the desired degree of smoothness of the coefficient functions, in particular $\Sb(\Cb)$ needs to be twice differentiable for $c_a(\Ic)$ to be continuous. 
Note that smooth solutions to the balance of momentum already require $\Sb$ to be $C^1$ and $\Phi$ to be $C^2$.
Also, Scheidler \cite{scheidler1996smoothness} provided coefficient values from derivatives of the stress with respect to particular deformations, unlike \eref{eq:derivative_coefficients}.

Smoothness and growth considerations affect the choice of NN activations.
For example, the St. Venant-like basis \eqref{eq:StV-E} incurs certain growth and asymptotic behavior.
Refactoring the coefficients as
\begin{equation} \label{eq:refactored_coefficients}
\tilde{c}_a = \| \Eb \|^n c_a
\end{equation}
can enforce asymptotic behavior near $\Cb \to \Ib$ as in \cref{jones2018machine}.
The orthonormal basis formulations also need special consideration due to the normalization which creates non-smoothness in the coefficients, as in \eref{eq:orthogonal_basis}.
An unnormalized basis, such as \eref{eq:unnormalized_basis}, avoids these issues.

%.........................................................................
\subsection{Summary of selected stress representations}
%.........................................................................

In \sref{sec:results} we compare a number of distinct formulations of TBNNs for hyperelastic response listed in \tref{tab:nn_variants}. 
Three are based on representing the strain energy potential $\Psi(\Ic)$ directly: 
(a) using the principal invariants $\Ic$ as inputs to a standard feed-forward dense neural network ({\it Rivlin-Pot}), 
(b) using the principal invariants with an input convex neural network ({\it Convex-Pot}), 
and (c) using spherical-deviatoric split invariants in a standard dense neural network ({\it Crisc-Pot}). 
For these models the derivative of the potential with respect to these invariants through automatic differentiation provides the stress response.
Six other models are based on coefficient-basis product formulations: (a) the customary power basis and coefficient functions in terms of the principal invariants ({\it Rivlin-Coeff}), 
(b) an input monotone neural network formulation of the coefficient functions with the power basis ({\it Mono-Coeff})
(c) the orthogonal basis with the Criscione invariants ({\it Crisc-Coeff}), 
(d) the orthogonal basis with the principal invariants ({\it Orthnorm-Coeff}), 
(e) an unnormalized orthogonal basis with more regular invariants ({\it Orth-Coeff}), 
and 
(f) a St.Venant-like basis with the principal invariants ({\it StV-Coeff}).
For these both the coefficient functions $c_a(\Ic)$ and the basis $\Bb_a$ are chosen.
Table \ref{tab:nn_variants} summarizes the differences in the TBNN variants.
In addition to these variations, we also explored how the method for calibration to stress data, e.g. via the coefficients found through regression or projection, or implicitly through direct calibration to stress, affects the model accuracy.

\begin{table}[h!]
\centering
\begin{tabular}{|c|ccc|ccc|}
\hline
& invariants & basis & coefficients & potential & convex & orthogonal basis  \\
\hline
Rivlin-Pot & \eref{eq:cayley_invariants} & \eref{eq:derivative_basis} & \eref{eq:derivative_coefficients}   & $\times$ &   &       \\
Convex-Pot & \eref{eq:cayley_invariants} & \eref{eq:derivative_basis} & \eref{eq:derivative_coefficients}  & $\times$ & $\times$ & \\
Crisc-Pot & \eref{eq:criscione_invariants} & \eref{eq:criscione_basis} & \eref{eq:criscione_stress} & $\times$  &  & $\times$  \\
Rivlin-Coeff & \eref{eq:cayley_invariants} & \eref{eq:rivlin_basis} & \eref{eq:tbnn} &  &  &  \\
Mono-Coeff & \eref{eq:cayley_invariants} & \eref{eq:monotone_stress} & \eref{eq:monotone_stress}  &   & $\times$ & \\
Crisc-Coeff & \eref{eq:criscione_invariants} & \eref{eq:criscione_basis} & \eref{eq:tbnn}  &  &  & $\times$  \\
Orthnorm-Coeff & \eref{eq:criscione_invariants} & \eref{eq:orthogonal_basis} & \eref{eq:tbnn} &  &  & $\times$  \\
Orth-Coeff & \eref{eq:smoother_invariants} & \eref{eq:unnormalized_basis} & \eref{eq:tbnn} &  &  & $\times$  \\
StV-Coeff & \eref{eq:cayley_invariants} & \eref{eq:StV-E} & \eref{eq:tbnn} &  &  &   \\
\hline
    \end{tabular}
    \caption{Tensor basis neural network variants.}
    \label{tab:nn_variants}
\end{table}

%%%%%%%%%%%%%%%%%%%%%%%%%%%%%%%%%%%%%%%%%%%%%%%%%%%%%%
\section{Data and training} \label{sec:data}
%%%%%%%%%%%%%%%%%%%%%%%%%%%%%%%%%%%%%%%%%%%%%%%%%%%%%%

For this study, we train the various NN models enumerated in Table \ref{tab:nn_variants} to stress data generated with classical hyperelastic models. 
In this section, we briefly discuss the classical data-generating models and give a detailed description of the data-generation process.   

%-----------------------------------------------------
\subsection{Data and training} \label{sec:complexity}
%-----------------------------------------------------
We remark that the complexity of the coefficient and potential functions is intrinsically connected to the stress measure, the basis function, and the invariants.
To emphasize this consider the second Piola-Kirchhoff stress given by 
\begin{equation}\label{eq:StressRepEx}
    \Sb = c_{0}^{C} \Ib + c_{1}^{C} \Cb + c_{-1}^{C} \Cb^{-1}.
\end{equation}
Naively,  one could presume that using the Kirchhoff stress tensor ${\boldsymbol{\tau}}$ and the left Cauchy-Green tensor $\Bb$ with an equivalent basis representation ($\Ib$, $\Bb$, $\Bb^{-1}$), i.e.
\begin{equation}
       {\boldsymbol{\tau}} = c^{B}_{0} \Ib + c^{B}_{1} \Bb + c^{B}_{-1} \Bb^{-1}
\end{equation}
the respective coefficients might be the same, e.g. $c^C_{a}= c_{a}^{B}$ for $a=0,1,-1$.
However, recalling that the Kirchhoff stress can be expressed as $ {\boldsymbol{\tau}} = \Fb \Sb \Fb^{T}$, \eref{eq:StressRepEx} can be rewritten as
\begin{equation}
    {\boldsymbol{\tau}} = c_{0}^{C} \Bb + c_{1}^{C} \Bb^{2} + c_{-1}^{C} \Ib.
\end{equation}
Hence, under the assumption that the eigenvalues are unique, we find that
\begin{equation}
        c^{B}_{0} \Ib + c^{B}_{1} \Bb + c^{B}_{-1} \Bb^{-1} = c_{0}^{C} \Bb + c_{1}^{C} \Bb^{2} + c_{-1}^{C} \Ib 
\end{equation}
which yields
\begin{equation}
    \begin{aligned}
       c^{B}_{0} =  c_{-1}^{C} - c_{1}^{C} I_{2} , \qquad 
       c_{1}^{B}   = c_{0}^{C} + c_{1}^{C} I_{1} , \qquad
       c^{B}_{-1}  = c_{1}^{C} I_{3}.
    \end{aligned}
\end{equation}
via the Cayley-Hamilton theorem \eqref{eq:cayley-hamilton}.
The complexity of the two sets of coefficient functions $\lbrace c^C_{0}, c^C_{1}, c^C_{-1} \rbrace$ and $\lbrace  c^{B}_{0}, c^{B}_{1}, c^{B}_{-1} \rbrace$ is therefore clearly different.
Using the Cayley-Hamilton theorem \eqref{eq:cayley-hamilton} to transform the model representation would also alter the complexity of the coefficient functions.
In order to make the following comparisons as fair as possible we have restricted ourselves to second Piola-Kirchhoff stress representations and data.
Note that the Piola transform would also affect the orthogonality of the basis.

We furthermore remark that additively separable energies that are based on the Valanis-Landel hypothesis \cite{treloar1974mechanics,jones1975properties,ogden1986recent} lead to more trivial calibrations, i.e. if 
\begin{equation}
    \Psi(I_{1},I_{2},I_{3}) = \Psi_{1}(I_{1}) + \Psi_{2}(I_{2}) + \Psi_{3}(I_{3})
\end{equation}
we can see from \eref{eq:cfunsICCinv} that this would result in 
\begin{equation}
    \begin{aligned}
            c_0(I_{1},I_{2}) &= 2 \left[ \partial_{I_1} \Psi_{1}(I_{1}) +  I_{1}  \partial_{I_2} \Psi_{2}(I_{2}) \right] \\
    c_1(I_{2}) &= - 2 \, \partial_{I_2} \Psi_{2}(I_{2}) \\
    c_{-1}(I_{3}) &= 2 I_{3} \partial_{I_3} \Psi_{3}(I_{3}).
    \end{aligned}
\end{equation} 
Hence, this leads to $c_{1}$ and $c_{-1}$ being functions of only one invariant and $c_{0}$ reduced to a function of two invariants. 
In order to avoid these simplifications we use only hyperelastic models that are not additively decomposable with regards to their inputs. 

Note the definition of the invariants can be engineered to reduce the complexity of the coefficient functions for a particular material dataset.
In a limiting case, the coefficient functions are themselves invariants and hence present the simplest representation in some sense, albeit one that is hard to discover \apriori from the measured data.
Representation complexity is particularly important in the low data regime which we explore.

%.....................................................
\subsection{Data models}
%.....................................................
Three well-known compressible hyperelastic models were selected to generate training data:  
(a) Mooney-Rivlin \cite{mooney1940theory,rivlin1948large},  
(b) a modified version of Carroll's hyperelastic law \cite{carroll2011strain, melly2021improved}, 
and (c) a Gent-type model \cite{gent1996new,pucci2002note}.
Each is expressed in terms of the invariants $I_{1}=\tr(\Cb)$, $I_{2}=\tr(\Cb^{-1}) \det (\Cb)$,  and $J=\sqrt{\det \Cb}$

The specific compressible Mooney-Rivlin model considered here has the strain energy function
\begin{equation}
\Psi = \theta_{1} \left( \frac{I_{1}}{J^{2/3}}- 3 \right) + \theta_{2} \left( \frac{I_{2}}{J^{4/3}} -3 \right) + \theta_{3} (J-1)^{2}\ ,
\end{equation}
which yields a second Piola-Kirchhoff stress of the form:
\begin{equation}
\begin{aligned}
    \Sb &= \underbrace{2 \left( \frac{\theta_{1}}{J^{2/3}} + I_{1} \frac{\theta_{2}}{J^{4/3}} \right)}_{c^*_{0}} \Ib \underbrace{- 2 \frac{\theta_2}{J^{4/3}}}_{c^*_{1}}  \Cb + \underbrace{J \left[-\frac{2}{3} \theta_{1} \frac{I_{1}}{J^{5/3}} - \frac{4}{3} \theta_{2} \frac{I_{2}}{J^{7/3}} + 2 \theta_{3} (J-1) \right] }_{c^*_{-1}}  \Cb^{-1} 
\end{aligned}
\end{equation}
We use (scaled) material parameters (
$\theta_{1}=0.92$ Pa, 
$\theta_{2}=2.37$ Pa and 
$\theta_{3}=10.001$ MPa) from fits to vulcanized rubber data, c.f. \cref{peng2021consistently}, for data generation.

Following \cref{melly2021improved}, a modified Carroll model is defined by the strain energy function
\begin{equation}
    \Psi = \theta_{1} \left( \frac{I_{1}}{J^{2/3}} - 3 \right) + \theta_{2} \left[ \left( \frac{I_{1}}{J^{2/3}} \right)^{4} - 81 \right] + \theta_{3} \left( \sqrt{\frac{I_{2}}{J^{4/3}}} - \sqrt{3} \right) + \theta_{4} (J-1)^{2} \ .
\end{equation}
This energy results in 
\begin{equation}
\begin{aligned}
    \Sb &= 
     \underbrace{2 \left[ \frac{\theta_{1}}{J^{2/3}} + 4 \theta_{2} \left( \frac{I_{1}}{J^{2/3}} \right)^{3} + I_1 \left( \theta_{2} \frac{1}{2 {J^{2/3} I_{2}}}  \right) \right]}_{c^*_0} \Ib 
    \underbrace{- 2 \theta_{2} \frac{1}{2 {J^{2/3} I_{2}}}}_{c^*_1} \, \Cb \\
    & + \underbrace{J \left( -\frac{2}{3} \theta_{1} \frac{I_{1}}{J^{4/3}} - \frac{8}{3} \theta_{2} \frac{I_{1}^{4}}{J^{11/3}} - \frac{2}{3} \theta_{3} \frac{\sqrt{I_{2}}}{J^{5/3}} + \theta_{4} (J-1) \right)}_{c^*_{-1}} \, \Cb^{-1}
\end{aligned}
\end{equation}
We use a scaled version of the material parameters reported in \cref{melly2021improved}, in particular 
    $\theta_{1} =151.09387$ GPa,
    $\theta_{2} =0.3028$ MPa,
    $\theta_{3} =68.33070$ GPa, and
    $\theta_{4} = 500$ TPa.

Lastly, we also utilize the response of a compressible version of the Gent+Gent model, as named in \cref{ogden2004fitting}, that is defined by the strain energy function
\begin{equation} \label{eq:gent_energy}
\Psi = -\frac{\theta_{1}}{2} J_{m} \log \left( 1 - \frac{I_{1}-3}{J_{m}} \right) - \theta_{2} \log \left( \frac{I_{2}}{J} \right) + \theta_{3} \left( \frac{1}{2} (J^{2}-1) - \log J \right) \ ,
\end{equation}
where we choose 
$\theta_1= 2.4195$ MPa, 
$\theta_{2}=1.8146$ MPa, 
$\theta_{3}=1.2097$ MPa and 
$J_{m}=77.931$.
This strain energy yields a second Piola-Kirchhoff stress of the form:
\begin{equation}
\Sb =    \underbrace{2 \left[ - \frac{\theta_{1}}{2} J_{m} \frac{1}{I_{1}-3-J_m} +  \theta_{2} \frac{I_{1}}{I_{2}}   \right]}_{c^*_0} \Ib 
    \underbrace{- 2 \theta_{2} \frac{1}{I_{2}}}_{c^*_1} \, \Cb 
    + \underbrace{J \, \theta_{3} (J - \frac{1}{J})}_{c^*_{-1}} \, \Cb^{-1} 
\end{equation}
The Gent+Gent model is not polyconvex; however, it is convex over a limited range where $I_1 < J_m + 3$.
For simplicity, we refer to this model simply as Gent.

Note that hereafter $c^*_a$ denote the true coefficients, which differ from the extracted coefficients $c_a$ near ill-conditioned solves, and the fitted NN coefficients $\hat{c}_a$.

%....................................................
\subsection{Training and validation} \label{sec:training}
%....................................................

\begin{figure}[h!] 
\centering
\includegraphics[width=0.9\textwidth]{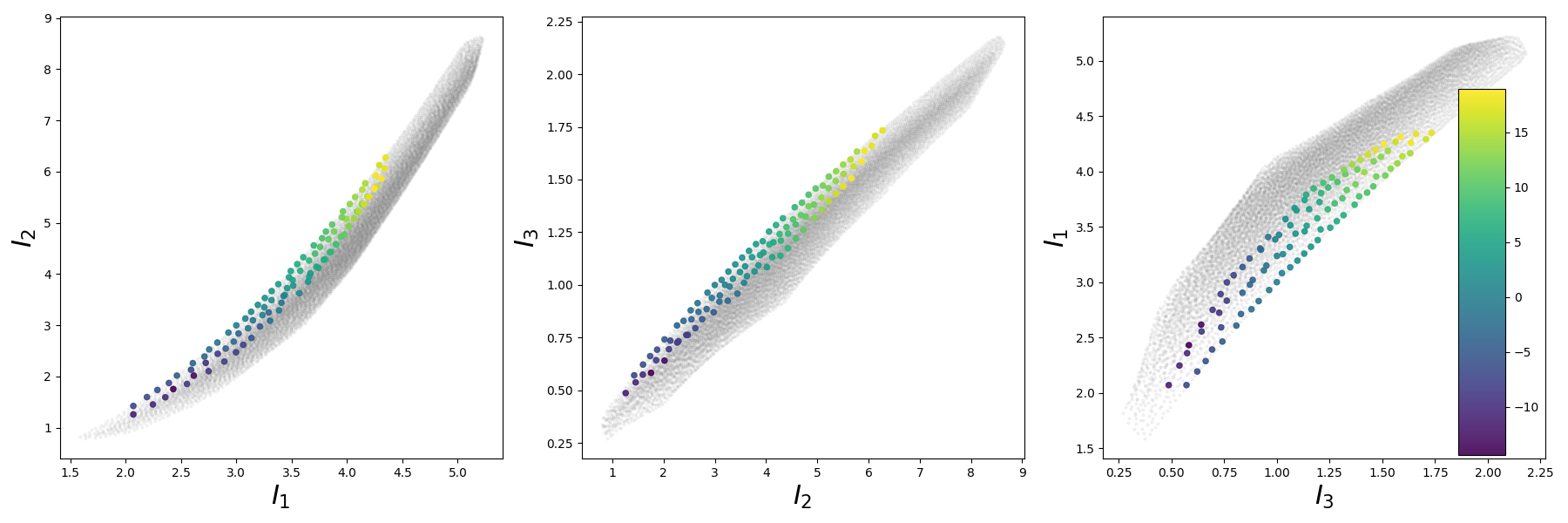}
\caption{Space-filling training/validation (colored by maximum stress component normalized by the mean for the Mooney-Rivlin data) and test samples in invariant space. }
\label{fig:SpaceFilling}
\end{figure}

For sampling, we define a nine-dimensional space around the undeformed configuration of the deformation gradient as
\begin{equation}
    \overline{F}_{ij} \in [F_{ij}^{L}, F_{ij}^{U}] = \begin{cases}
        1 - \delta \leq 1 \leq 1+\delta, & \text{when} \, \,  $i=j$, \\
        -\delta \leq 0 \leq \delta, & \text{otherwise}.
    \end{cases}
\end{equation}
We then define a training region with $\delta=0.2$ and a test region with $\delta=0.3$ and use the space-filling sampling technique proposed in \cref{fuhg2022physics} to generate $100$ training points and $10,000$ test points that fill their respective spaces. 
\Fref{fig:SpaceFilling} shows the spread of these samples in invariant space ($I_{1}, I_{2}, J$). 
Then given the triple $(I_{1},I_{2},I_{3}=J^{2})$ we can reconstruct the right Cauchy-Green tensor as
\begin{equation}
    \begin{aligned}
        \Cb = \begin{bmatrix}
            \frac{1}{3} I_{1} - 2\sqrt{H} \cos \left( \frac{\pi- \beta}{3}\right) & 0 & 0 \\
            0 & \frac{1}{3} I_{1} - 2\sqrt{H} \cos \left( \frac{\pi+ \beta}{3}\right) & 0 \\
            0 & 0 & \frac{1}{3} I_{1} - 2\sqrt{H} \cos \left( \frac{\beta}{3}\right)
        \end{bmatrix}
    \end{aligned}
\end{equation}
where
\begin{equation}
    H= \frac{1}{9} (I_{1}^{2}- 3 I_{2}), \qquad G = \frac{1}{3} I_{1} I_{2} - I_{3} - \frac{2}{27} I_{1}^{3}, \qquad \beta = \arccos \left( - \frac{G}{2 H^{3/2}} \right)
\end{equation}
from which we can obtain the values of the invariants and the basis of all the investigated stress representations of \sref{sec:formulation}. 
When training the model we use an $80/20$ split to obtain $20$ validation data points. 

\begin{figure}[h!] 
\centering
\includegraphics[width=0.9\textwidth]{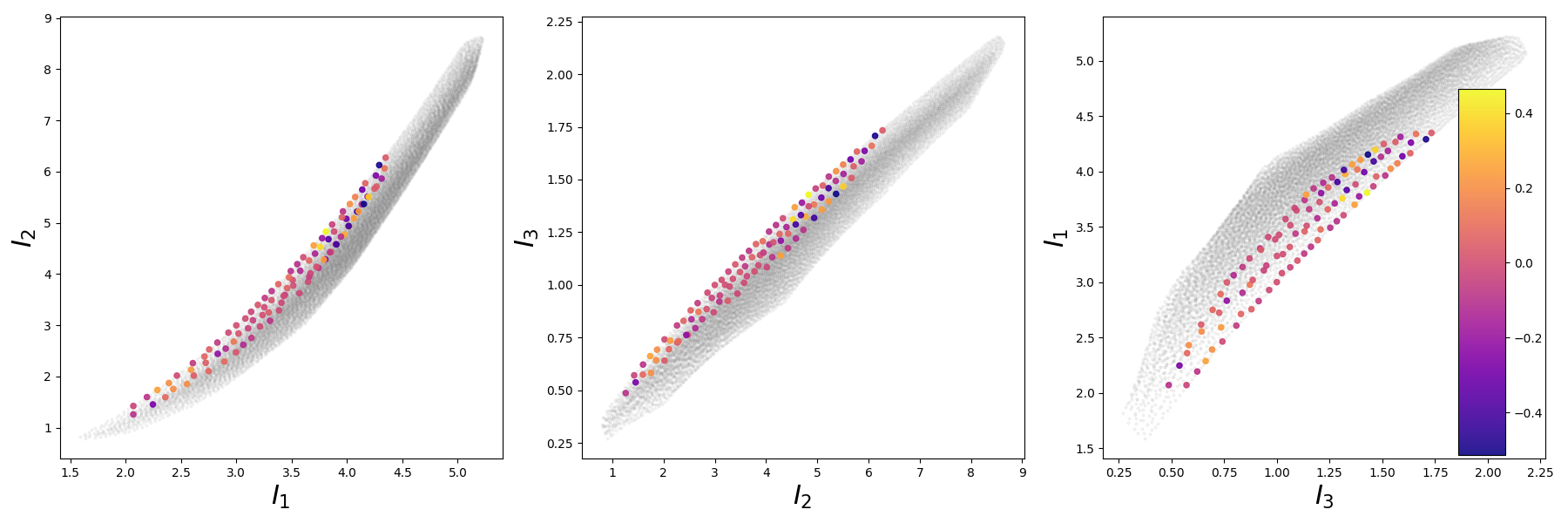}
\caption{Effective stress noise in percent of Frobenius norm error over the test data of a Mooney-Rivlin dataset. }
\label{fig:InvariantSpaceNoise}
\end{figure}

After obtaining a set of training and testing data we added noise to the resulting coefficient values to disrupt symmetries and analytic functional forms. 
In particular, we take the coefficient values $c_{0}, c_{1}, c_{-1}$ corresponding to the representation
\begin{equation}
    \Sb = c_{0} \Ib + c_{1} \Cb + c_{-1} \Cb^{-1}
\end{equation}
for every data point, and define a noisy version $\tilde{c}_{a} = c_{a}+ \mathcal{N}(0, 0.02 \, |c_{a}|)$ for $a=0,1,-1$ which then gives a noisy stress
\begin{equation}
    \tilde{\Sb} = \tilde{c}_{0} \Ib + \tilde{c}_{1} \Cb + \tilde{c}_{-1} \Cb^{-1} \ .
\end{equation}
This $\tilde{\Sb}$ was then used as the target stress to obtain the coefficients for all other models, e.g. the Criscione model.
Hence, we generate $100$ noisy training samples that have the same invariants as the noiseless counterparts and use the same noiseless data for the test set. 
An example of a generated noisy test set is shown in \Fref{fig:InvariantSpaceNoise} for the Mooney-Rivlin model.

All the tensor basis neural network models \cite{ling2016machine} were implemented in {\it PyTorch} \cite{paszke2019pytorch}. 
Potential models were formed from a multilayer, densely connected feedforward neural network (NN) with a single output
\begin{equation}
\Phi = \NN(\Ic)
\end{equation}
where the coefficients $c_a = \partial_{I_a} \Phi$ are obtained through automatic differentiation.
Summation with the known basis provides the stress
\begin{equation}
\Sb = \sum_a c_a \Bb_a
\end{equation}
The coefficient-based models utilized a monolithic NN with 3 outputs
\begin{equation}
c_a = \NN_a(\Ic)
\end{equation}
and the same summation to form the stress.
To be consistent all TBNN models consisted of 3 layers with 30 neurons per layer and a {\it Softplus} activation function \cite{glorot2011deep}. 
\aref{app:potentialNN} and \aref{app:coeffNN} provide additional details of the implementation of potential and coefficient-based TBNNs, respectively.

The training loss was formulated on the mean squared error of either the stress components or the coefficients 
\begin{equation}
\text{MSE} =  \lambda_\Sb \sum_{i=1}^{N_\text{train}} \| \Sb_i - \hat{\Sb}_{i} \|^2  + \lambda_c \sum_{i=1}^{N_\text{train}} \sum_{a=1}^3 \| [c_a]_i- [\hat{c}_a]_{i} \|^2 \ ,
\end{equation}
since these are available from representative volume element (RVE)/experimental data, whereas the strain energy is less accessible.
Although mixing both losses proved useful in preliminary studies, all reported data is from models trained with either $\lambda_\Sb = 1, \lambda_c = 0$ or $\lambda_\Sb = 0, \lambda_c = 1$.
Note the coefficients scale as $| c_a | \sim \| \Sb |\ \| \Bb_a \|^{-1}$ so the coefficient-based loss can suffer from numerical conditioning issues.
All models were trained for $50,000$ epochs using the {\it Adam}  optimizer \cite{kingma2014adam} with a constant learning rate of $0.001$. 

We compared the performance of the models using the normalized mean squared error of the stress over the $10,000$ testing data points
\begin{equation}
\text{RMSE} =  \sqrt{\frac{\sum_{i=1}^{N_\text{test}} \| \Sb_i - \hat{\Sb}(\Cb_{i}) \|^2 }{ \sum_i \|  {\Sb}_{i} \|^2 }}
\end{equation}
where $\hat{\Sb}_{i}$ is the predicted stress and $\Sb_i$ represents the ground truth at data point $i$.

%%%%%%%%%%%%%%%%%%%%%%%%%%%%%%%%%%%%%%%%%%%%%%%%%
\section{Results} \label{sec:results}
%%%%%%%%%%%%%%%%%%%%%%%%%%%%%%%%%%%%%%%%%%%%%%%%%

First, we survey the test losses of the models enumerated in \tref{tab:nn_variants}.
Then we undertake detailed investigations of why the theoretically equivalent representations do or do not perform well by examining where the largest errors occur and how the predictions compare to held-out data.

%.......................................................................
\subsection{Comparison of test losses} \label{sec:loss_comp}
\begin{figure}[h!] 
\centering
\includegraphics[width=0.95\textwidth]{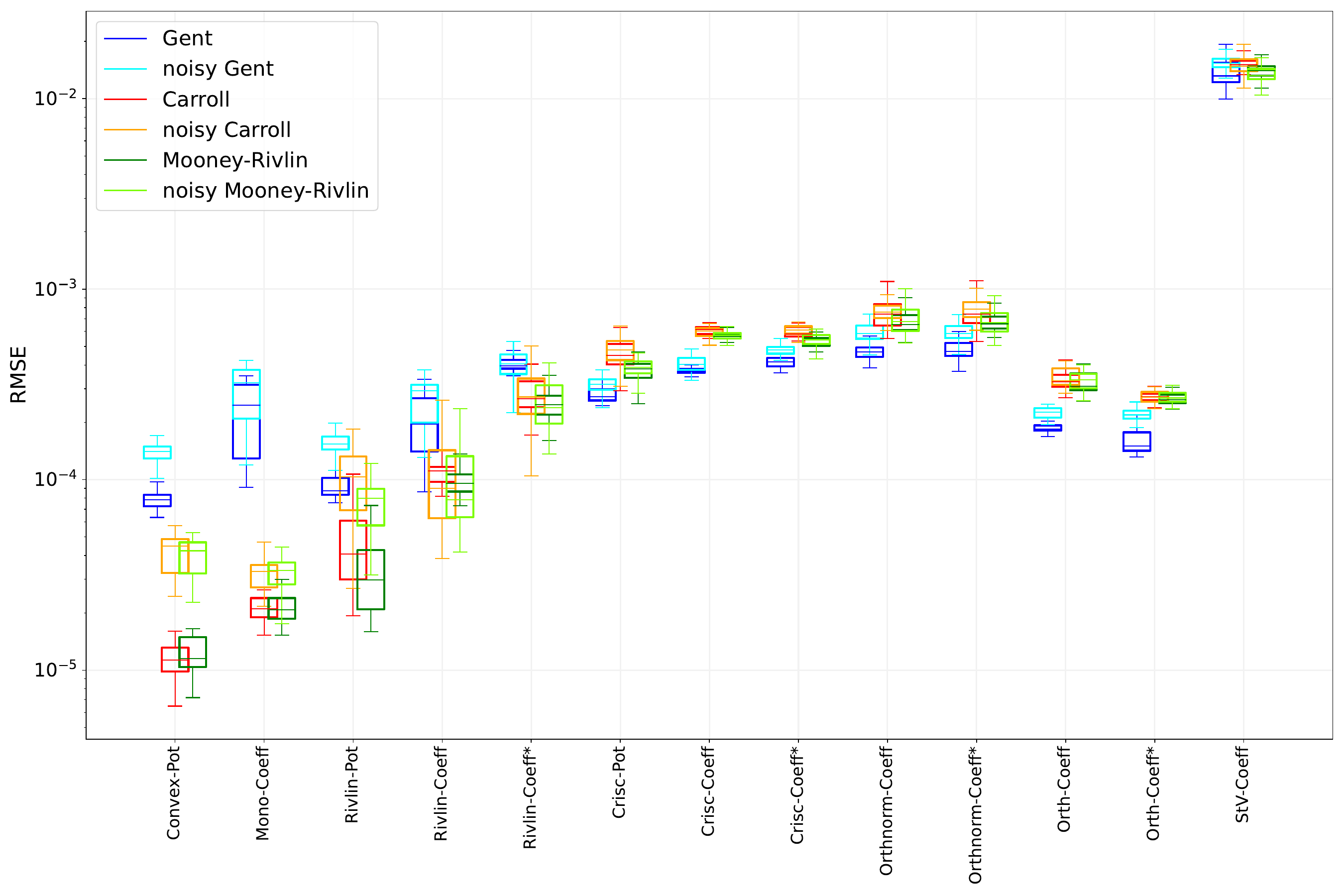}
\caption{Comparison of TBNN variants summarized in \tref{tab:nn_variants} over test sets for the three different data models with and without noise.
Note that coefficient-based variants with $^*$ were trained to stress data, and had to discover the coefficient values. 
Also the $\Eb$-based variant of the St.Venant basis \eref{eq:StV-E} had errors well above the displayed range so only the $\Cb$-based variant \eref{eq:StV-C} is shown.}
\label{fig:test_error_comparison}
\end{figure}

%.......................................................................
For each TBNN model described in \tref{tab:nn_variants} we assembled an ensemble of 30 parameterizations using random initialization of the NN parameters and shuffling the training/validation subsets of the 100 training points.
\Fref{fig:test_error_comparison} shows the range of RMSE test errors for the six datasets from the 3 traditional models described in \sref{sec:data}, each with and without noise.
Clearly, the various theoretically equivalent TBNN formulations perform differently and each of the datasets evoke different errors.
Overall the polyconvex (Conv-Pot) and monotonic (Mono-Coeff) models appear to perform the best, although the standard potential-based (Rivlin-Pot) model has comparable performance to Mono-Coeff.
The coefficient-based Rivlin-Coeff has considerably higher test errors than the potential-based Rivlin-Pot, despite Rivlin-Pot relying on automatic differentiation.
The Criscione and other orthogonal models perform worse than the convex, monotonic and Rivlin models but they do better on the Gent data than on the other two datasets.
We observe that the Gent model has a qualitatively different functional form than the selected Mooney-Rivlin and Carroll data-generating models, e.g. the presence of log terms in the energy \eref{eq:gent_energy}.
The Orth model with smoother invariants performs the best of the orthogonal basis models and is an anomaly in that it trains better indirectly to stress than to the extracted coefficients.
This may be due to the conditioning issues with solving for coefficients of a power basis, mentioned in \sref{sec:training}.
The St. Venant model (both $\Cb$ and $\Eb$ based) is an outlier with large errors likely due to the mismatch in the growth of the tensor basis and the data, which necessitates more complex coefficient functions.
Although small, the differences in calibration techniques can have an effect. 
Training to stress, instead of extracted coefficients, can regularize the trained coefficient functions, since stress is smoother than the coefficient functions as per the Man-Serrin theorem discussed in \sref{sec:addl_constraints}. 
Also training to stress can discourage a potential reinforcement of bias from individually trained coefficient functions that need to coordinate to form an accurate stress.
Training to coefficients, on the other hand, removes the potentially ill-conditioned linear algebra implicit in training to stress.
Projection of data onto expected bases may also remove discrepancies as with the noisy datasets.

The test errors seem to be largely dominated by testing the models in extrapolation, more insight will be given in the following sections.

%.......................................................................
\subsection{Locations of worst errors} \label{sec:error_locations}
%.......................................................................

\begin{figure}[h!]
\centering
\begin{subfigure}[b]{0.95\linewidth} \centering
\includegraphics[width=0.99\textwidth]{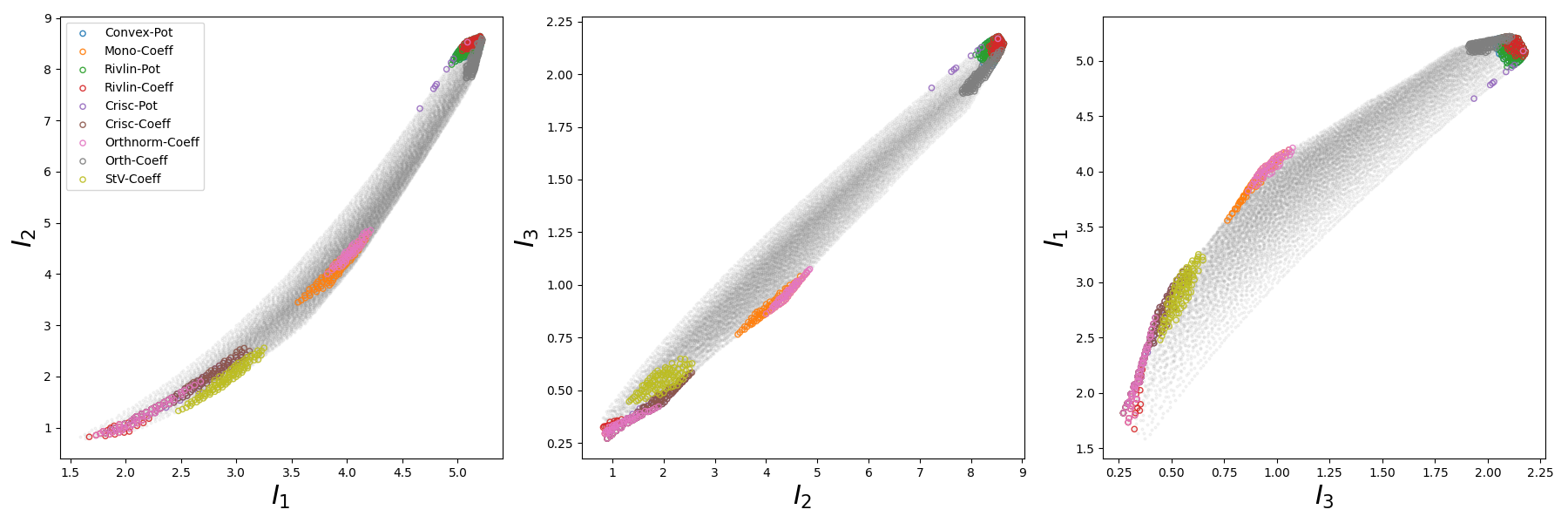}
\subcaption{Without noise}
\end{subfigure}
\begin{subfigure}[b]{0.95\linewidth} \centering
\includegraphics[width=0.99\textwidth]{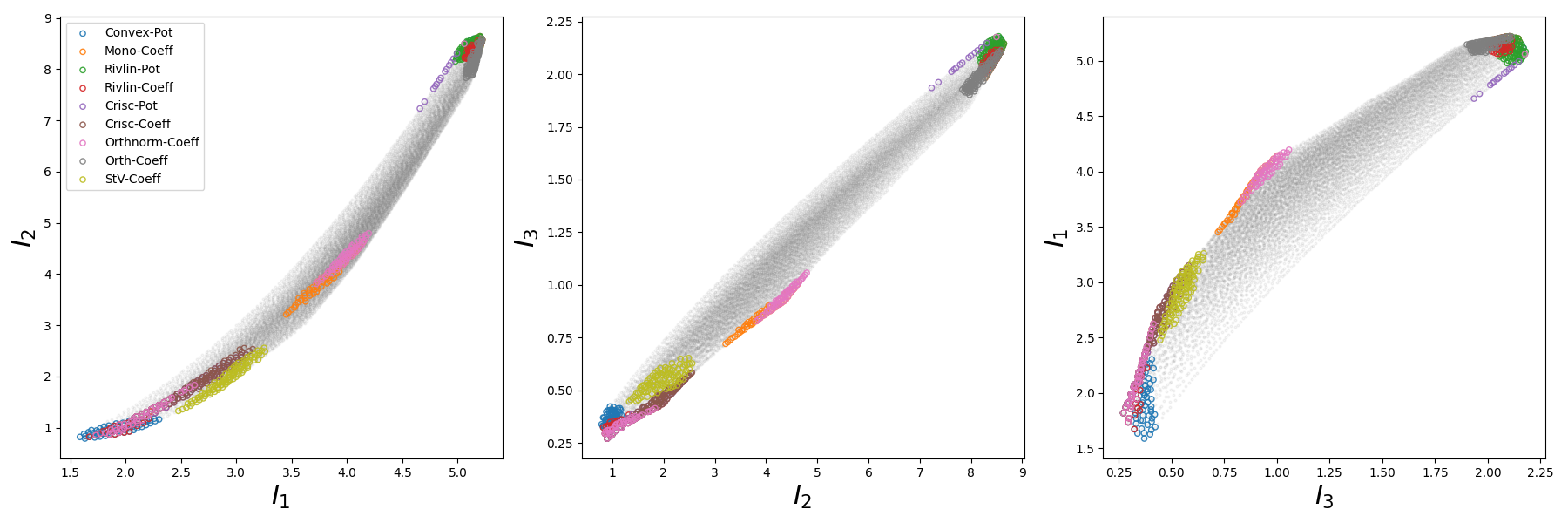}
\subcaption{With noise}
\end{subfigure}
\caption{Mooney-Rivlin data, worst 1\% errors for each model.} \label{fig:worst_mooneyrivlin}
\end{figure}

\begin{figure}[h!]
\centering
\begin{subfigure}[b]{0.95\linewidth} \centering
\includegraphics[width=0.99\textwidth]{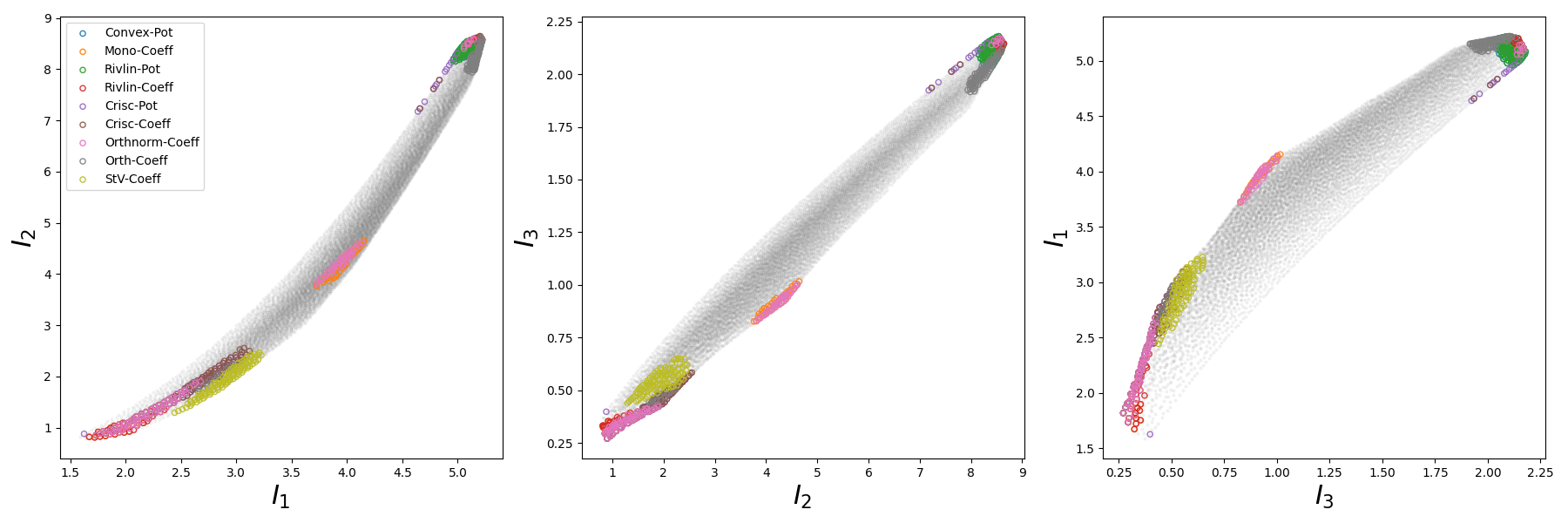}
\subcaption{Without noise}
\end{subfigure}
\begin{subfigure}[b]{0.95\linewidth} \centering
\includegraphics[width=0.99\textwidth]{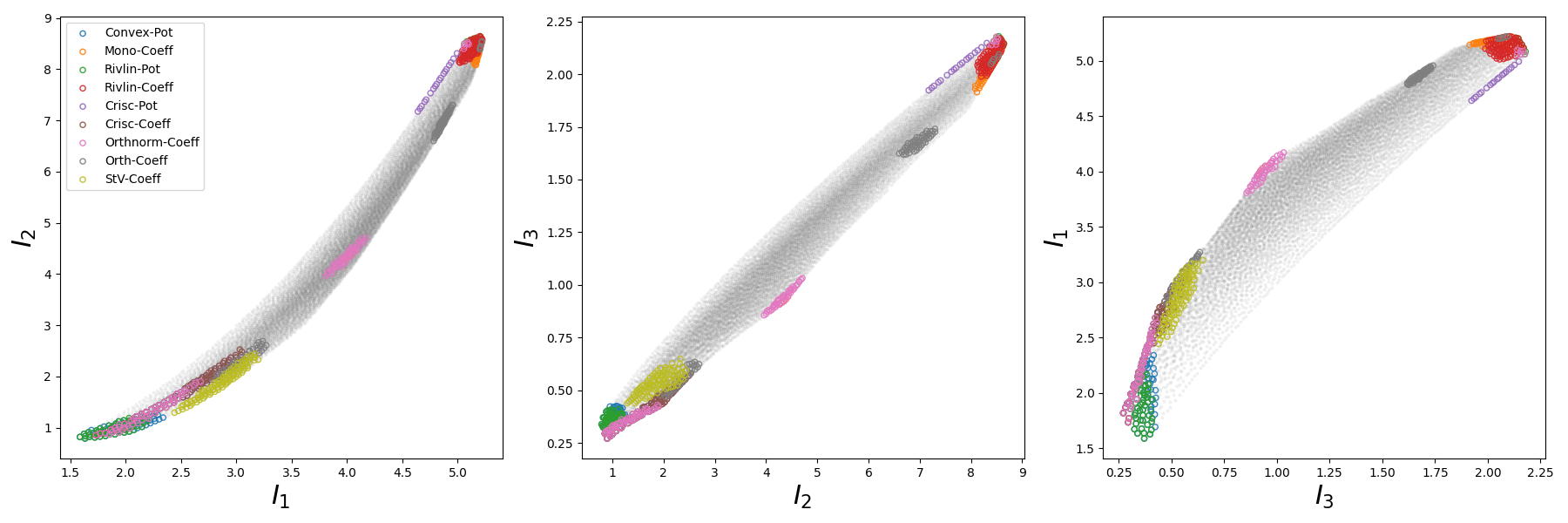}
\subcaption{With noise}
\end{subfigure}
\caption{Modified Carroll data, worst 1\% errors for each model.} \label{fig:worst_carroll}
\end{figure}

\begin{figure}[h!]
\centering
\begin{subfigure}[b]{0.95\linewidth} \centering
\includegraphics[width=0.99\textwidth]{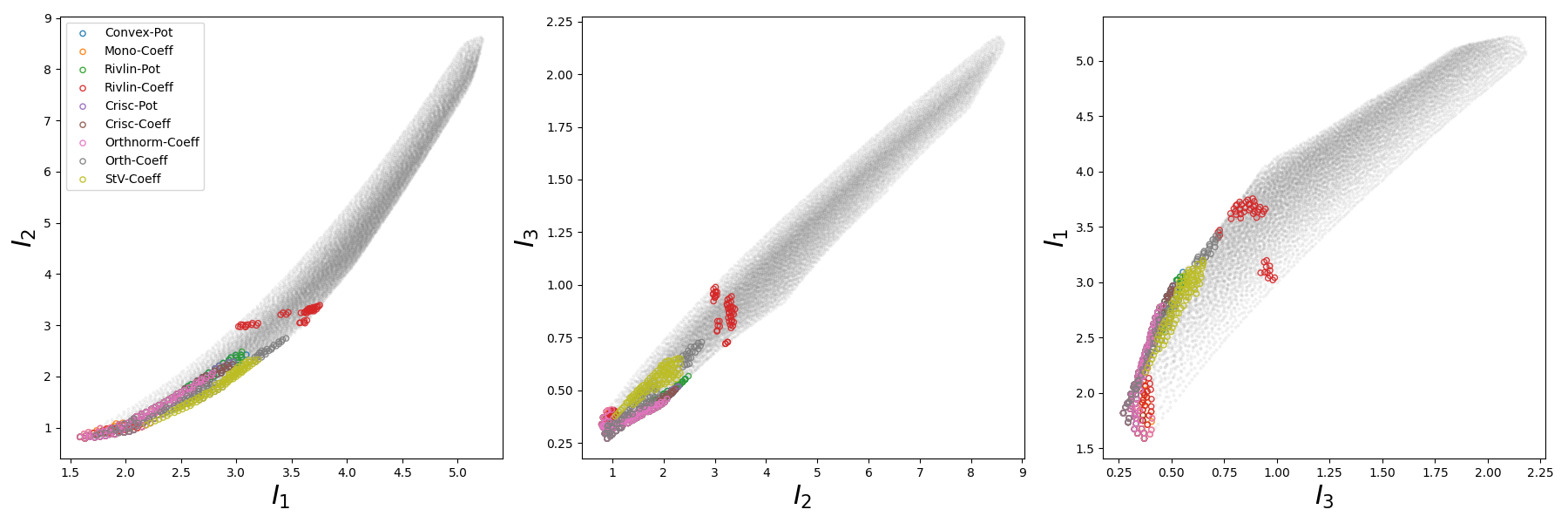}
\subcaption{Without noise}
\end{subfigure}
\begin{subfigure}[b]{0.95\linewidth} \centering
\includegraphics[width=0.99\textwidth]{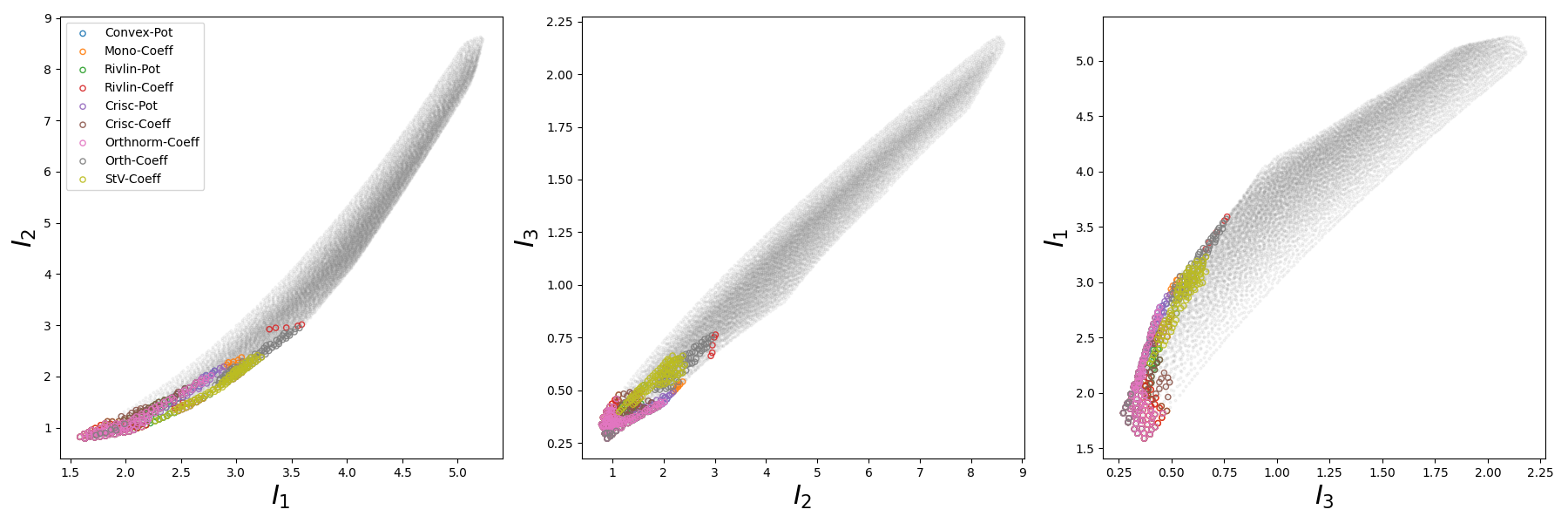}
\subcaption{With noise}
\end{subfigure}
\caption{Gent data, worst 1\% errors for each model.} \label{fig:worst_gent}
\end{figure}

The worst 1\% errors of the 10,000 sample test set for each model are shown in \Fref{fig:worst_mooneyrivlin} for the Mooney-Rivlin data,  in \Fref{fig:worst_carroll} for the modified Carroll data, and in \Fref{fig:worst_gent} for the Gent data.
For reference, the undeformed state is at $(I_1=3,I_2=3,I_3=1)$ which is inside the hull of sample points shown in these figures.
Generally speaking, for most models, the worst errors are at the boundary of the test locus where they are forced into extrapolation.
Note that $I_3$ is associated with volumetric deformation, $I_1$ can be interpreted at the linearization of $I_3$, while $I_2$ is sensitive to shear and deviatoric deformations.

Examining \Fref{fig:worst_mooneyrivlin}, the Convex-Pot, Rivlin-Pot, and Orth-Coeff have the largest errors where the invariants (and eigenvalues of $\Cb$) are large, while the Crisc-Coeff and the similar Orthnorm-Coeff have largest errors in the low region.
The Crisc-Pot has relatively large errors at both extremes.
Of the better-performing models, the Mono-Coeff formulation is an outlier since it has its worst errors in the midrange of the invariants, albeit still at the boundary.
Likewise, the Orthnorm-Coeff has high error in the midrange, as well as the low range, while the St. Venant model performs particularly poorly in the low to midrange.
The patterns are relatively unchanged with discrepancy added by noise, with the exception of the worst errors transitioning to the lower range for the most accurate model, Convex-Pot.
The errors for the Carroll data shown in \Fref{fig:worst_carroll} largely resemble those for the Mooney-Rivlin data, although for this data the Mono-Coeff and Orth-Coeff models seem more sensitive to noise. 
They, together with the Convex-Pot, shift their worst error locations with noise.
The errors for the Gent data, shown in \Fref{fig:worst_gent}, however, present different patterns.
For this case, all models perform worst where the invariants are small, which can be ascribed to the Gent model being ill-defined where $I_1$ becomes less than a value determined by the parameter $J_m$. 
There are also scattered worst error locations in the midrange for the Rivlin-Coeff model.

\begin{figure}[h!]
\centering
\begin{subfigure}[b]{0.4\linewidth} \centering
\includegraphics[width=0.9\textwidth]{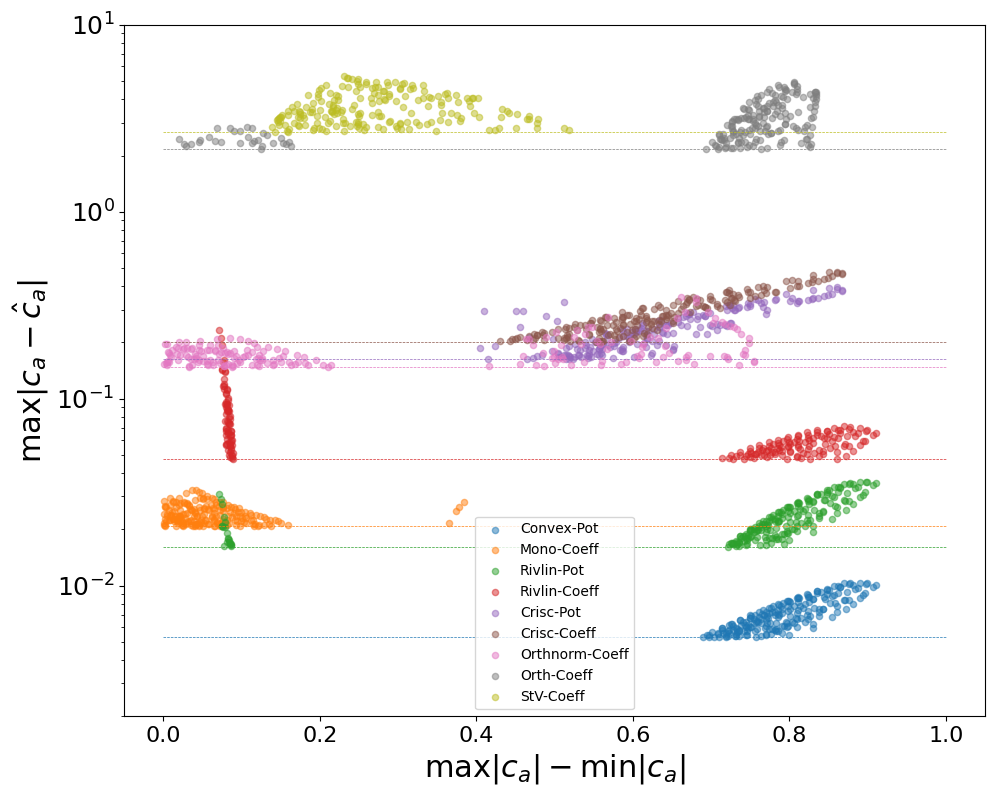}
\subcaption{Without noise}
\end{subfigure}
\begin{subfigure}[b]{0.4\linewidth} \centering
\includegraphics[width=0.9\textwidth]{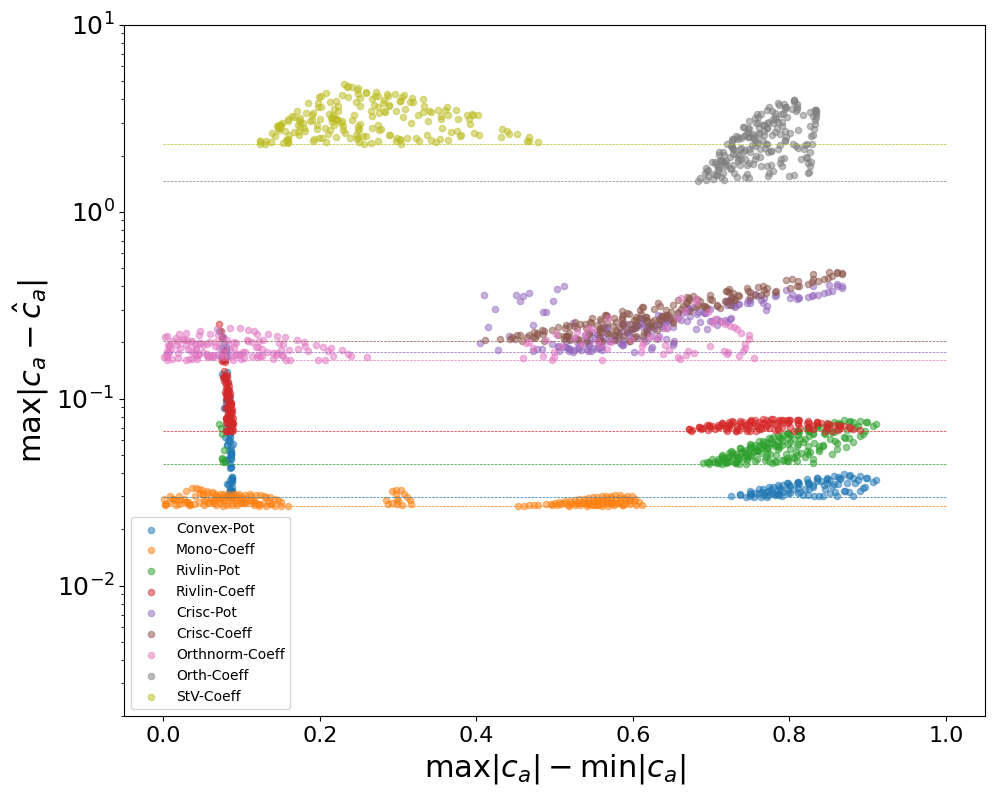}
\subcaption{With noise}
\end{subfigure}
\caption{Maximal errors for each model for the Mooney-Rivlin data. The horizontal lines indicate where the data has been clipped.} \label{fig:corr_mooneyrivlin}
\end{figure}

\begin{figure}[h!]
\centering
\begin{subfigure}[b]{0.4\linewidth} \centering
\includegraphics[width=0.9\textwidth]{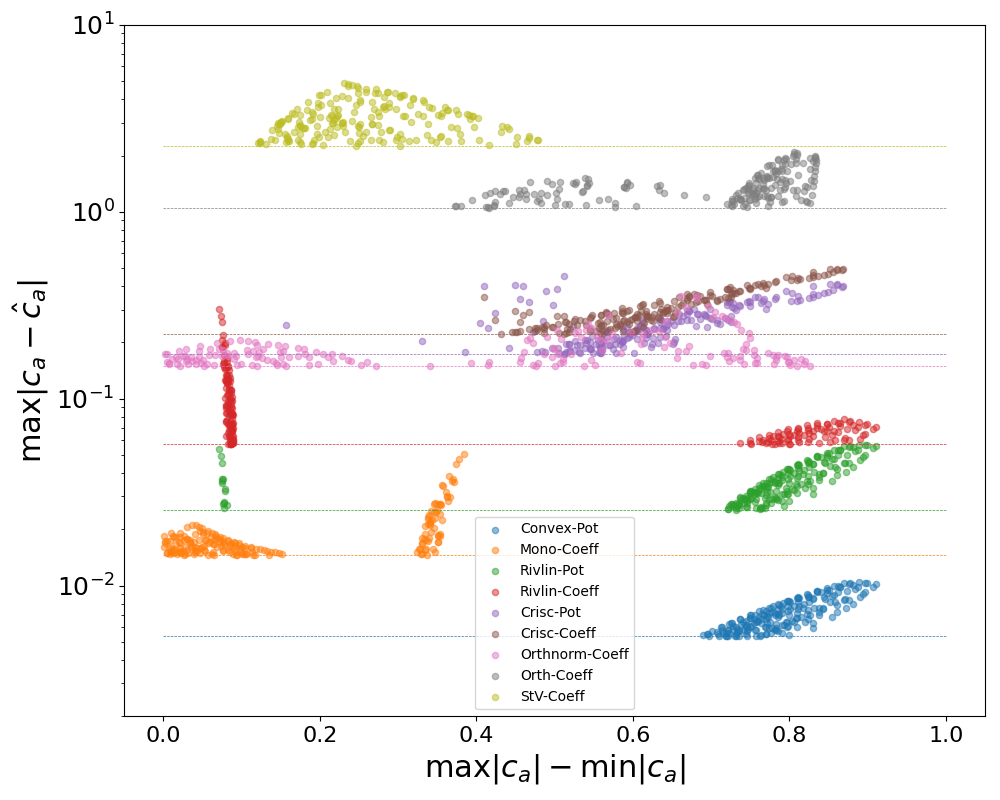}
\subcaption{Without noise}
\end{subfigure}
\begin{subfigure}[b]{0.4\linewidth} \centering
\includegraphics[width=0.9\textwidth]{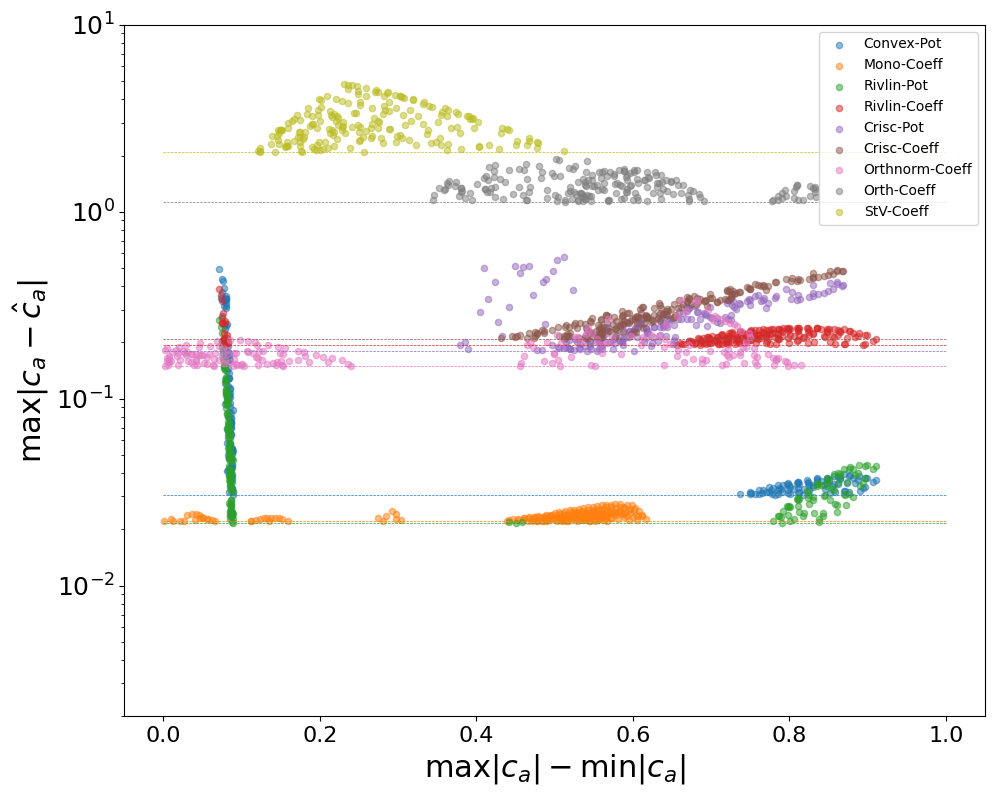}
\subcaption{With noise}
\end{subfigure}
\caption{Maximal errors for each model for the modified Carroll data. The horizontal lines indicate where the data has been clipped. } \label{fig:corr_carroll}
\end{figure}

\begin{figure}[h!]
\centering
\begin{subfigure}[b]{0.4\linewidth} \centering
\includegraphics[width=0.9\textwidth]{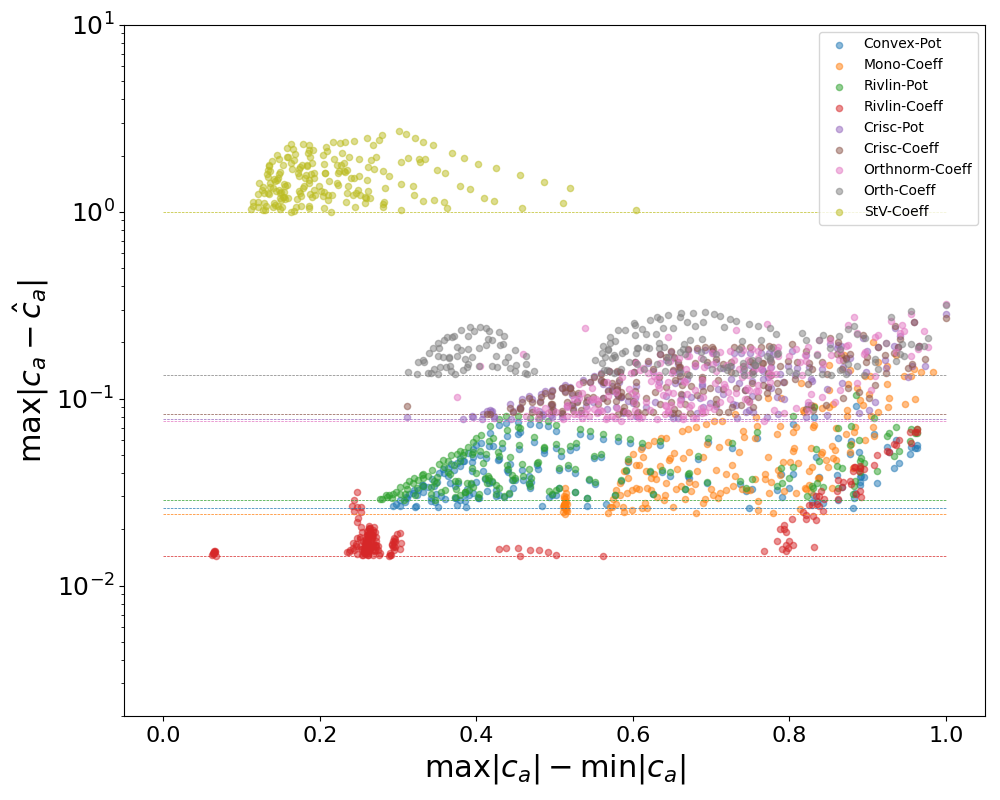}
\subcaption{Without noise}
\end{subfigure}
\begin{subfigure}[b]{0.4\linewidth} \centering
\includegraphics[width=0.9\textwidth]{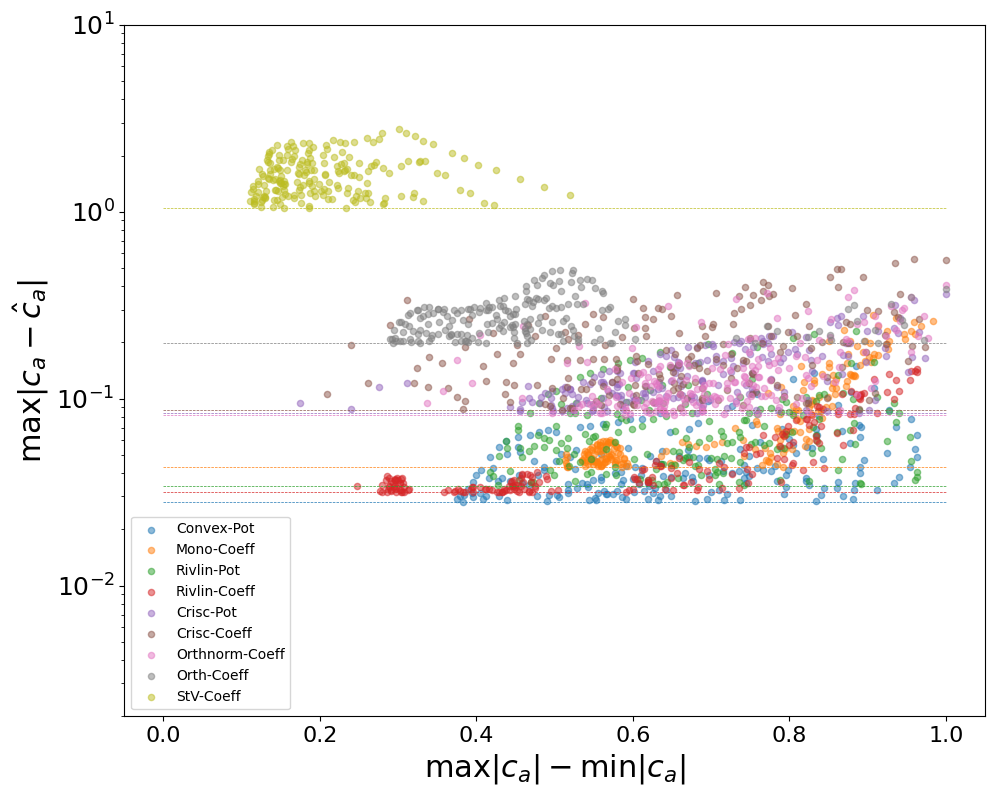}
\subcaption{With noise}
\end{subfigure}
\caption{Maximal errors for each model for the Gent data. The horizontal lines indicate where the data has been clipped.} \label{fig:corr_gent}
\end{figure}

\Fref{fig:corr_mooneyrivlin}, \Fref{fig:corr_carroll}, and \Fref{fig:corr_gent} 
 provide another view of the error patterns and corroborate the observations from the previous plots.
These figures illustrate the correlation of the maximum errors with the difference in the largest and smallest (Rivlin-Ericksen) coefficient values for a particular data point in the test set. 
Large differences in the coefficient values are associated with less symmetric deformations.
For each model, the figures show how the worst errors shift as a function of the difference in the coefficient values. 
For the Mooney-Rivlin data, only the worst-performing model, StV-C, has a single locus of maximal errors. 
Of the best-performing models, Convex-Pot, Rivlin-Pot, and Rivlin-Coeff have similar patterns that remain stable after the injection of noise.
The Mono-Coeff model, on the other hand, changes the locations of where the worst errors occur relative to the difference in the coefficient values and also has the worst errors on par with Convex-Pot.
Again the patterns for the Carroll data are similar to the Mooney-Rivlin data, while the Gent data present qualitatively different patterns.
For the Gent data, all models, except the worst performing StV-C, have overlapping loci of worst errors that do not change appreciably with noise.
This is perhaps due to discrepancies between the relatively simple TBNN models and the Gent data.

Next, we aim to test the performance of the  TBNN variants to generalization.
We conjecture that the differences in generalization performance can be attributed to how much of the complexity of the stress-strain mapping is intrinsically provided by the nonlinearity of the stress representation bases. 
Meaning, if the nonlinearity of the bases in the chosen representation is approaching the nonlinearity of the stress-strain mapping then the coefficient can be described by simpler lower-order functions. 
If the coefficients are lower-order functions, i.e. constant, linear, or quadratic, then accurately extrapolating this behavior with a TBNN is going to require less training data and a less complex NN to be accurate. 

\begin{figure}[h!]
\centering
\begin{subfigure}[b]{0.3\linewidth} \centering
\includegraphics[scale=0.18]{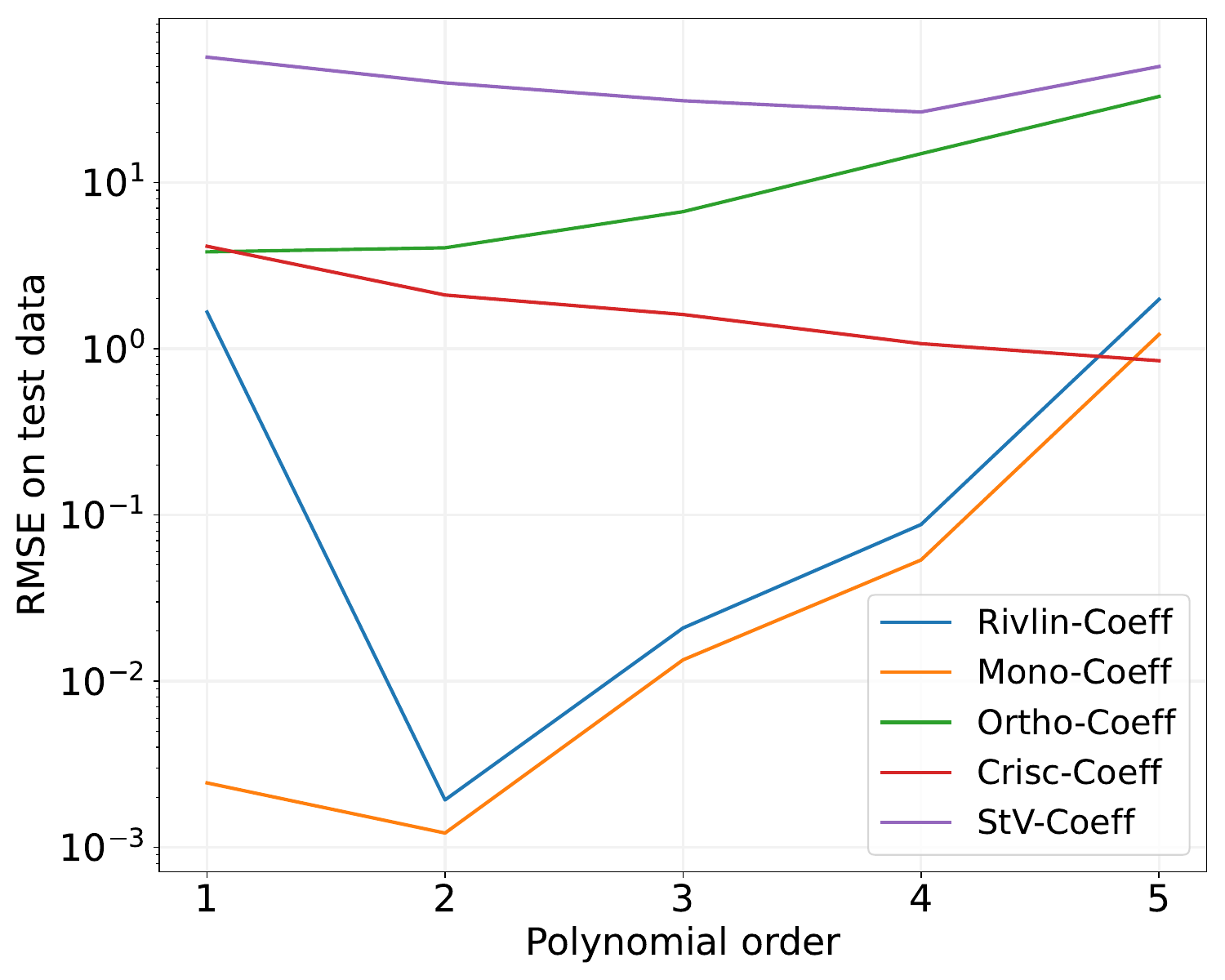}
\subcaption{Mooney-Rivlin data}
\end{subfigure}
\begin{subfigure}[b]{0.3\linewidth} \centering
\includegraphics[scale=0.18]{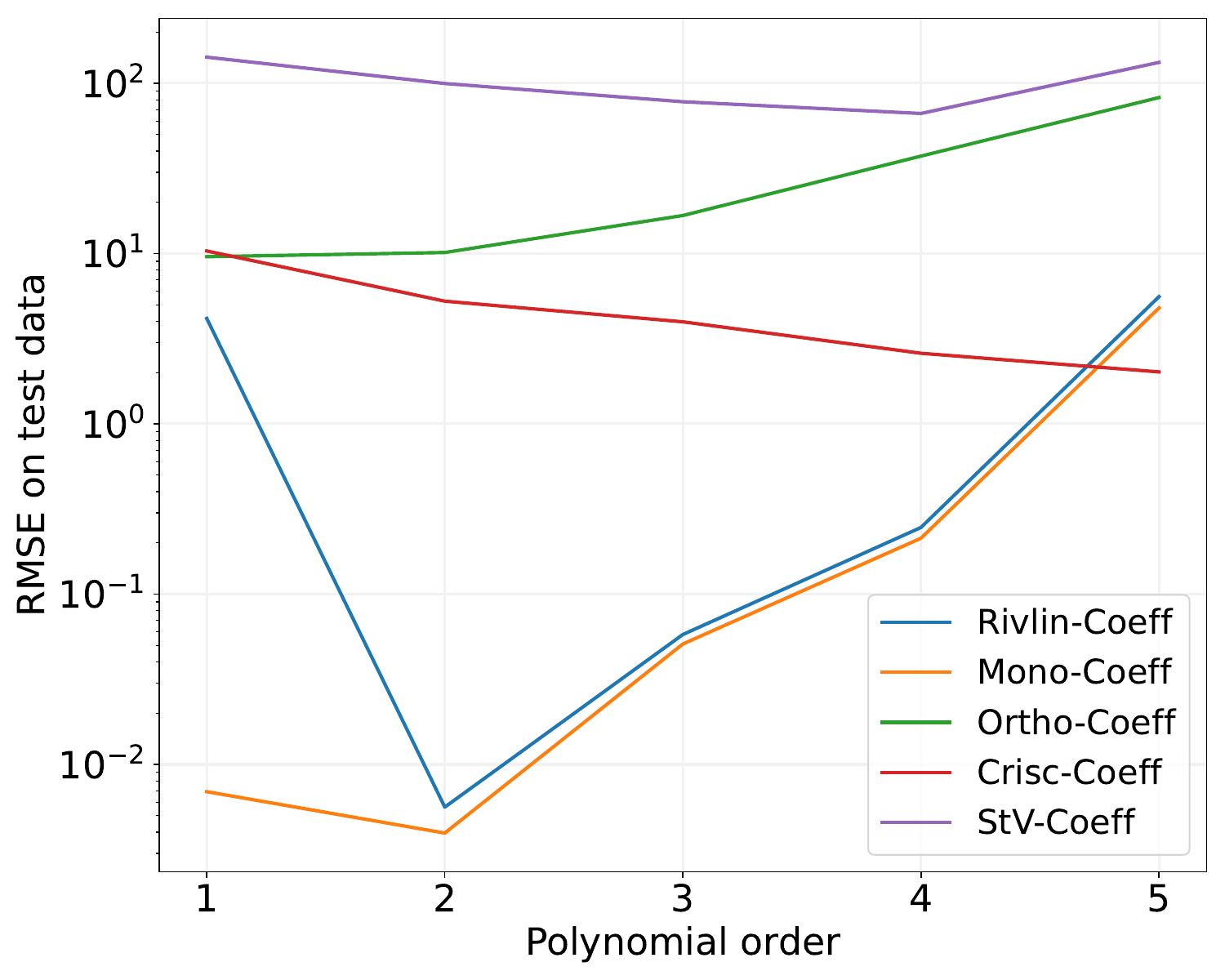}
\subcaption{Modified Carroll data}
\end{subfigure}
\begin{subfigure}[b]{0.3\linewidth} \centering
\includegraphics[scale=0.18]{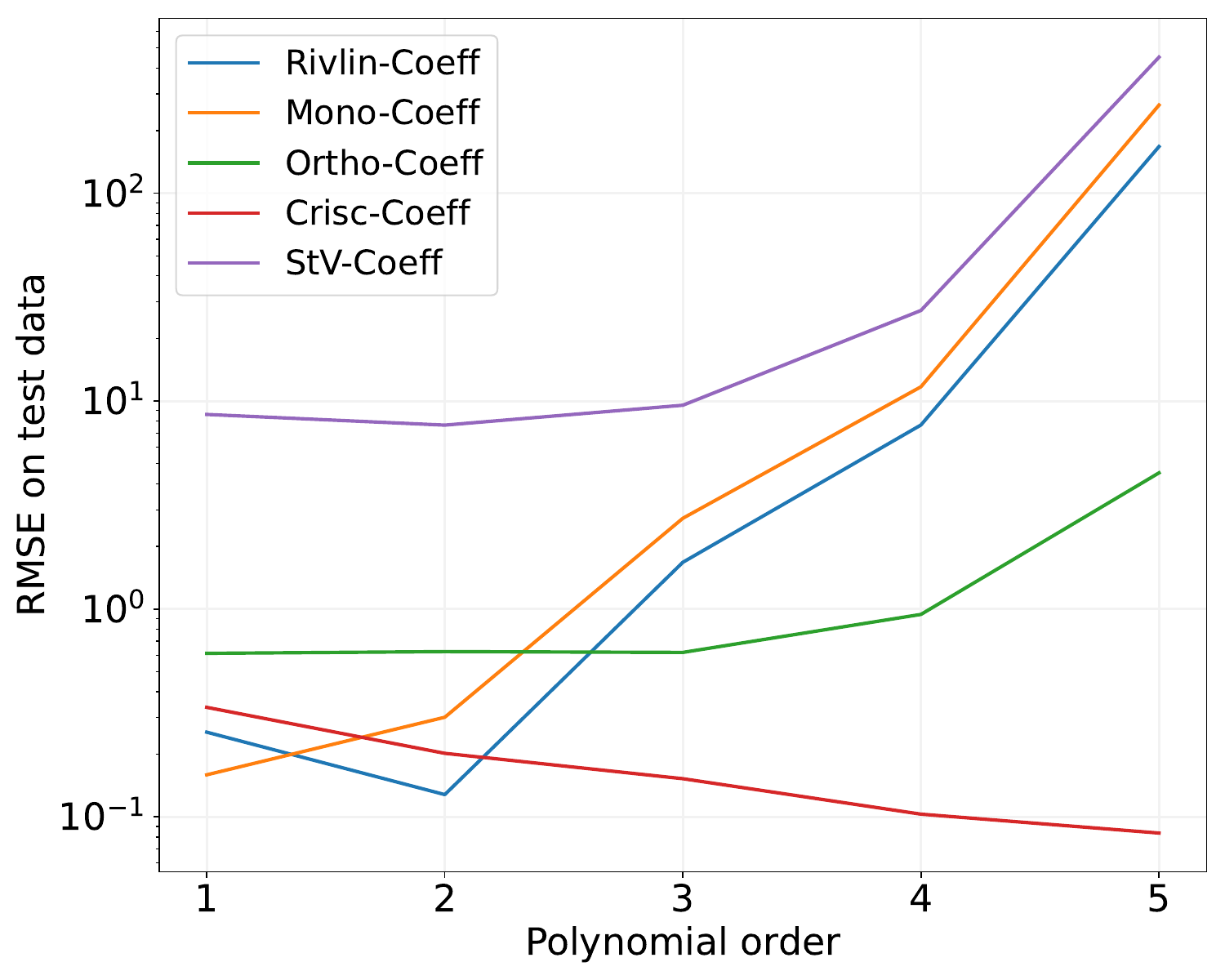}
\subcaption{Gent data}
\end{subfigure}
\caption{RMSE between predicted coefficient values and reference coefficient values using a polynomial of $n$-th order as the regressor which was trained on the training data and tested on the test data set.} \label{fig:NearFieldError}
\end{figure}

\begin{figure}[h!]
\centering
\begin{subfigure}[b]{0.3\linewidth} \centering
\includegraphics[scale=0.18]{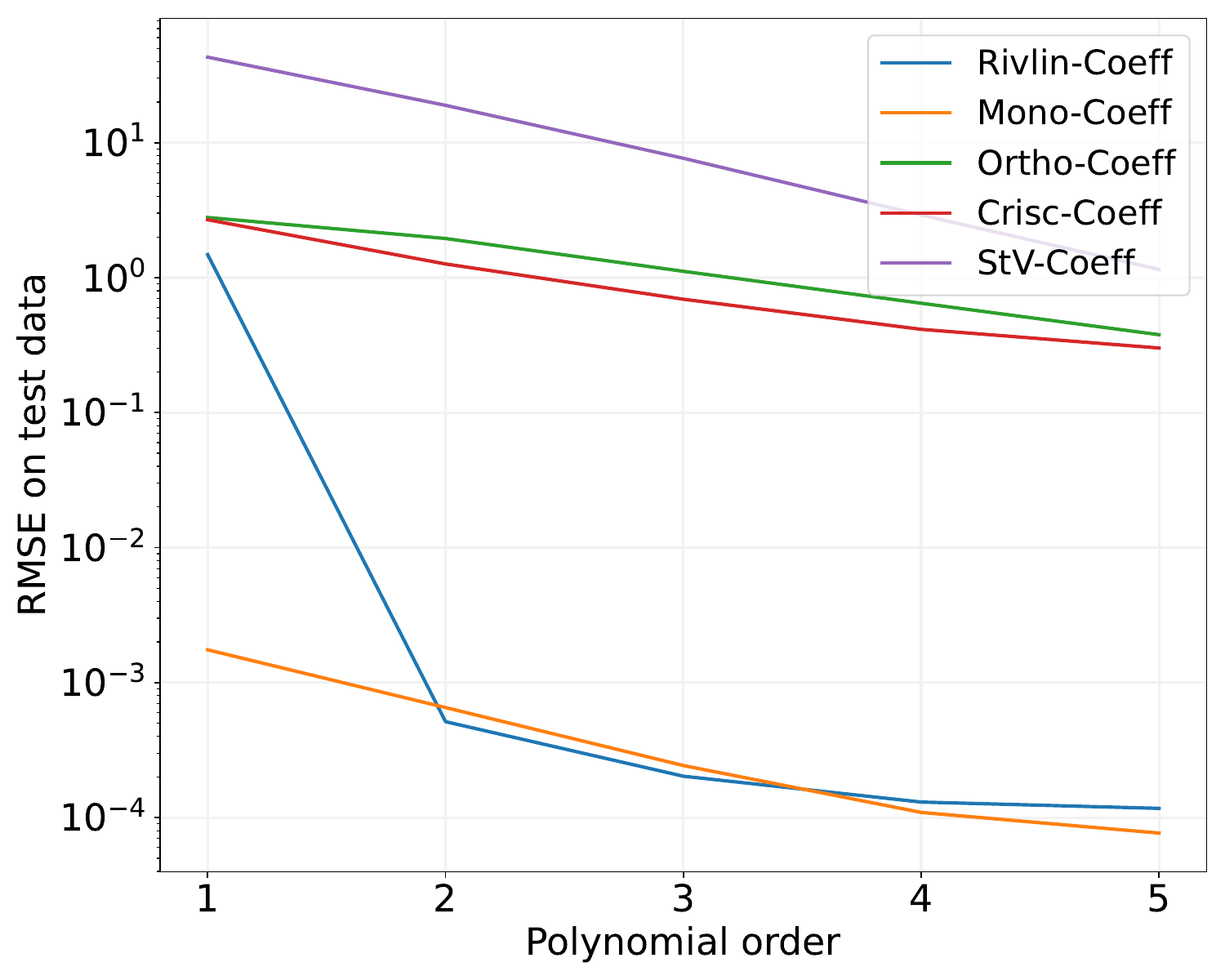}
\subcaption{Mooney-Rivlin data}
\end{subfigure}
\begin{subfigure}[b]{0.3\linewidth} \centering
\includegraphics[scale=0.18]{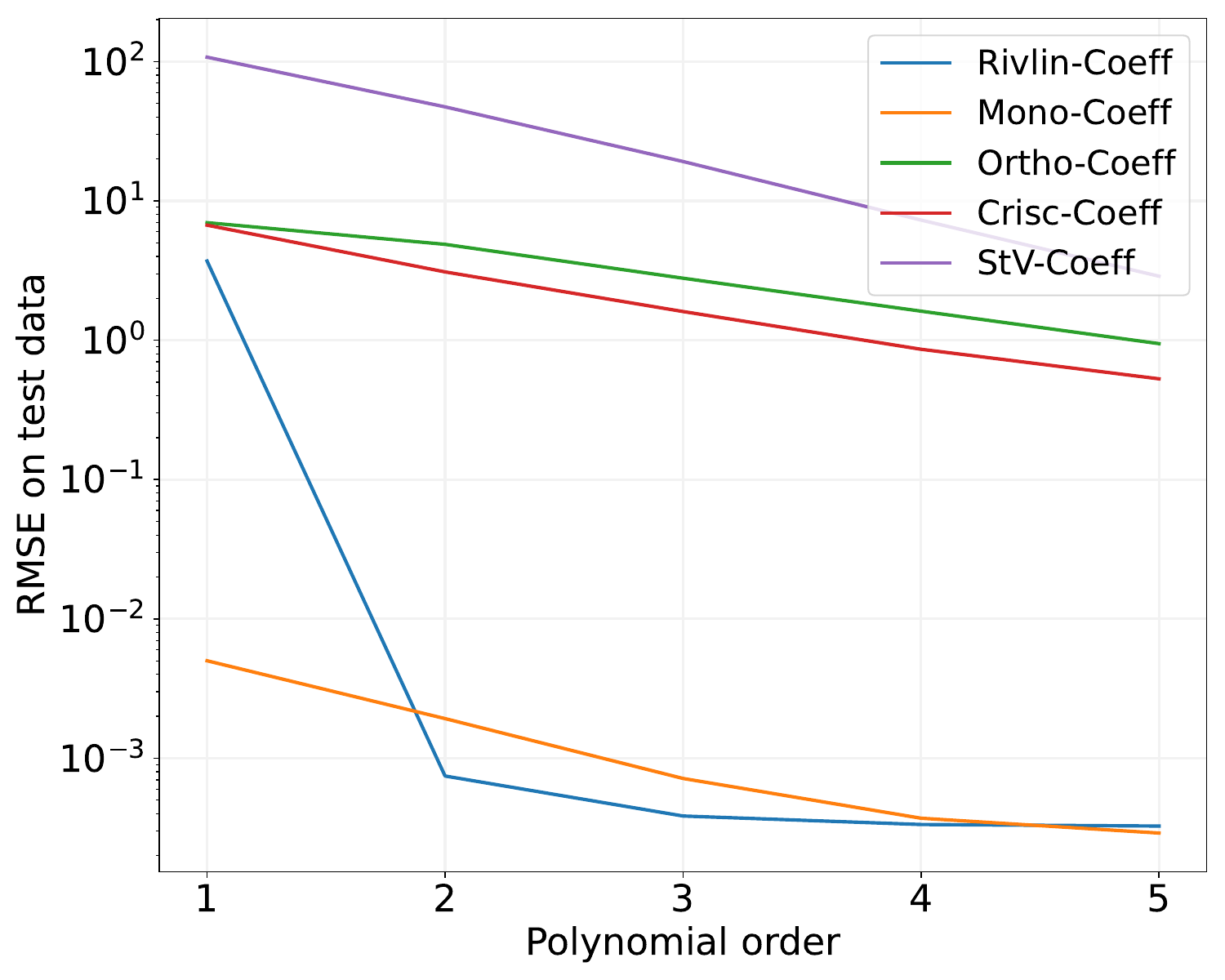}
\subcaption{Modified Carroll data}
\end{subfigure}
\begin{subfigure}[b]{0.3\linewidth} \centering
\includegraphics[scale=0.18]{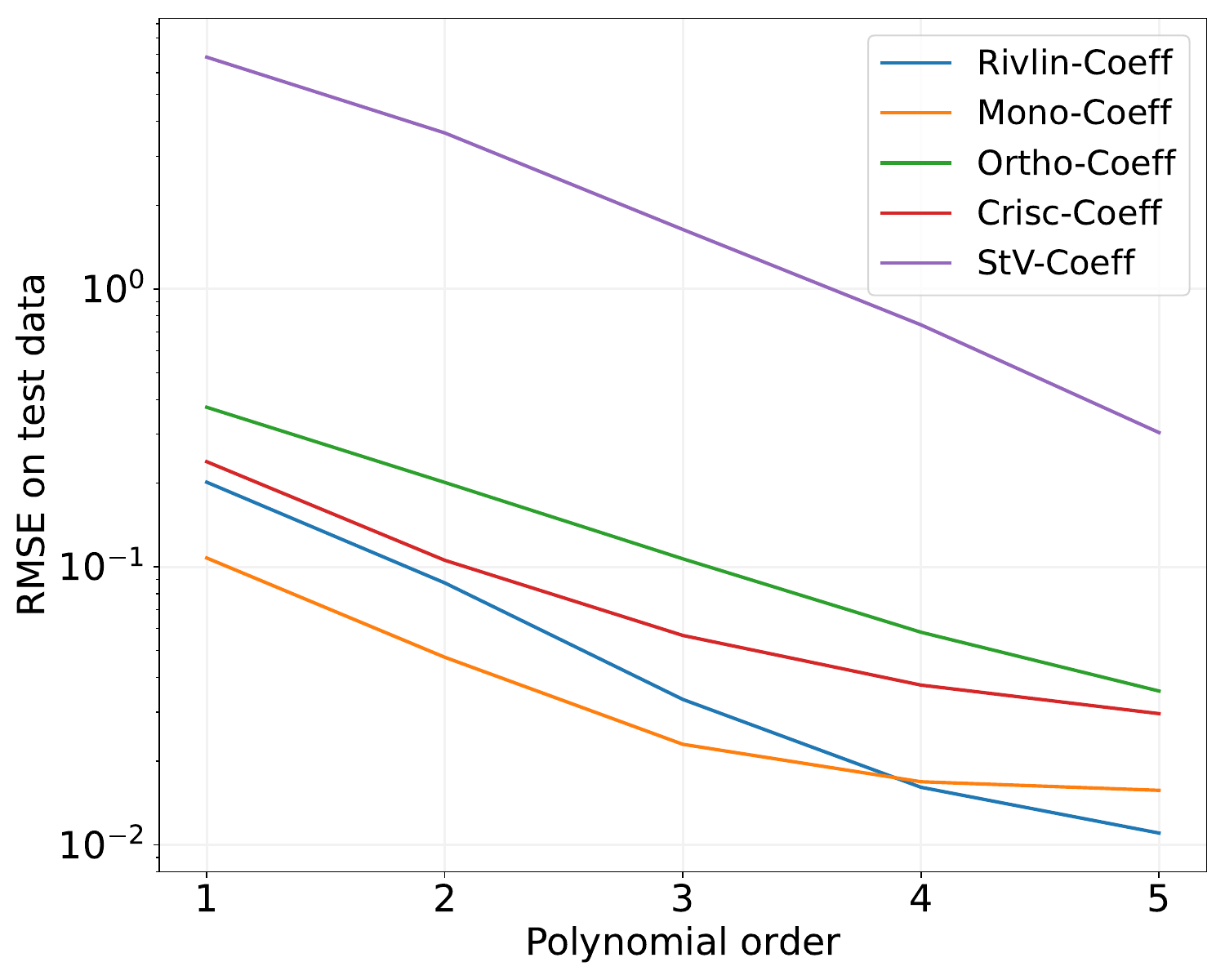}
\subcaption{Gent data}
\end{subfigure}
\caption{RMSE between predicted coefficient values and reference coefficient values using a polynomial of $n$-th order as the regressor which was trained on all available data (test data) and tested on the same data.} \label{fig:FarFieldError}
\end{figure}

To check this hypothesis we offer the following approach. Consider the mapping $\mathcal{M}_{IC}$ from the respective invariants to the respective coefficient values, e.g 
\begin{equation}
    \mathcal{M}_{IC}: \mathcal{I} \rightarrow [c_{1}, c_{2},c_{3}].
\end{equation}
Let this mapping be approximated by a polynomial regressor of $n$-th order with interaction terms denoted by $\hat{\mathcal{M}}^{n}_{IC}(\mathcal{I} )$, e.g. for $n=2$
\begin{equation}\label{eq:PolyRegCoeff}
    \hat{\mathcal{M}}^{2}_{IC}(\mathcal{I} ) = \hat{\bm{c}}^{2}_{IC} = \bm{\alpha}_{0} +\bm{\alpha}_{1} I_{1}+\bm{\alpha}_{2} I_{2} +\bm{\alpha}_{3} I_{3}+\bm{\alpha}_{4} I_{1}^{2} + \bm{\alpha}_{5} I_{1} I_{2}+ \bm{\alpha}_{6} I_{2}^{2}+ \bm{\alpha}_{7} I_{2} I_{3} + \bm{\alpha}_{8} I_{3}^{2} 
\end{equation}
where $\bm{\alpha}_{i} \in \mathbb{R}^{3}$. 
To check the potential complexity of this mapping we look at two scenarios. 
First, how polynomial regression fitted on the training data predicts the test data, and second, how well $\hat{\mathcal{M}}^{n}_{IC}(\mathcal{I} )$ predicts the test data coefficients if it was trained on the test data. 
The root-mean-squared error (RMSE) between the reference and predicted test data coefficients of the former is shown in
\Fref{fig:NearFieldError} for the three noiseless data cases. 
Surprisingly, polynomials of second order generalize the best for Rivlin-Coeff and Mono-Coeff.
This leads us to the conjecture that the complexity of the coefficient functions of Rivlin-Coeff and Mono-Coeff are generally lower than the other representation. This seems to correlate with the results of the TBNN generalization errors c.f. \Fref{fig:test_error_comparison}. 
The coefficient error between the regressor of an increasing polynomial order, trained on the test data and evaluated on the test data is displayed in \Fref{fig:FarFieldError}. 
It can be seen that, compared to the previous case (\Fref{fig:test_error_comparison}), Mono-Coeff and Rivlin-Coeff still have the lowest errors but more significantly that the complexity of the coefficient functions seems to have changed, i.e. while second order polynomials where best for models trained on the training data, now an increasing polynomial order seems to help to accurately fit the coefficient functions.
Note that the upward trends after initial low discrepancy fits in some of the data in \Fref{fig:NearFieldError} could possibily be attributed to overfitting of the polynomials to the 100-point training dataset.

Next, we aim to gauge what the contribution of the basis representation is on the accuracy. From the output of the polynomial regression of \eqref{eq:PolyRegCoeff} an $n$-th order polynomial prediction of the stress can be obtained
\begin{equation}
    \hat{\Sb}^{n}_{IC} = \sum_{a=1}^{3} \, \hat{c}^{n}_{a, IC} \Bb_{a}.
\end{equation}
We furthermore assume an alternative mapping $\mathcal{M}_{IS}$ from the invariants of the representation to the symmetric components of the stress 
\begin{equation}
   \mathcal{M}_{IS}: \mathcal{I} \rightarrow [S_{11}, S_{12},S_{13}, S_{22}, S_{23}, S_{33]}]
\end{equation}
for which we build a similar $n$-th order polynomial referred to as $\hat{\mathcal{M}}_{IS}^{n}$.
Then, we can find the difference between the RMSEs of $\hat{\Sb}^{n}_{IC}$ and $\hat{\mathcal{M}}_{IS}^{n}$ evaluated on  the test data, i.e.
\begin{equation*}
   \Delta \text{RMSE}^{n}(\mathcal{I}^{test}) = \text{RMSE}(\hat{\mathcal{M}}_{IS}^{n}(\mathcal{I}^{test})) - \text{RMSE}(\hat{\Sb}^{n}_{IC}(\mathcal{I}^{test})).
\end{equation*}
Simplistically, this difference between the stress errors between these two regressors will help us judge the role and contribution of the basis generators, i.e. if $\Delta \text{RMSE}^{n}>0$ then the bases have a positive contribution to the prediction which means that the RMSE of obtaining the stress from a linear combination of coefficients and bases $\hat{\Sb}^{n}_{IC}=\sum_{a=1}^{3} \, \hat{c}^{n}_{a, IC} \Bb_{a}$ is lower than the mapping from invariants to stress directly. This would tell us that the basis components take complexity out of the system. On the other hand if $\Delta \text{RMSE}^{n}<0$ the basis components make the mapping between invariants and stress more complex.

\begin{figure}[h!]
\centering
\begin{subfigure}[b]{0.3\linewidth} \centering
\includegraphics[scale=0.18]{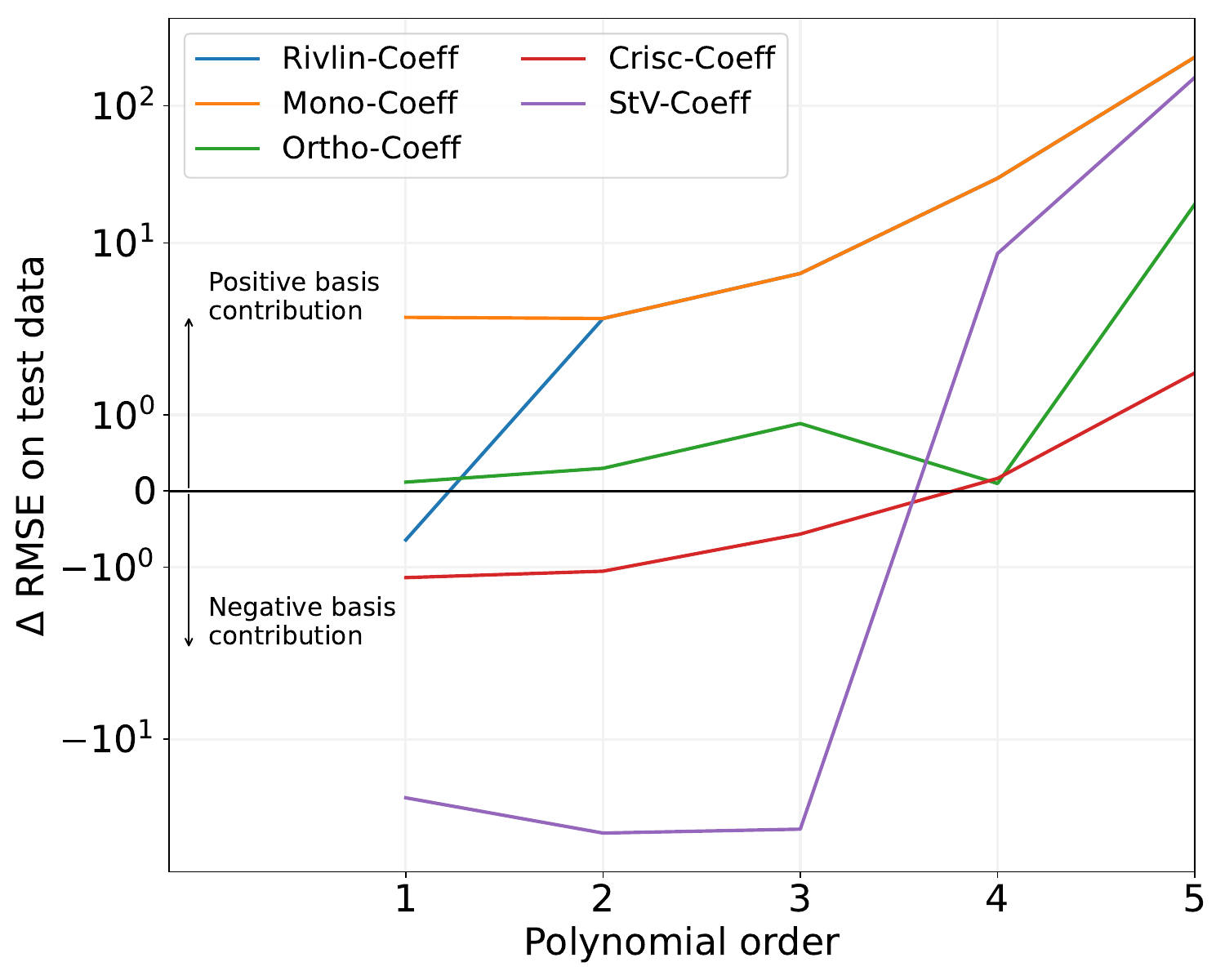}
\subcaption{Mooney-Rivlin data}
\end{subfigure}
\begin{subfigure}[b]{0.3\linewidth} \centering
\includegraphics[scale=0.18]{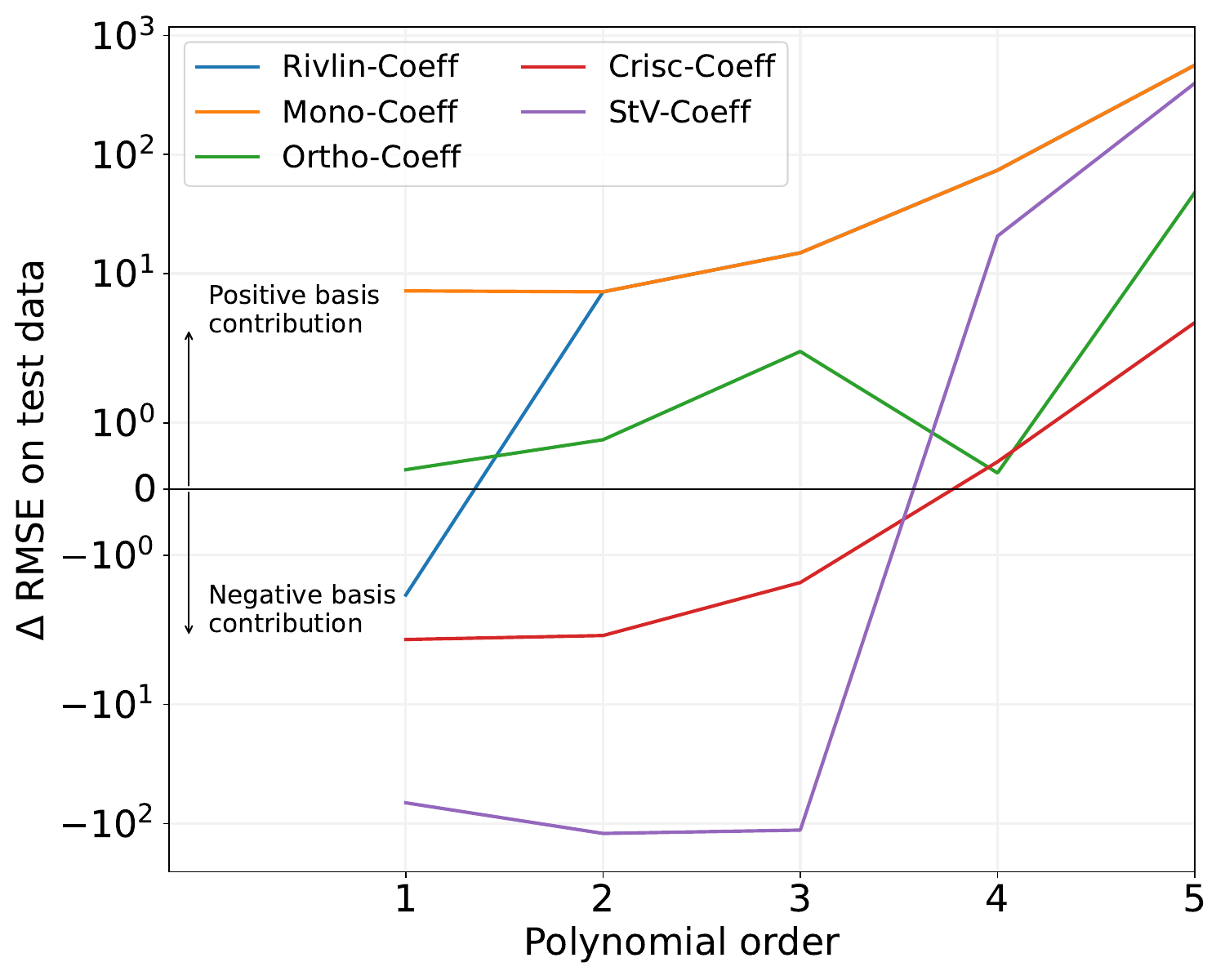}
\subcaption{Modified Carroll data}
\end{subfigure}
\begin{subfigure}[b]{0.3\linewidth} \centering
\includegraphics[scale=0.18]{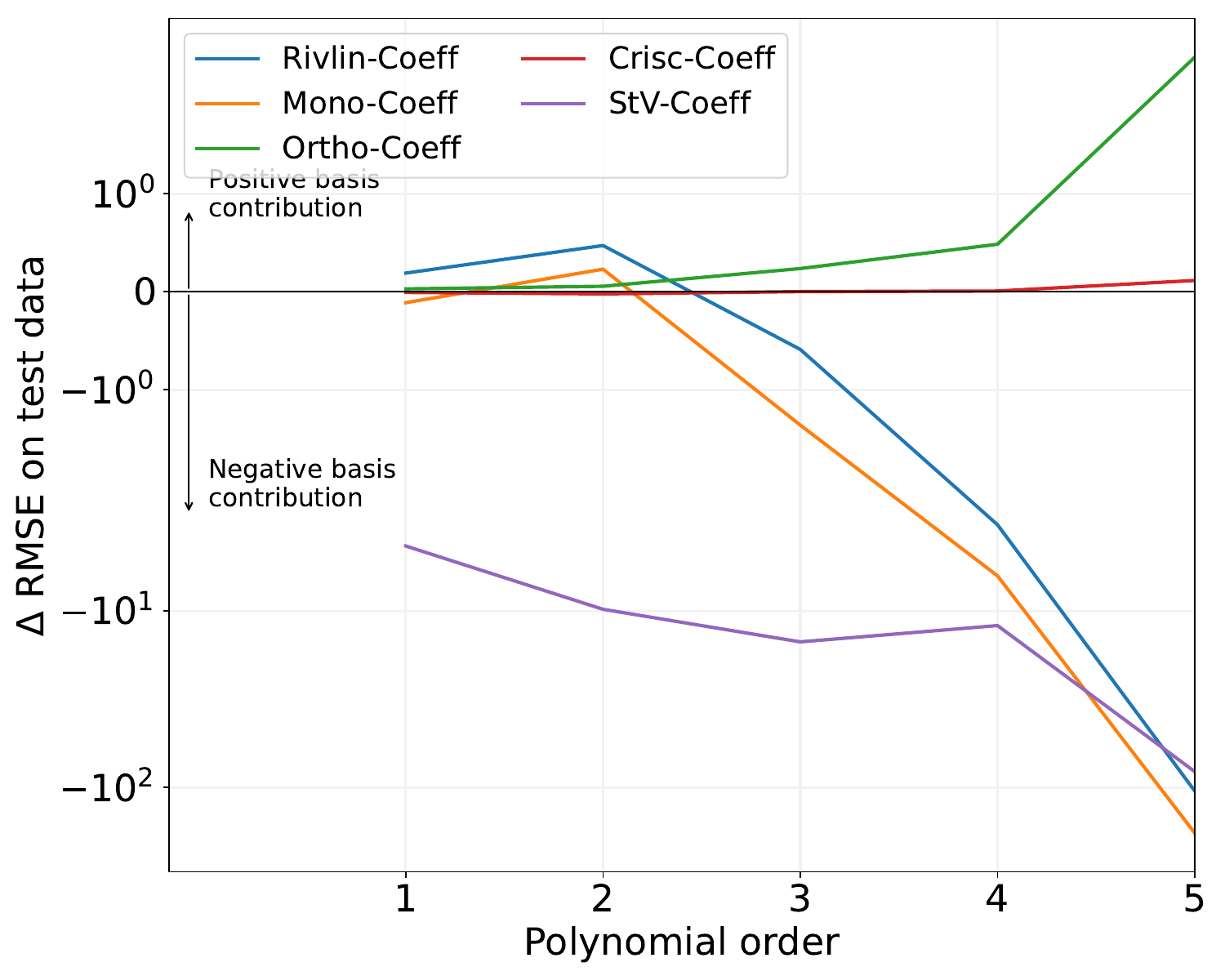}
\subcaption{Gent data}
\end{subfigure}
\caption{  The RMSE difference between the stress prediction of an n-th order polynomial trained from the invariants to coefficients $\hat{\Sb}^{n}_{IC}$ and an n-th order polynomial trained from the invariants directly to the stress $\hat{\mathcal{M}}_{IS}^{n}$, i.e. $\Delta \text{RMSE}^{n}(\mathcal{I}^{test}) = \text{RMSE}(\hat{\mathcal{M}}_{IS}^{n}(\mathcal{I}^{test})) - \text{RMSE}(\hat{\Sb}^{n}_{IC}(\mathcal{I}^{test}))$.  Both models were trained on the \textbf{training} data and tested on the \textbf{test} data. } \label{fig:NearFieldErrorDiff}
\end{figure}

\begin{figure}[h!]
\centering
\begin{subfigure}[b]{0.3\linewidth} \centering
\includegraphics[scale=0.18]{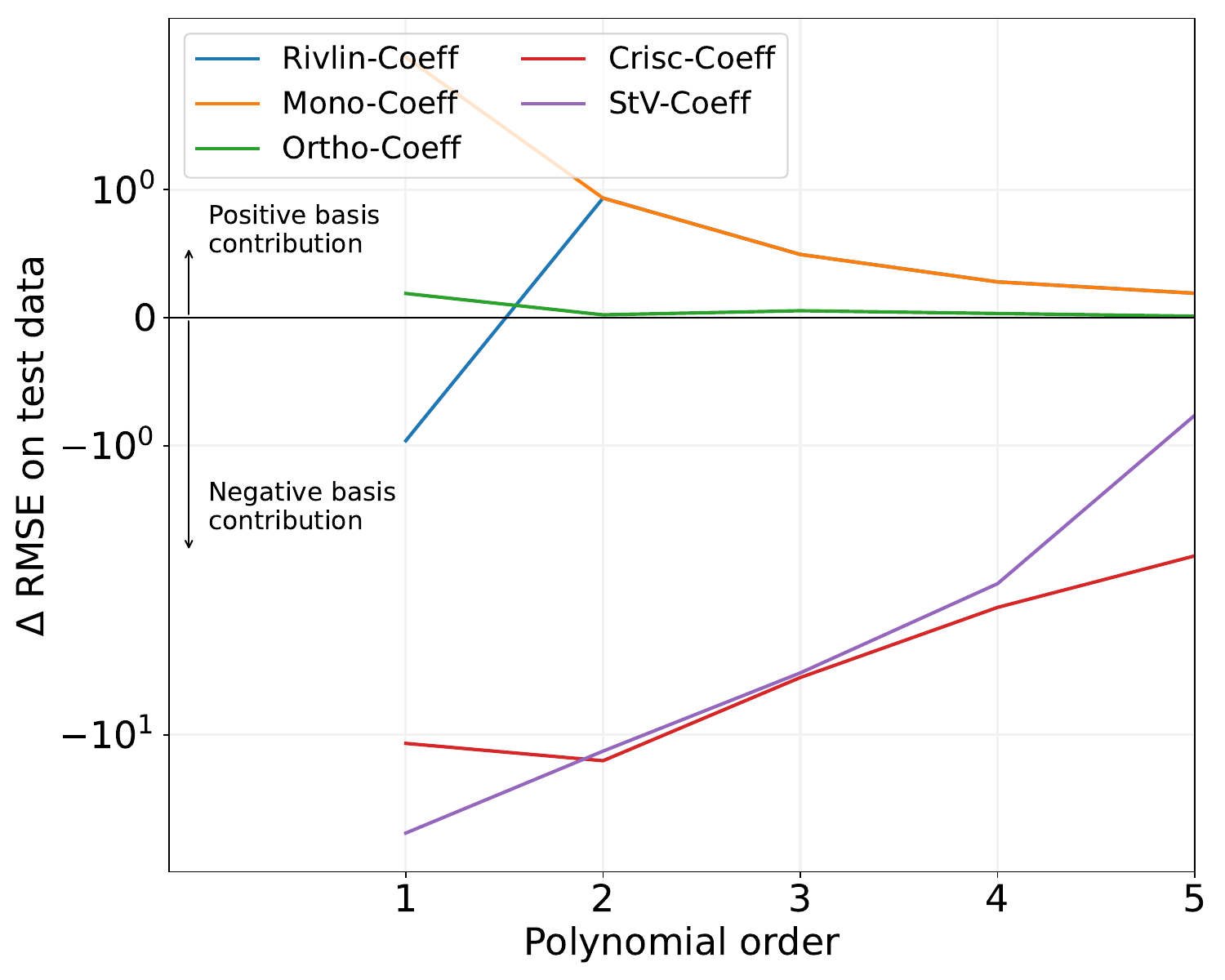}
\subcaption{Mooney-Rivlin data}
\end{subfigure}
\begin{subfigure}[b]{0.3\linewidth} \centering
\includegraphics[scale=0.18]{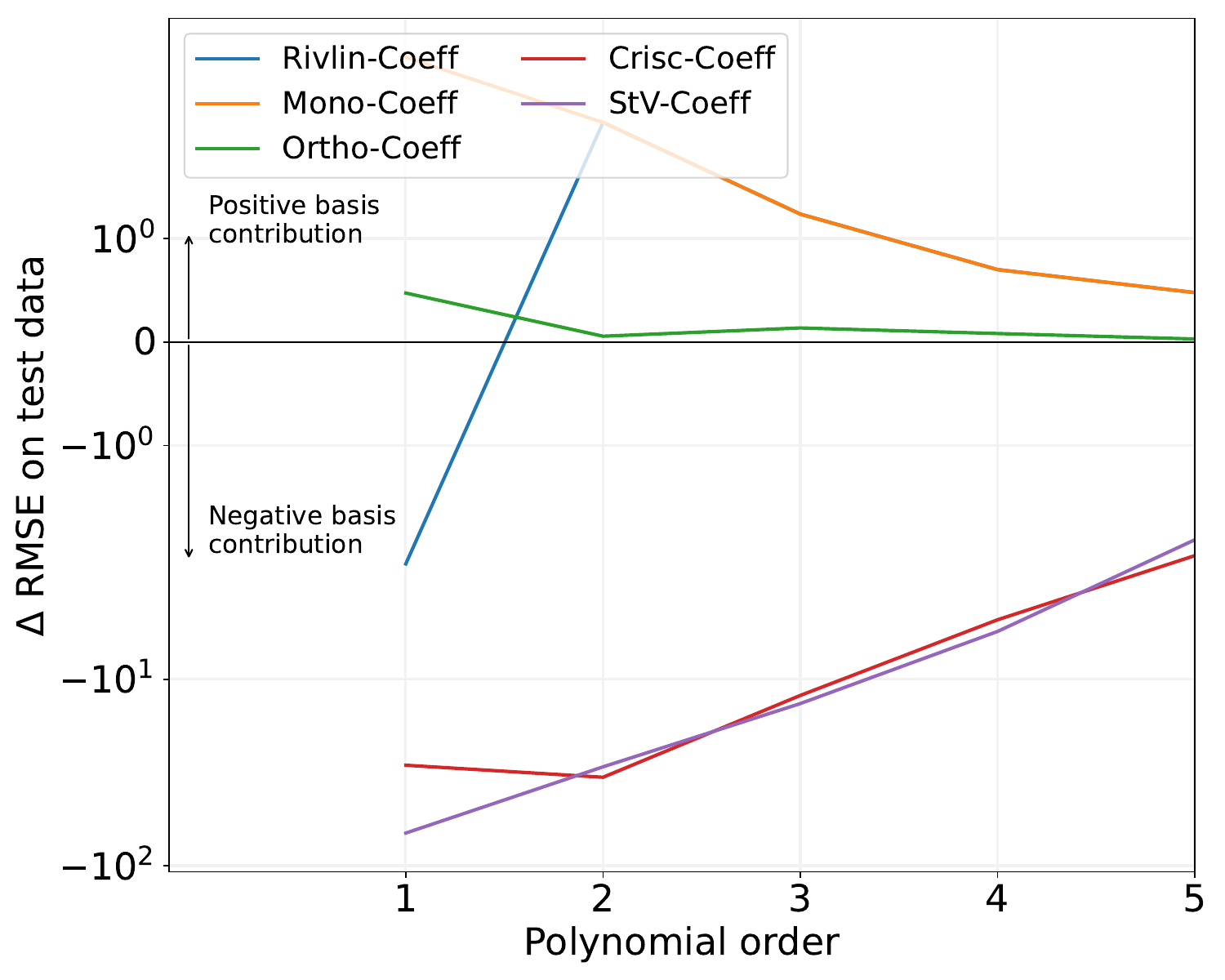}
\subcaption{Modified Carroll data}
\end{subfigure}
\begin{subfigure}[b]{0.3\linewidth} \centering
\includegraphics[scale=0.18]{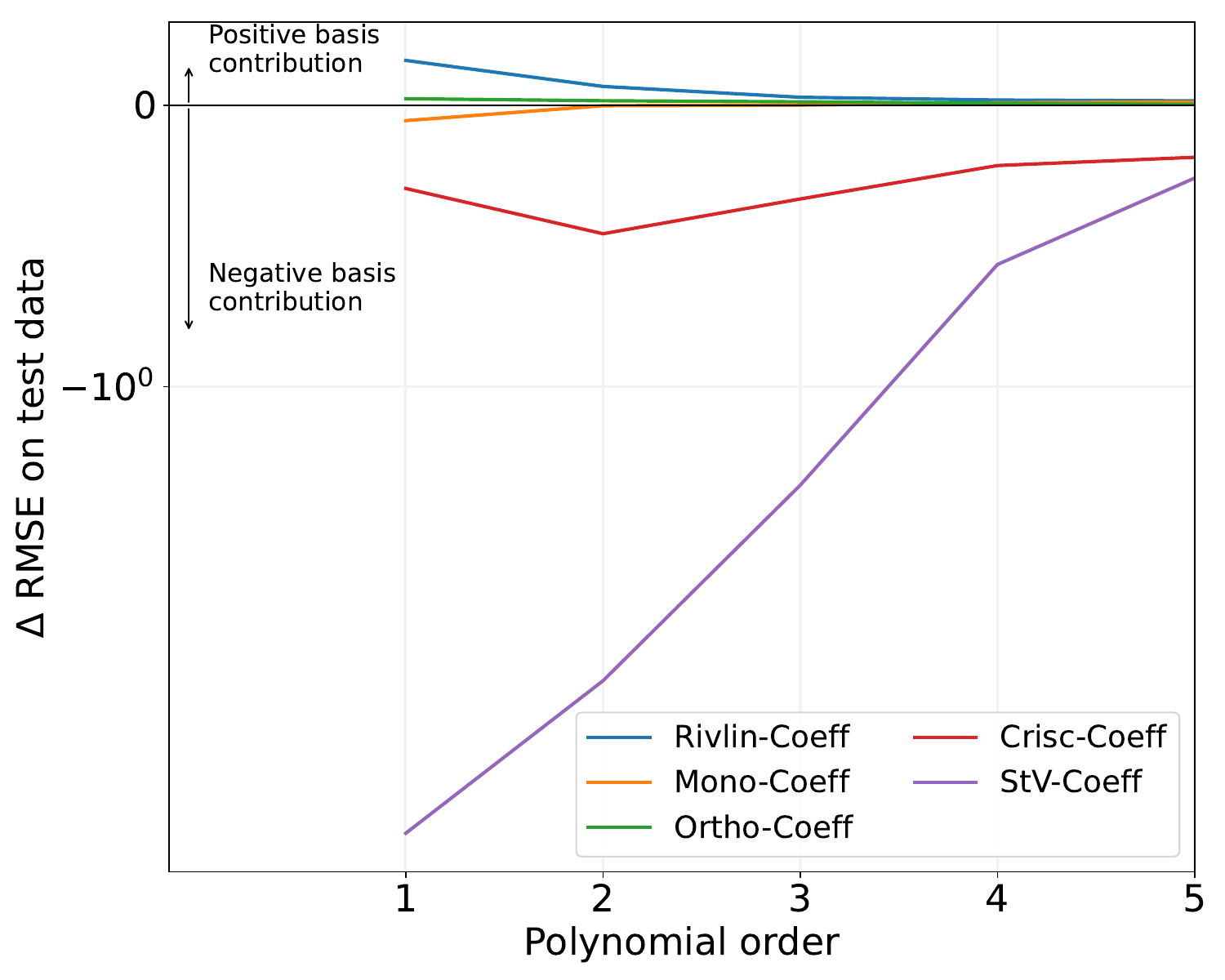}
\subcaption{Gent data}
\end{subfigure}
\caption{The RMSE difference between the stress prediction of an n-th order polynomial trained from the invariants to coefficients $\hat{\Sb}^{n}_{IC}$ and an n-th order polynomial trained from the invariants directly to the stress $\hat{\mathcal{M}}_{IS}^{n}$, i.e. $\Delta \text{RMSE}^{n}(\mathcal{I}^{test}) = \text{RMSE}(\hat{\mathcal{M}}_{IS}^{n}(\mathcal{I}^{test})) - \text{RMSE}(\hat{\Sb}^{n}_{IC}(\mathcal{I}^{test}))$.  Both models were trained on all available \textbf{test} data and tested on the same data.} \label{fig:FarFieldErrorDiff}
\end{figure}

In \Fref{fig:NearFieldErrorDiff} we compare the RMSE-difference of models trained on the training data and evaluated on the test data while
\Fref{fig:FarFieldErrorDiff} highlights the RMSE difference when the models were trained and tested on the test data. It can be seen that the basis components of the Mono-Coeff and Rivlin-Coeff representations generally help in reducing the complexity of the invariants-stress mapping, especially for lower polynomial order while the opposite is true for the remaining representations that were investigated.

Overall we believe that this investigation of the data through the lens of a polynomial regressor suggests that our hypothesis is valid. The TBNN has better generalization capabilities for some of the stress representations because the complexity of the mapping is reduced owing to nonlinearities introduced by the basis components.

%.......................................................................
\subsection{Interpolation} \label{sec:interpolation} 
\begin{figure}[h!]
\begin{subfigure}[b]{0.3\linewidth}
        \centering
    \includegraphics[scale=0.18]{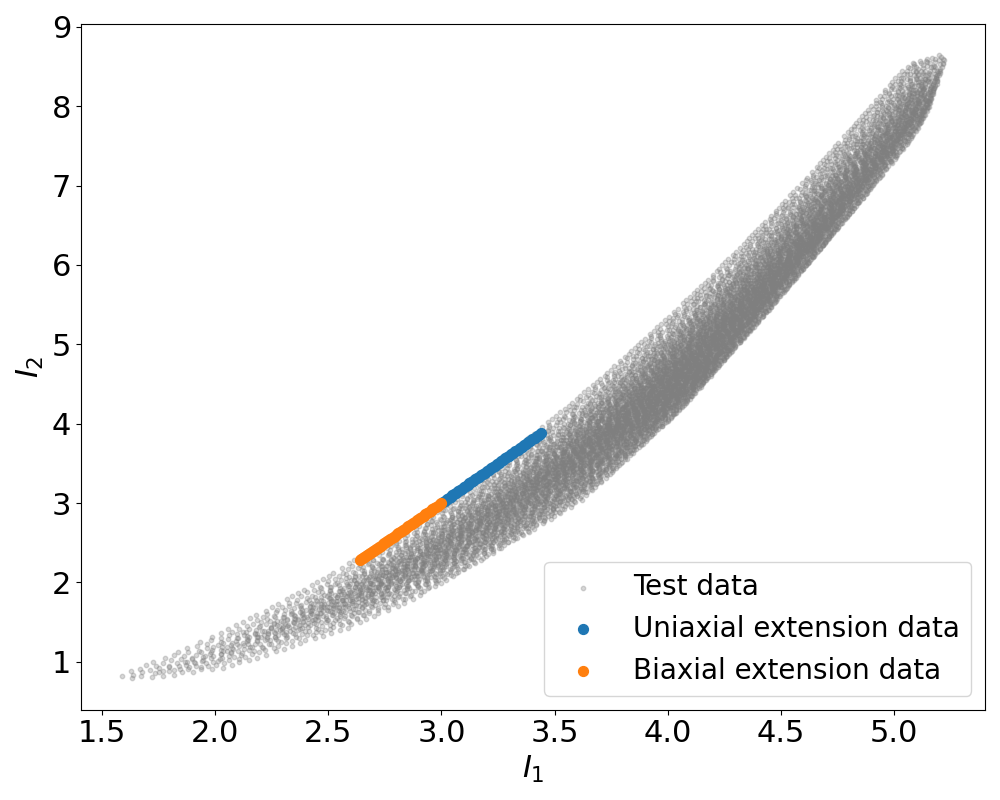}
    \subcaption{}
\end{subfigure}
\begin{subfigure}[b]{0.3\linewidth}
        \centering
\includegraphics[scale=0.18]{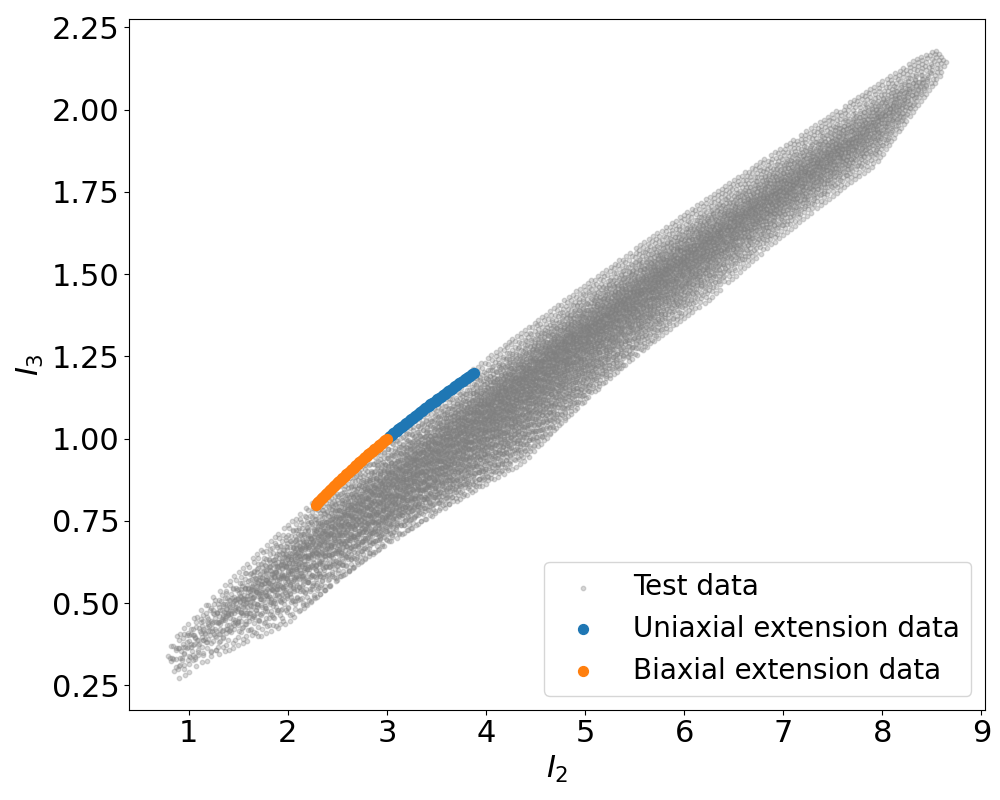}
    \caption{}
\end{subfigure}
\begin{subfigure}[b]{0.3\linewidth}
        \centering
\includegraphics[scale=0.18]{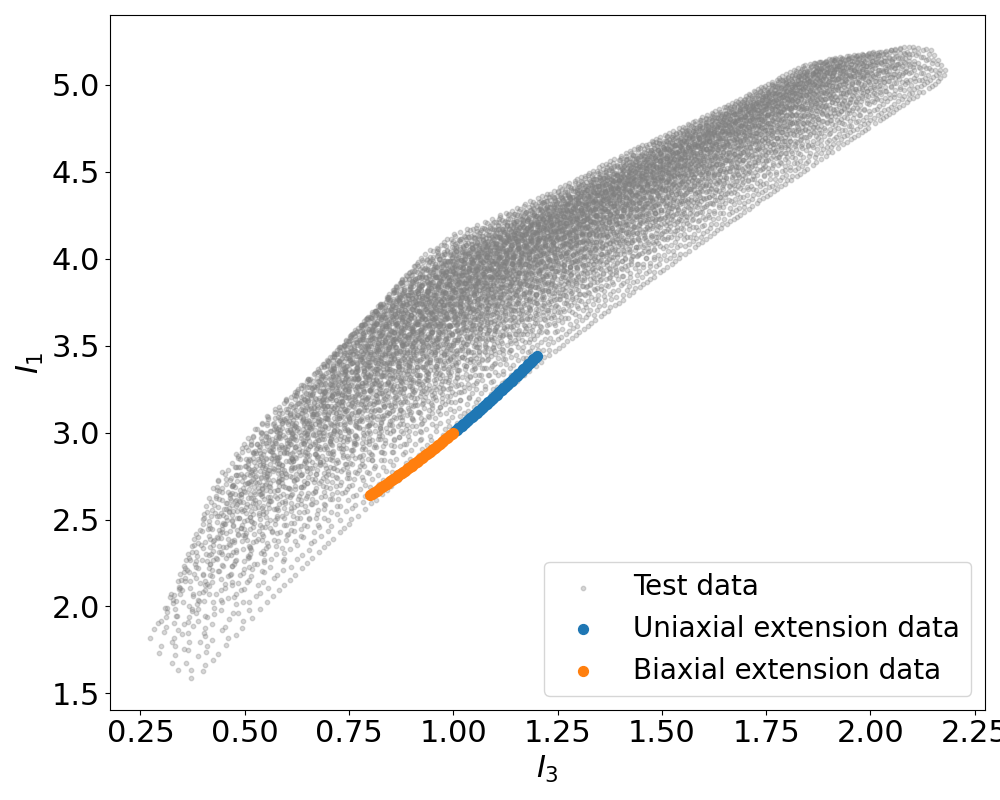}
    \caption{}
\end{subfigure}
    \caption{Uniaxial and Biaxial extension parameterization in invariant space given by $\gamma$} \label{fig:pathUniBia}
\end{figure}

%.......................................................................
The TBNN models have the ability to form smooth extensions to coefficient functions for high symmetry loading.
Consider the parameterized invariants 
\cite{currie2004attainable}
\begin{equation}\label{eq:UniBiaPath}
\begin{aligned}
        I_{1}(\gamma) &= 3 - 2\, \gamma  + \gamma^{2} \\
        I_{2}(\gamma) &= 3 - 4 \gamma  + 2\, \gamma^{2} \\
        I_{3}(\gamma) &= (1- \, \gamma)^{2} 
\end{aligned}
\end{equation}
with $\gamma \in [-0.2, 0.2]$. A uniaxial extension can be observed for $\gamma<0$ while $\gamma>0$ yields an {(equi)biaxial} extension. This path is highlighted in \Fref{fig:pathUniBia} in the projected invariant space and is inside the training region even though not explicitly part of the training data set.
We specifically focus on this path due to the fact that it is characterized by two coalescent principal stretches $\lambda_{2}=\lambda_{1}$ for $\gamma< 0$ and $\lambda_{3}=\lambda_{2}$ for $\gamma>0$ where $\lambda_{1} \leq \lambda_{2} \leq \lambda_{3}$.
Here $\lambda_a = \sqrt{\epsilon}_a$ are the principal stretches.
As described earlier and as seen in \aref{app:multiplicity}, this means that the solution matrix for the coefficients of some of the presented stress representations is ill-conditioned and special schemes are needed to be able to solve for the coefficients. We remark that:
\begin{enumerate}
    \item Depending on the loading path, these schemes lead to discontinuities  near the reference state $\Cb = \Ib$. In particular, this is the case for the Rivlin-Coff ($\Sb=c_{1} \Ib + c_{2} \Cb + c_{3} \Cb^{-1}$) and St.V-Coeff ($\Sb = c_{1} \Cb + c_{2}\Cb^{2} + c_{3} \Cb^{3}$) representations  for the path described in \eref{eq:UniBiaPath}.
    \item Due to the space-filling way the training data was generated the principal strains of all training data points are unique, apart from the undeformed configuration. This means that the trained models have not been trained on coefficients that came as a result of the special schemes described in \aref{app:multiplicity}. 
    Hence, examining the predicted coefficients in this high-symmetry case provides an interesting and interpretable test case.
\end{enumerate}

\begin{figure}[h!]
\begin{subfigure}[b]{0.45\linewidth}
\centering
\includegraphics[scale=0.25]{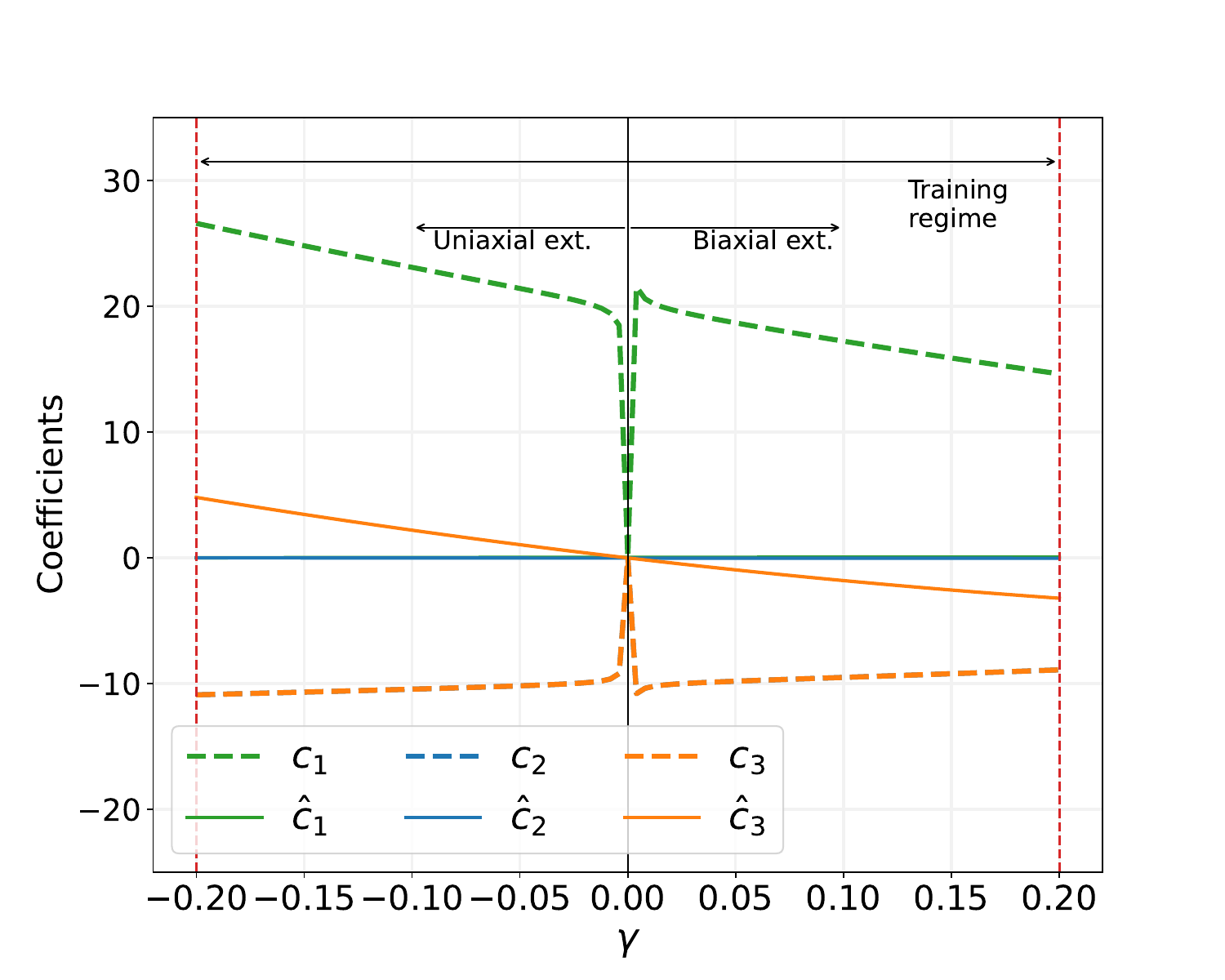}
\caption{Rivlin-Coeff coefficients}
\end{subfigure}
\begin{subfigure}[b]{0.45\linewidth}
\centering
\includegraphics[scale=0.25]{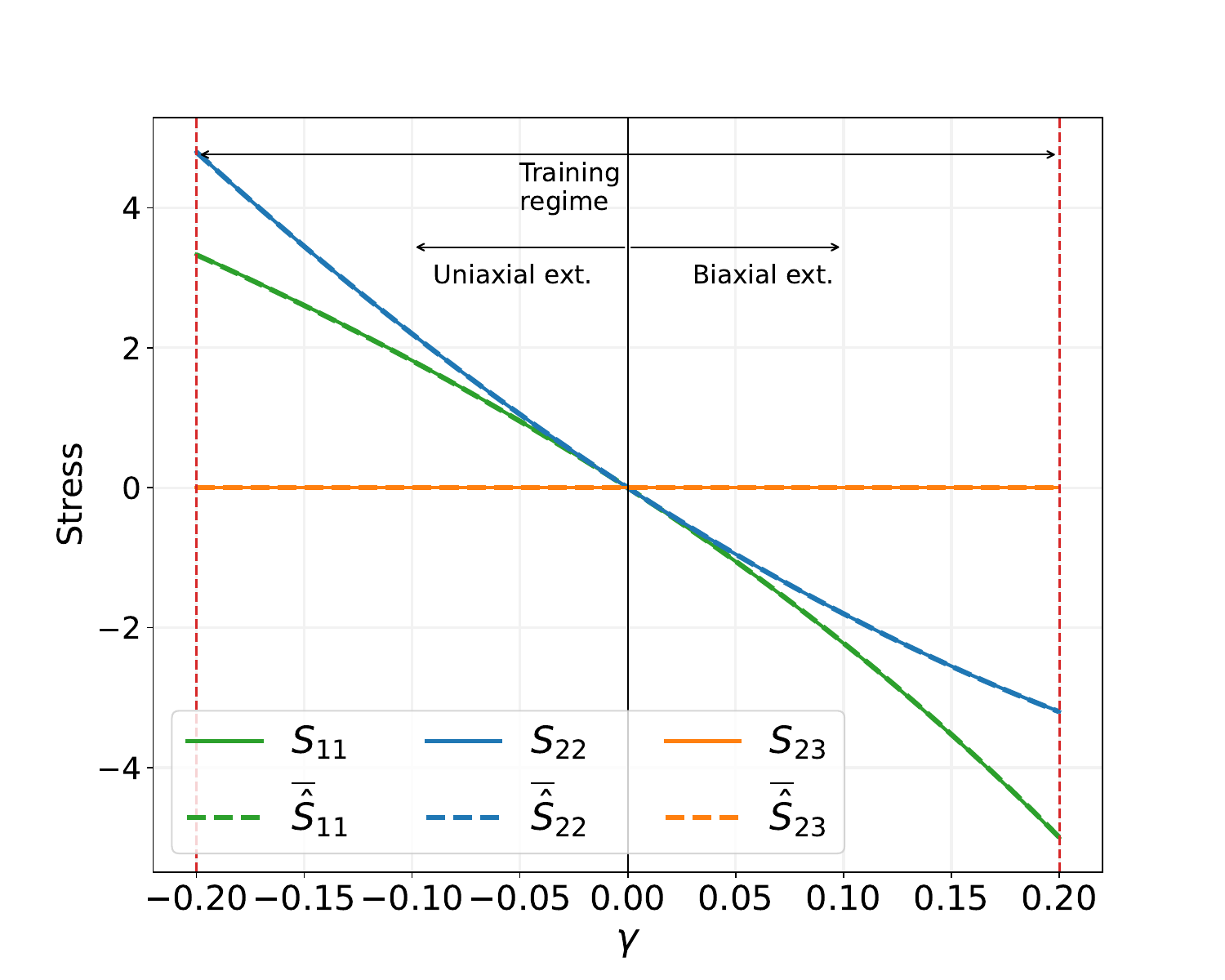}
\caption{Rivlin-Coeff stress}
\end{subfigure}

\begin{subfigure}[b]{0.45\linewidth}
\centering
\includegraphics[scale=0.25]{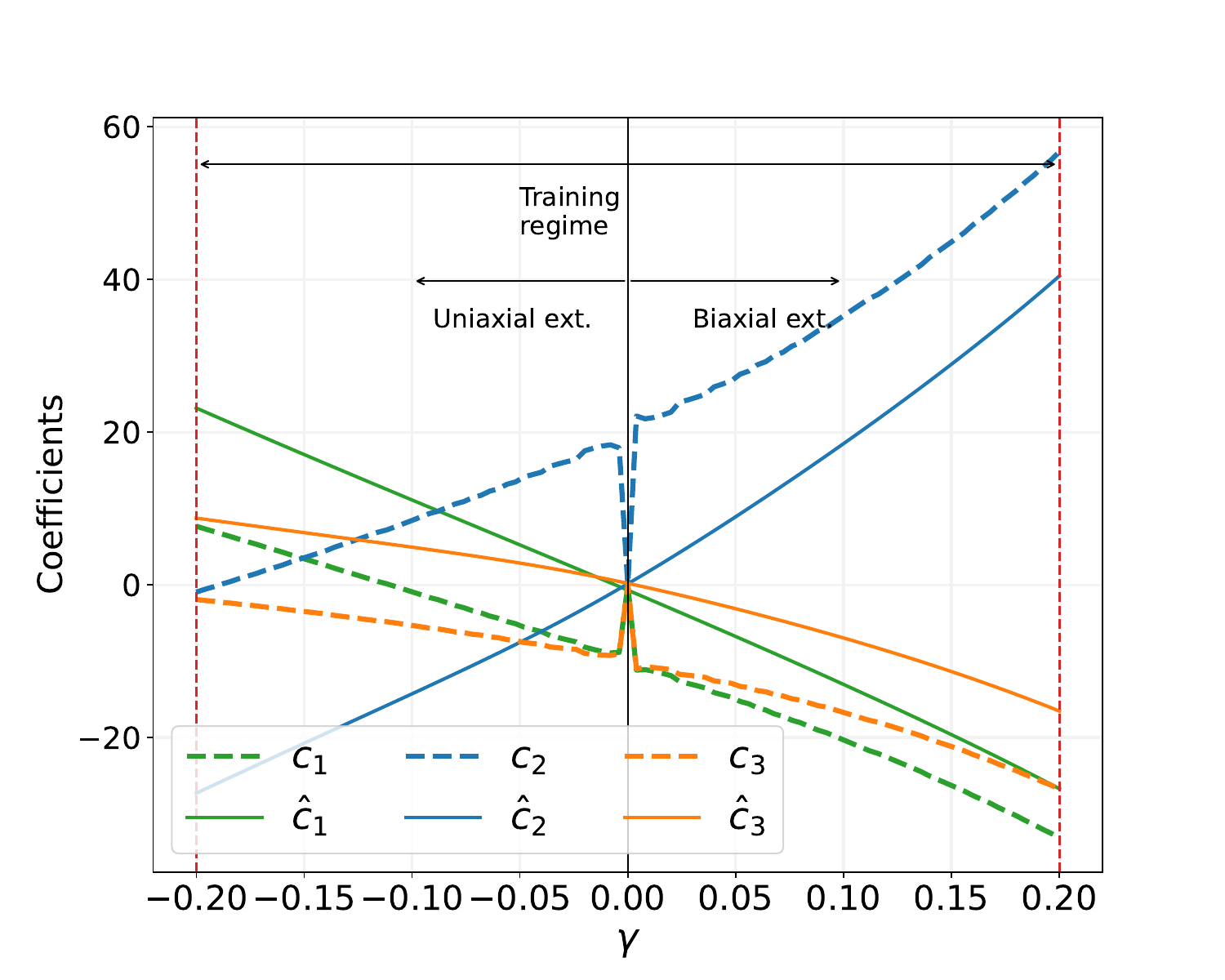}
\caption{St.V-Coeff coefficients}
\end{subfigure}
\begin{subfigure}[b]{0.45\linewidth}
\centering
\includegraphics[scale=0.25]{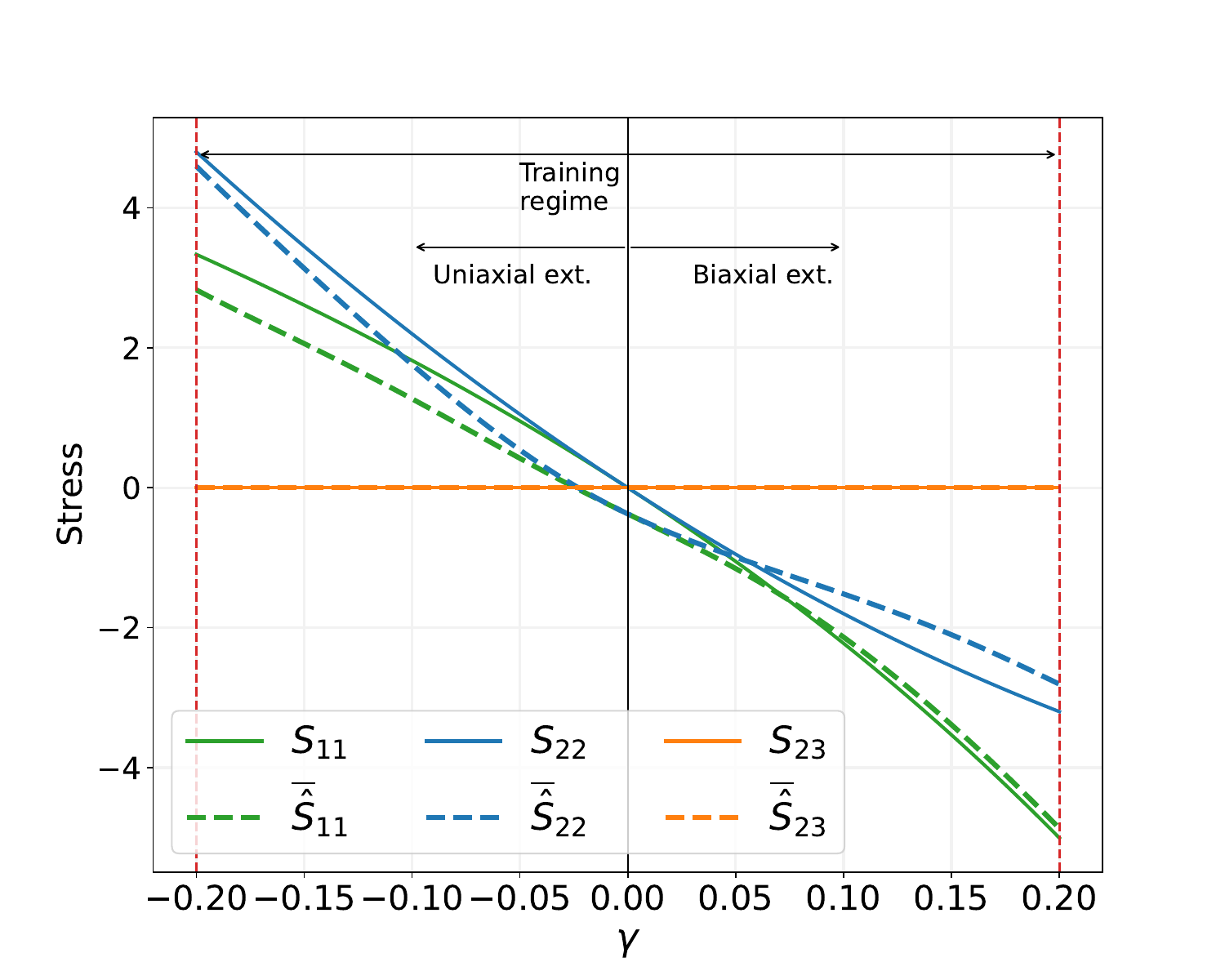}
\caption{St.V-Coeff stress}
\end{subfigure}

\caption{Reference and predicted coefficients around the undeformed configuration for uniaxial ($\gamma<0$) and biaxial ($\gamma>0$) extension for the Rivlin-Coeff and St.V-Coeff representations obtained from the Mooney-Rivlin model.
}\label{fig:DiscCoeffUni}
\end{figure}

\Fref{fig:DiscCoeffUni} shows the trained coefficients and stress predictions for the Rivlin-Coeff and St.V-Coeff models over $\gamma$.
Clearly, the extracted coefficient functions near the reference state $\Cb = \Ib$ become discontinuous; however, the built-in continuity of the NN enables an approximate continuous extension.
This approximation $\hat{c}_a$ is different than the coefficients $c_a$ extracted from the equation system, altered to accommodate multiplicity, but still yields an accurate stress representation for Rivlin-Coeff and a sufficient one for StV-Coeff.
It seems that the smooth extension avoids large errors that would be incurred if the extracted coefficients were approximated.

\begin{figure}[h!]
\begin{subfigure}[b]{0.45\linewidth}
\centering
\includegraphics[scale=0.25]{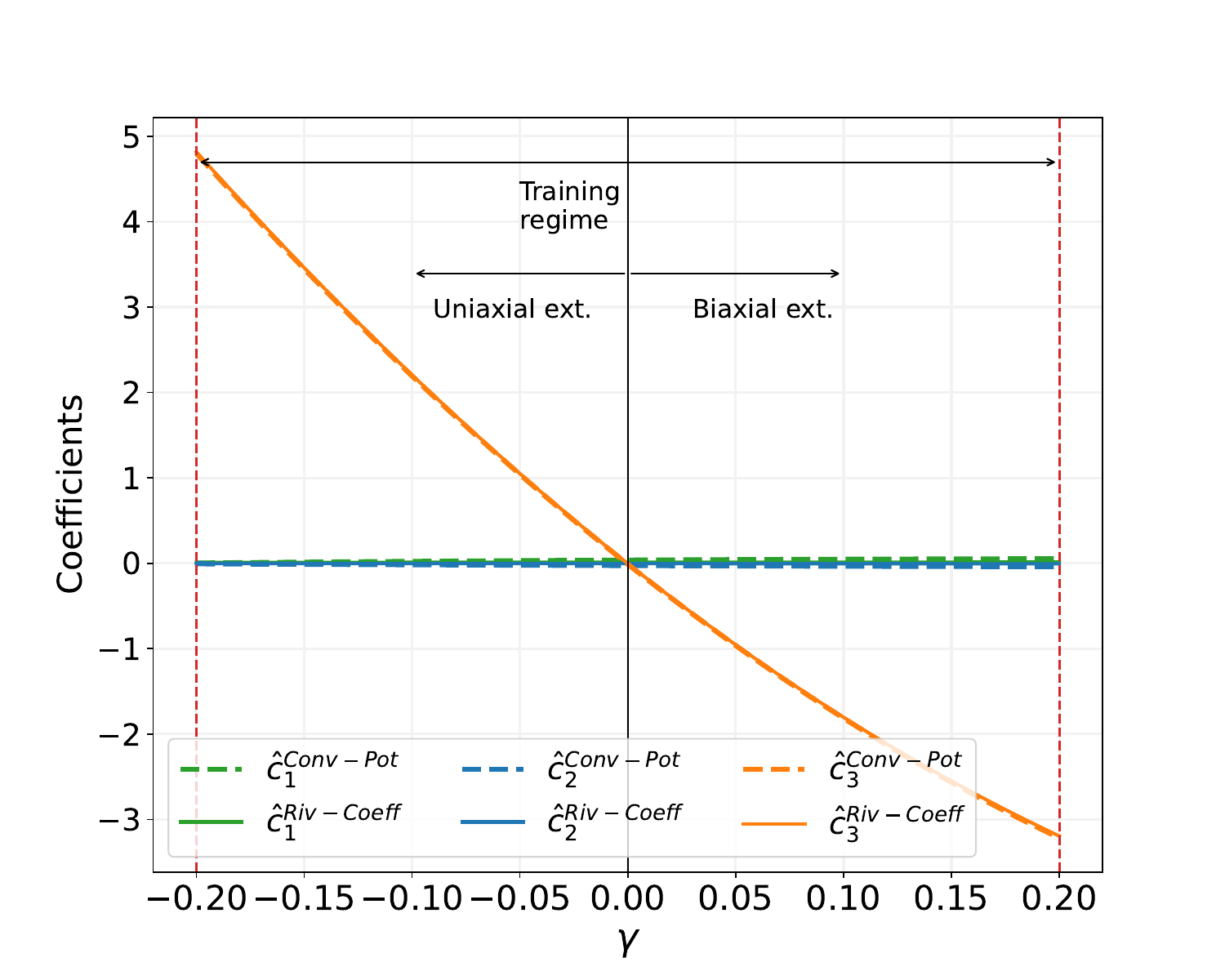}
\caption{Rivlin-Coeff and Convex-Pot coefficients}
\end{subfigure}
\begin{subfigure}[b]{0.45\linewidth}
\centering
\includegraphics[scale=0.25]{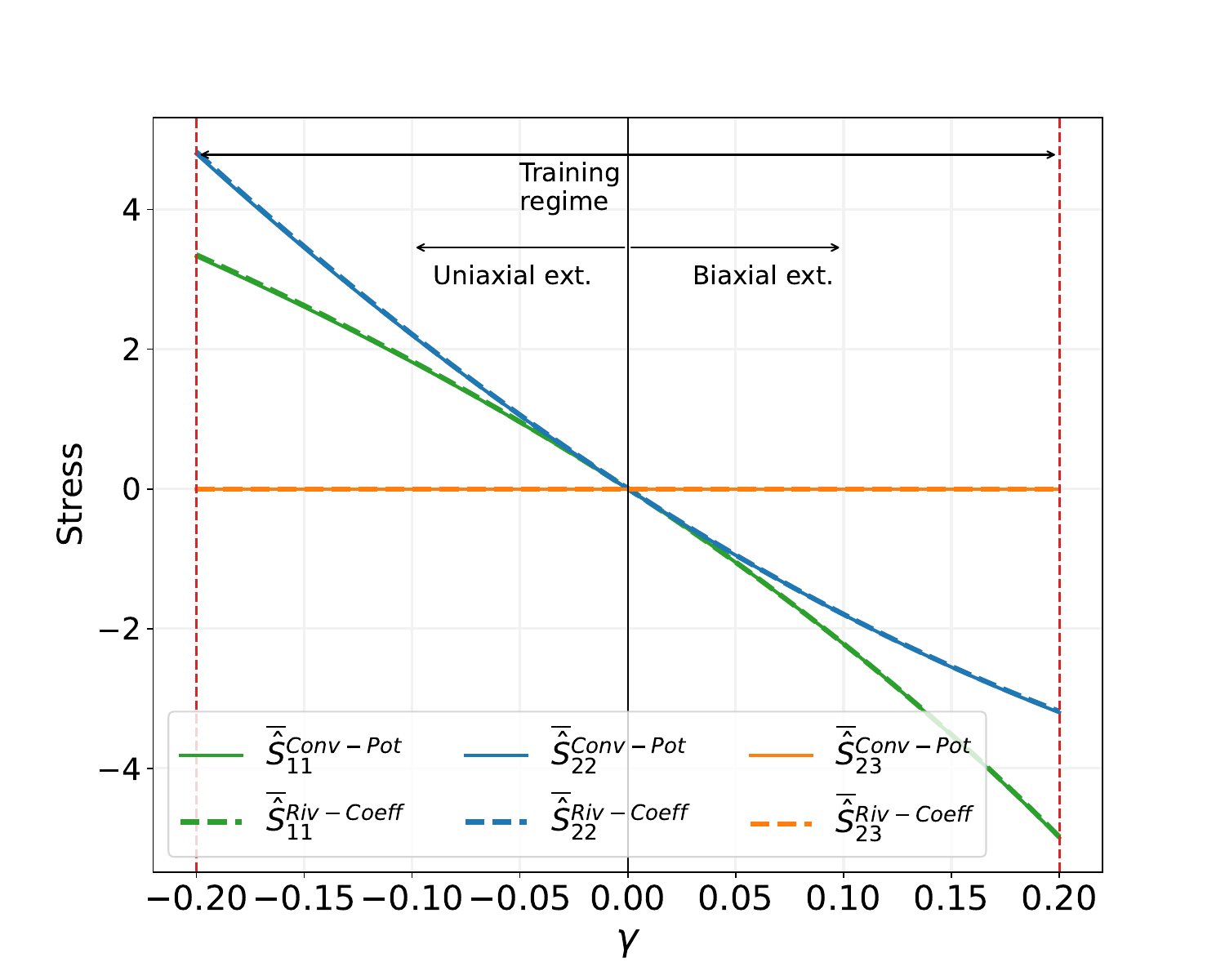}
\caption{Rivlin-Coeff and Convex-Pot stress}
\end{subfigure}
\caption{Reference and predicted coefficients around the undeformed configuration for uniaxial ($\gamma<0$) and biaxial ($\gamma>0$) extension for the Rivlin-Coeff and Convex-Pot representations obtained from the Mooney-Rivlin model.
}\label{fig:DiscCoeffUniRivICNN}
\end{figure}

\begin{figure}[h!]
\begin{subfigure}[b]{0.45\linewidth}
\centering
\includegraphics[scale=0.25]{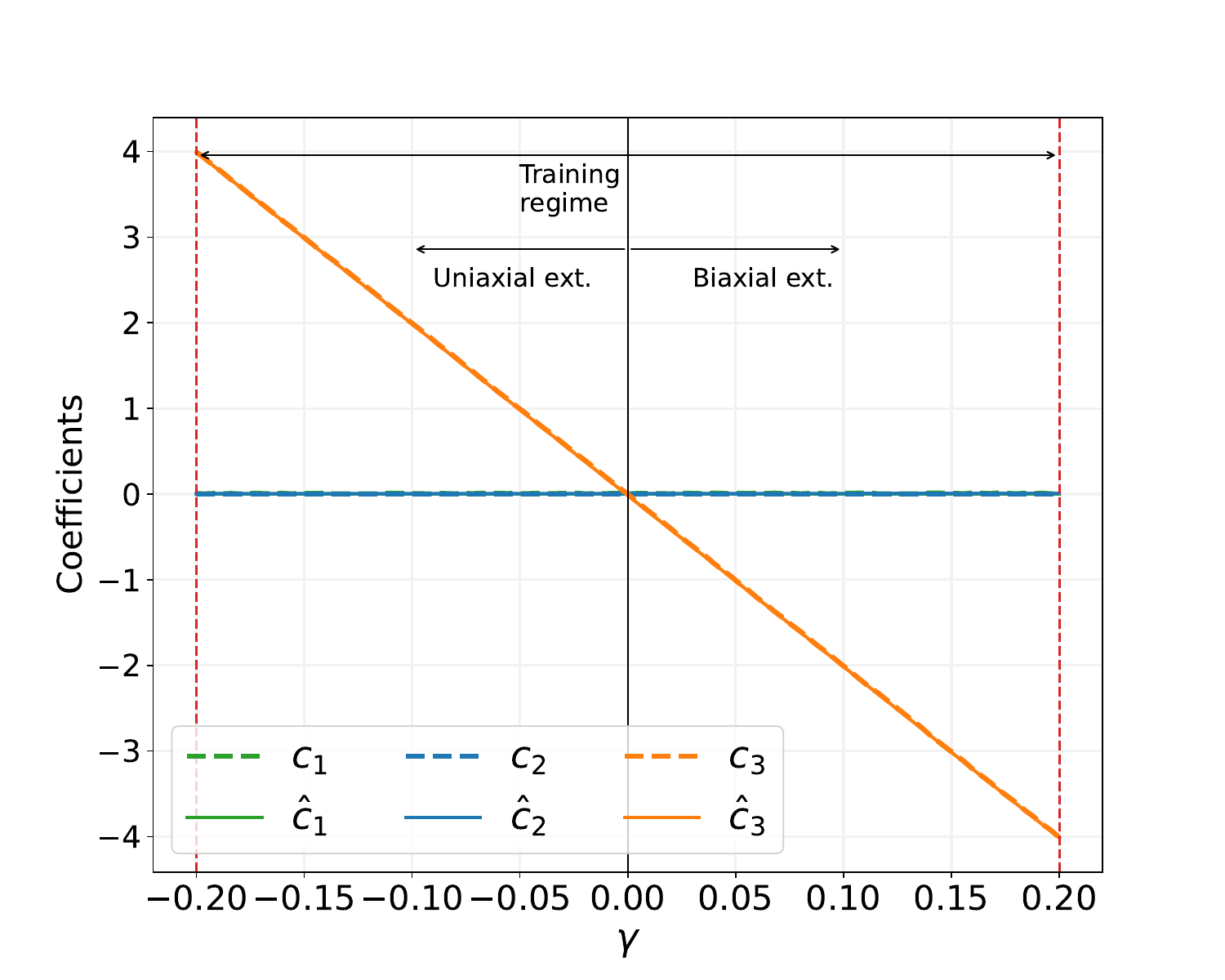}
\caption{Mono-Coeff coefficients}
\end{subfigure}
\begin{subfigure}[b]{0.45\linewidth}
\centering
\includegraphics[scale=0.25]{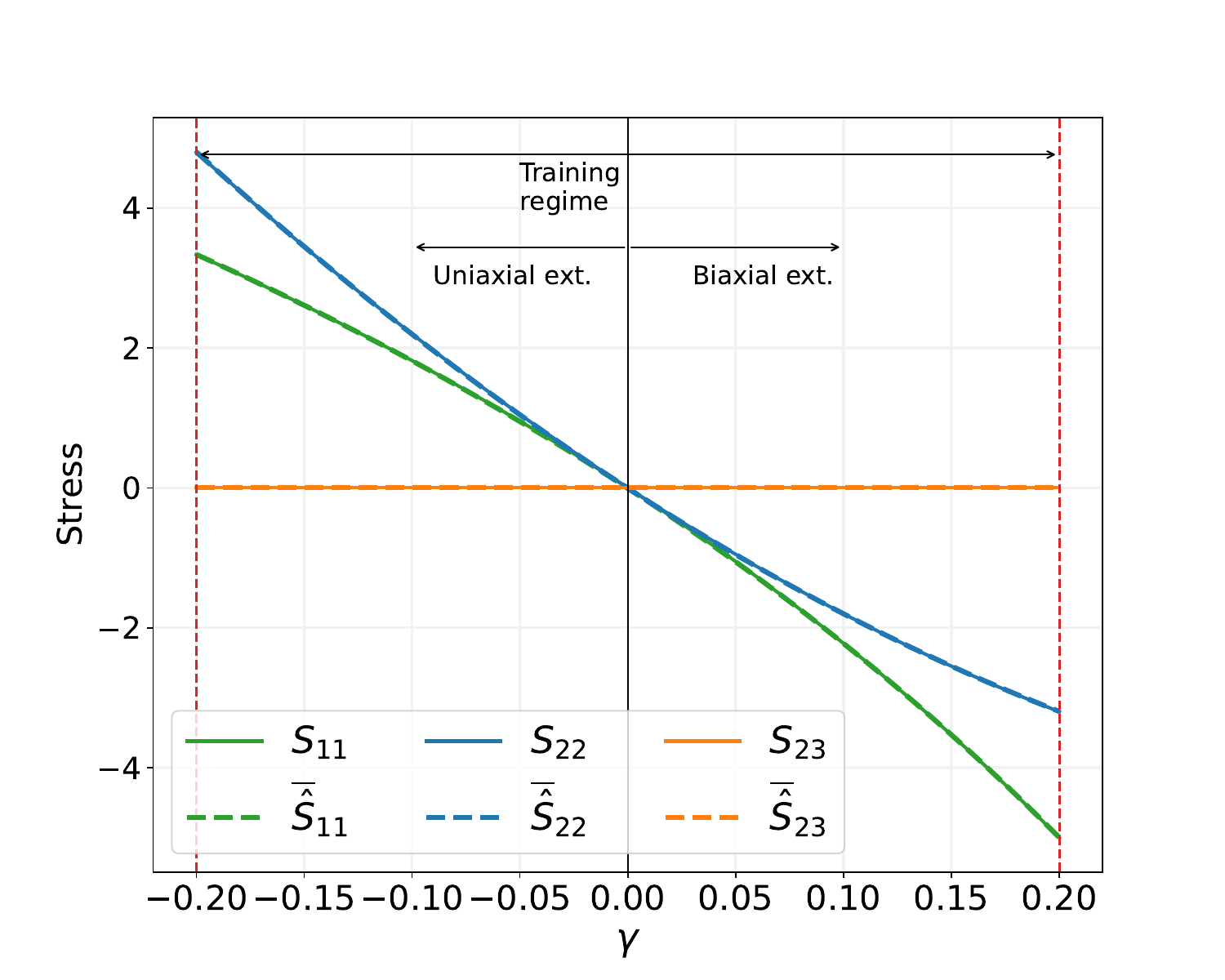}
\caption{Mono-Coeff stress}
\end{subfigure}

\caption{Reference and predicted coefficients around the undeformed configuration for uniaxial ($\gamma<0$) and biaxial ($\gamma>0$) extension for the Mono-Coeff representation obtained from the Mooney-Rivlin model.
}\label{fig:NotDiscCoeffUni}
\end{figure}

Remarkably, the resulting predicted coefficients for Rivlin-Coeff are practically equivalent to the derived coefficients from the potential prediction of Convex-Pot, \Fref{fig:DiscCoeffUniRivICNN}. This is interesting since Convex-Pot has not been trained on the coefficients.
We furthermore remark that (for this loading case) not all of the representations show a discontinuous coefficient behavior as a result of the coalescent principal strains. For example for Mono-Coff, whose coefficient matrix was also ill-conditioned and had to be adapted, the coefficient behavior is smooth over $\gamma$, see \Fref{fig:NotDiscCoeffUni}.
In this case, the NN coefficients are basically identical to the extracted coefficients obtained from the altered equation system.

%.......................................................................
\subsection{Extrapolation} \label{sec:extrapolation}
\begin{figure}[h!]
\begin{subfigure}[b]{0.3\linewidth}
        \centering
    \includegraphics[scale=0.18]{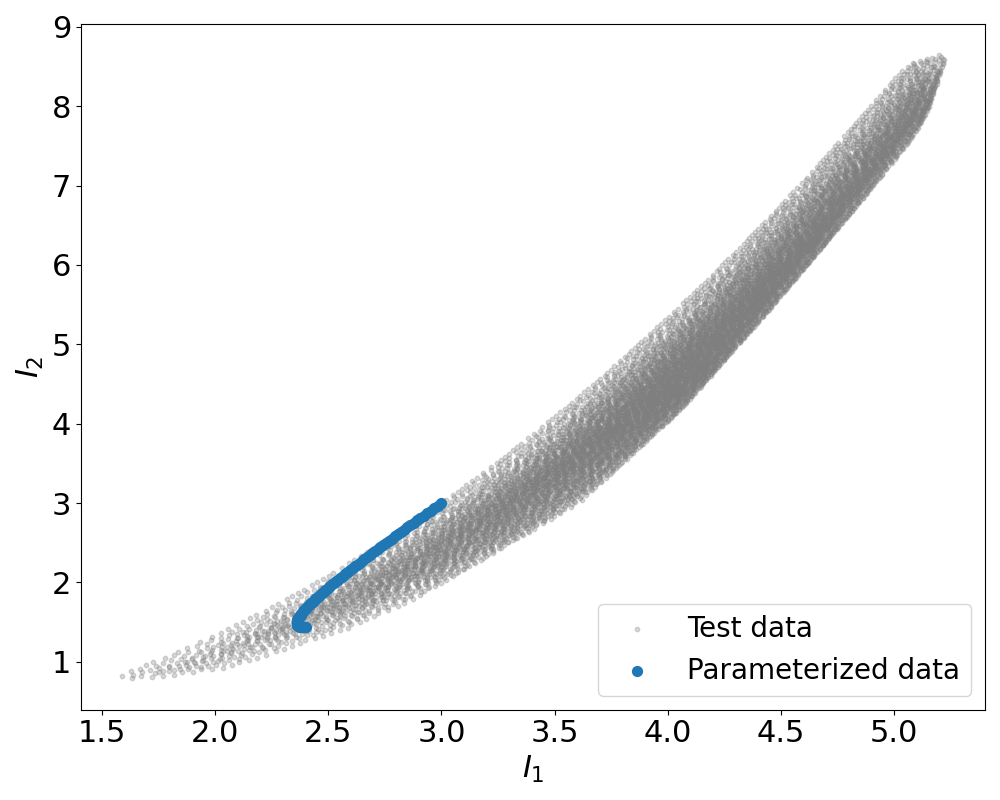}
    \subcaption{}
\end{subfigure}
\begin{subfigure}[b]{0.3\linewidth}
        \centering
\includegraphics[scale=0.18]{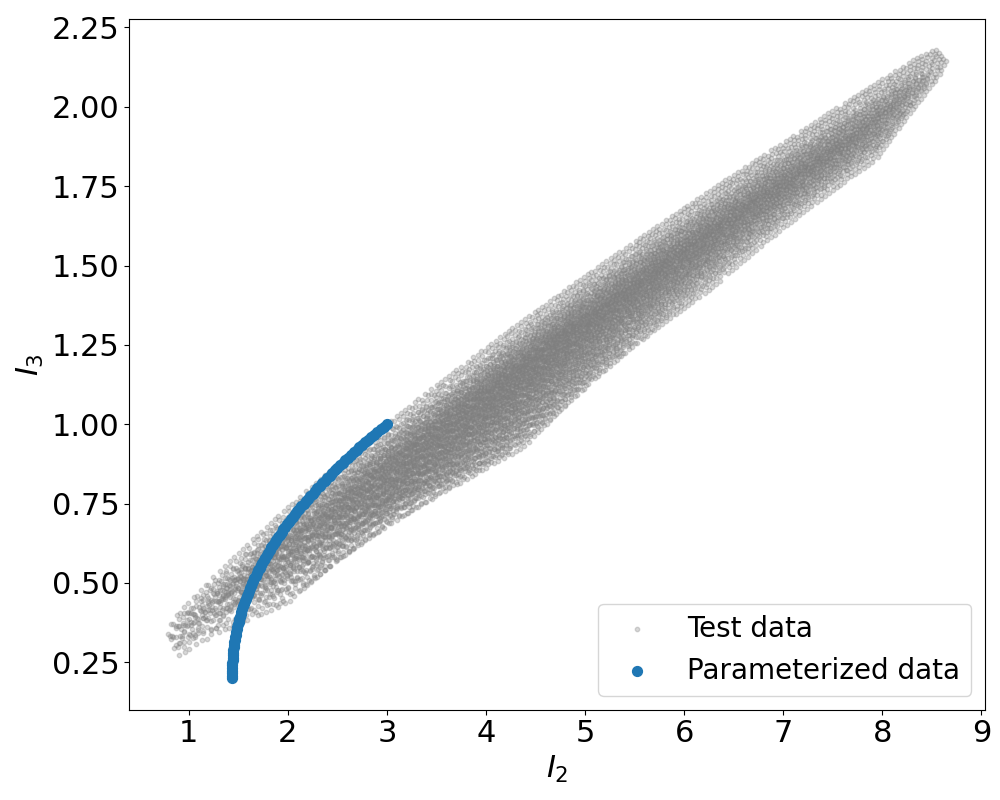}
    \caption{}
\end{subfigure}
\begin{subfigure}[b]{0.3\linewidth}
        \centering
\includegraphics[scale=0.18]{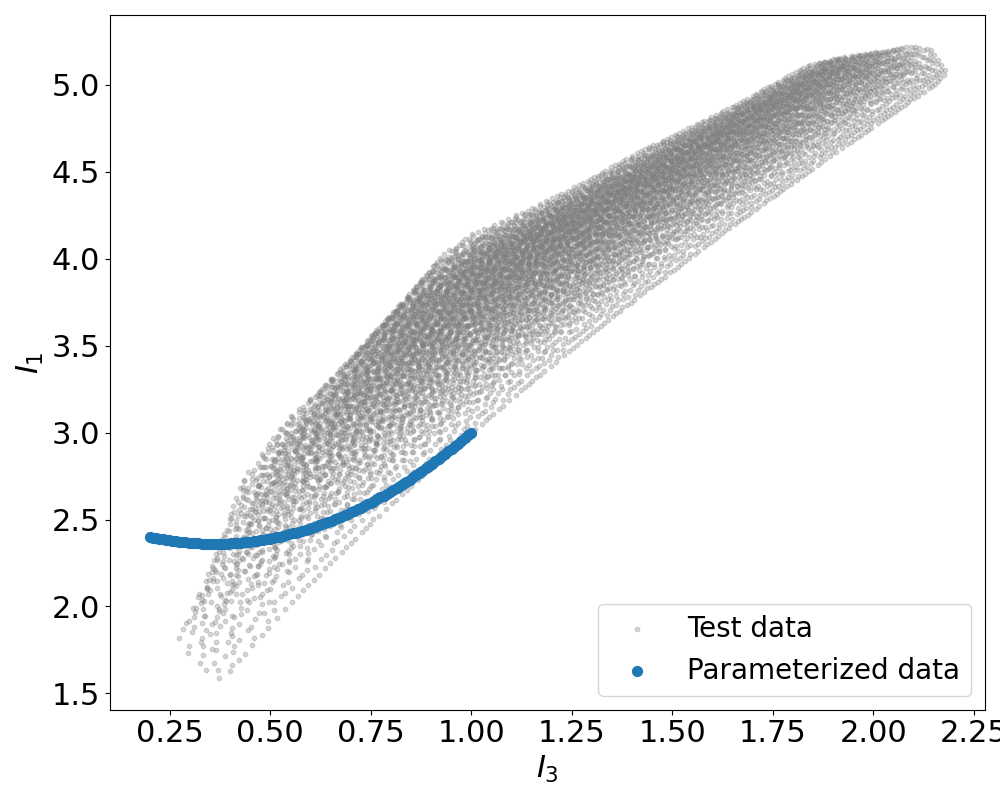}
    \caption{}
\end{subfigure}
    \caption{Parameterization in invariant space given by $\gamma$} \label{fig:path}
\end{figure}

\begin{figure}[h!]
\begin{subfigure}[b]{0.48\linewidth}
        \centering
\includegraphics[scale=0.3]{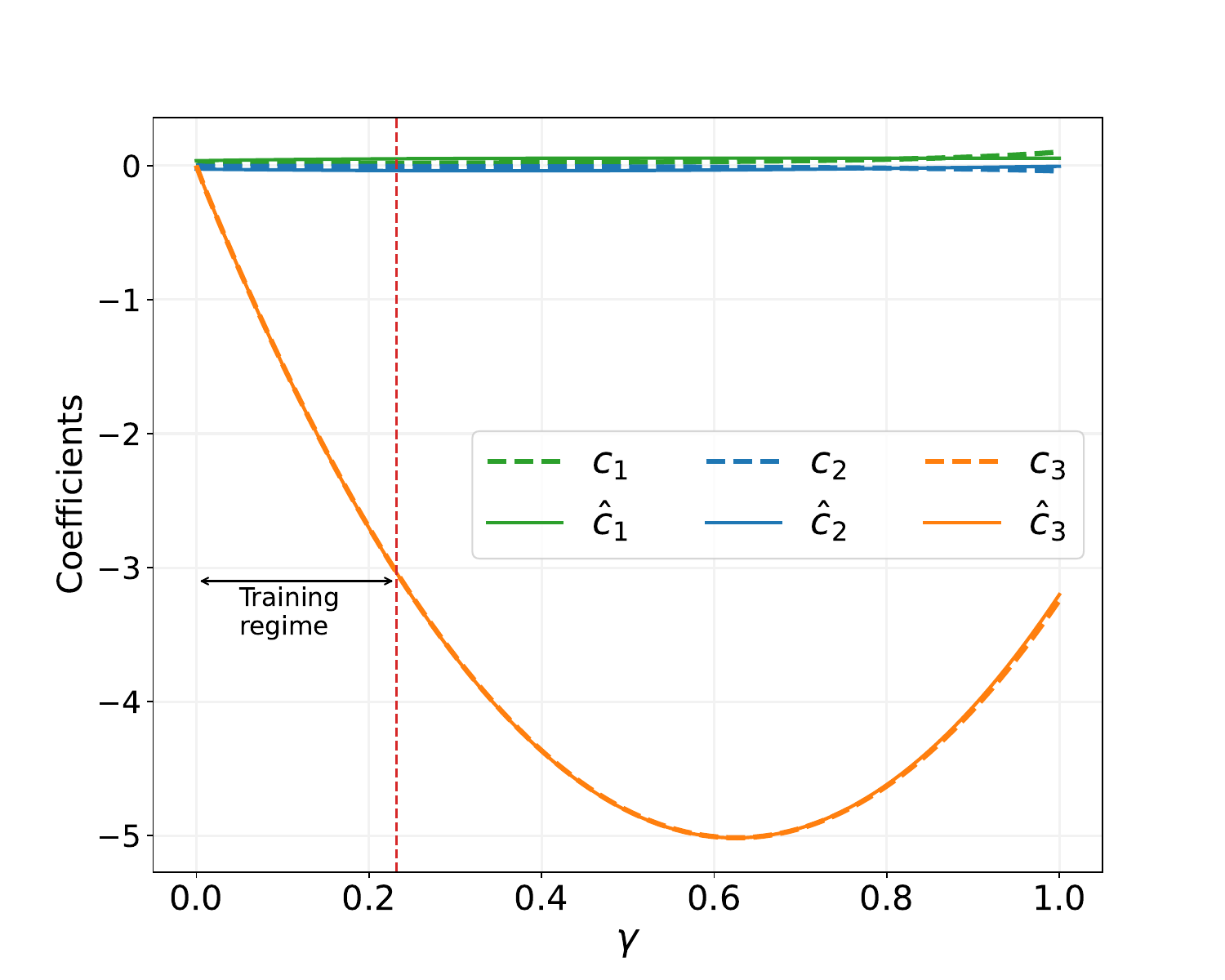}
    \caption{Convex-Pot}
\end{subfigure}
\begin{subfigure}[b]{0.48\linewidth}
        \centering
    \includegraphics[scale=0.3]{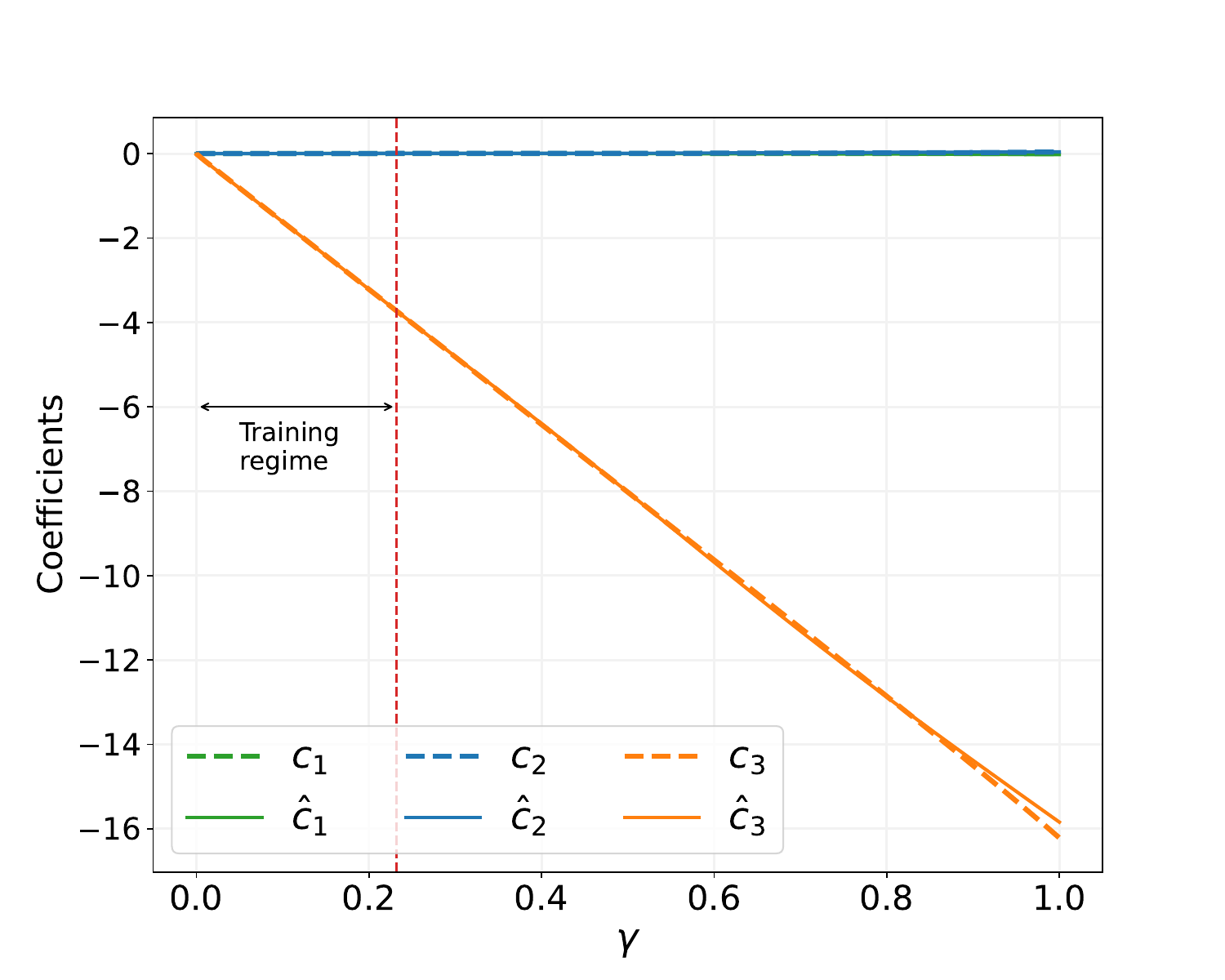}
    \subcaption{Mono-Coeff}
\end{subfigure}
\begin{subfigure}[b]{0.48\linewidth}
        \centering
    \includegraphics[scale=0.3]{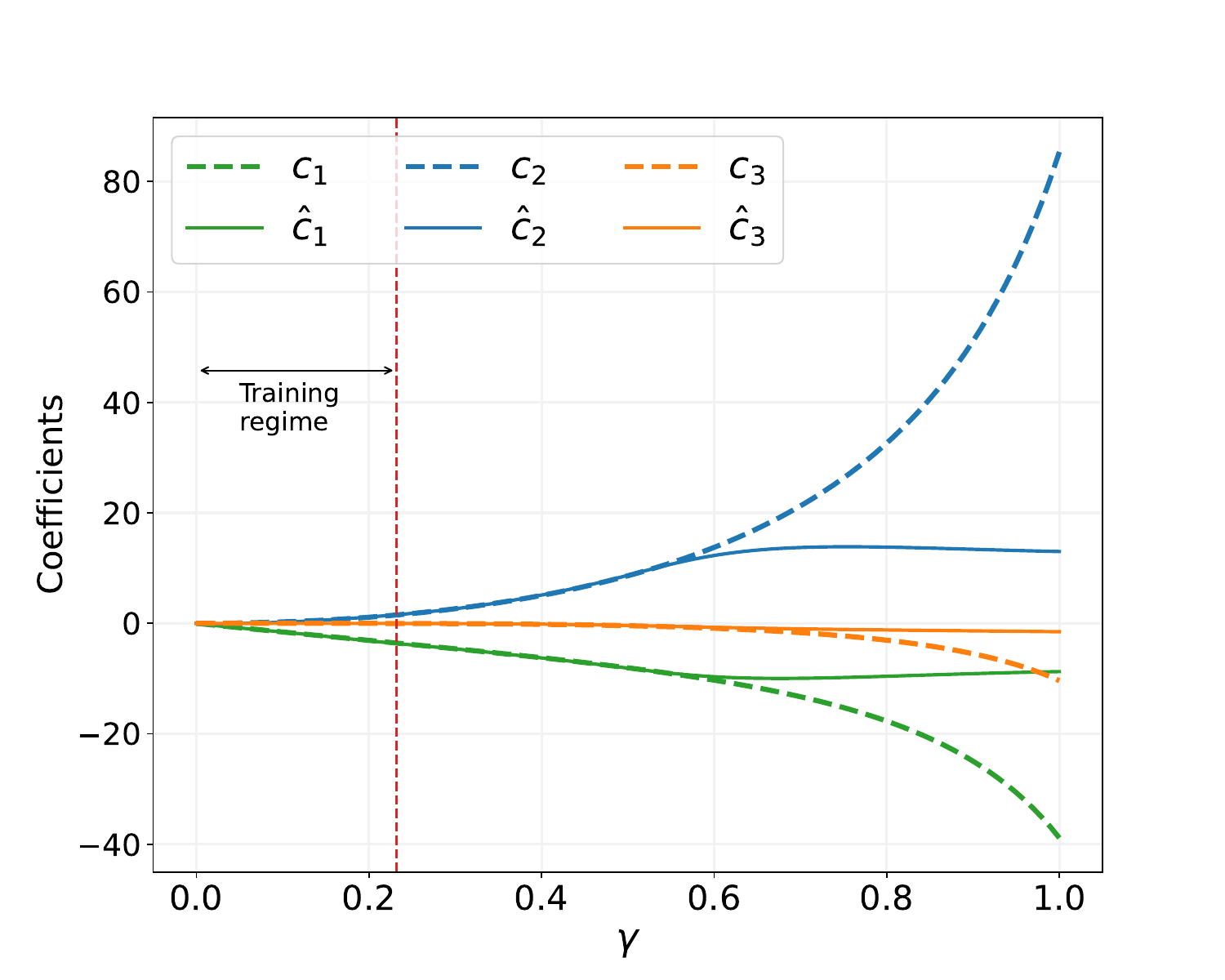}
    \subcaption{Crisc-Pot}
\end{subfigure}
\begin{subfigure}[b]{0.48\linewidth}
        \centering
\includegraphics[scale=0.3]{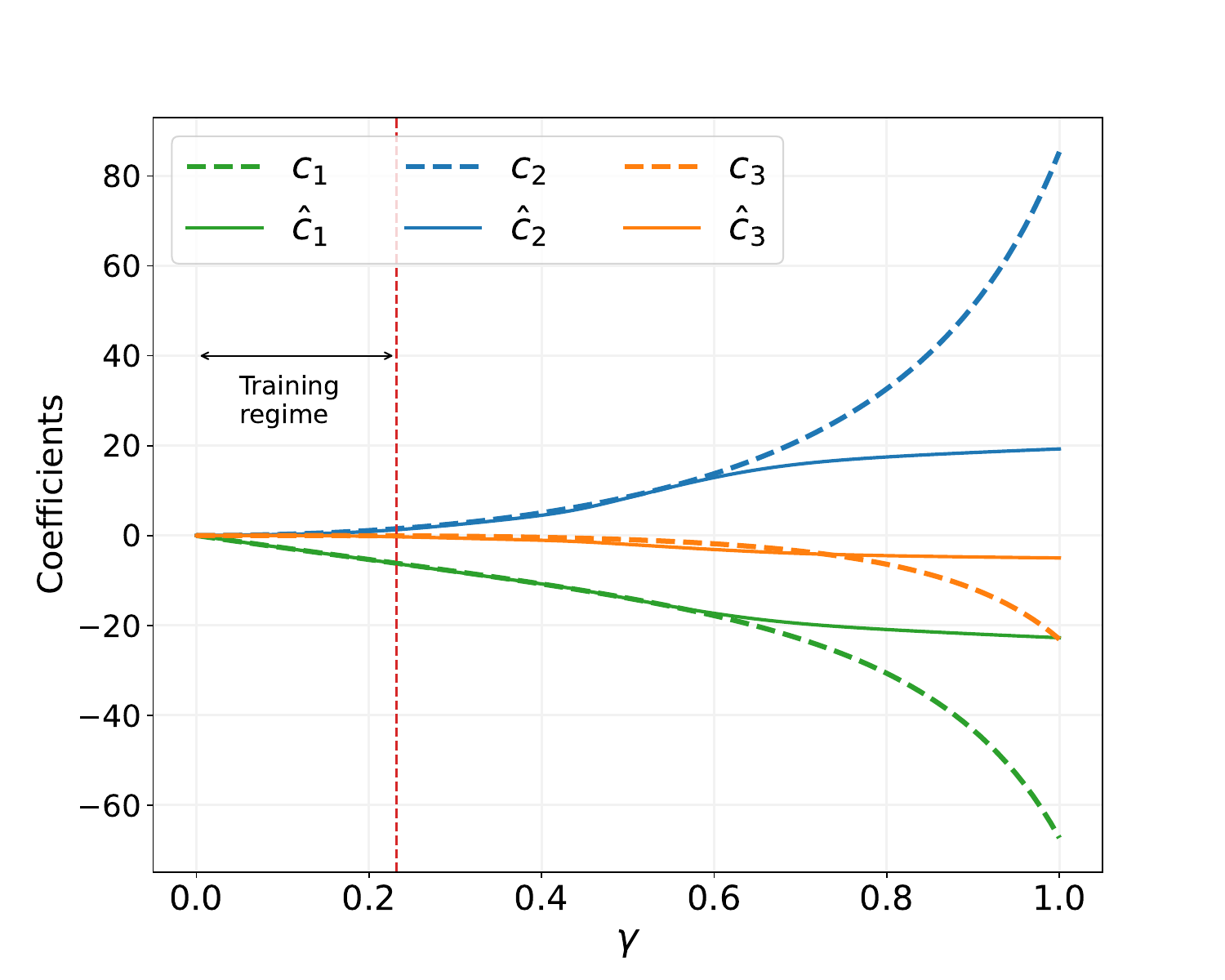}
    \caption{Ortho-Coeff}
\end{subfigure} 
\begin{subfigure}[b]{1.0\linewidth}
        \centering
\includegraphics[scale=0.3]{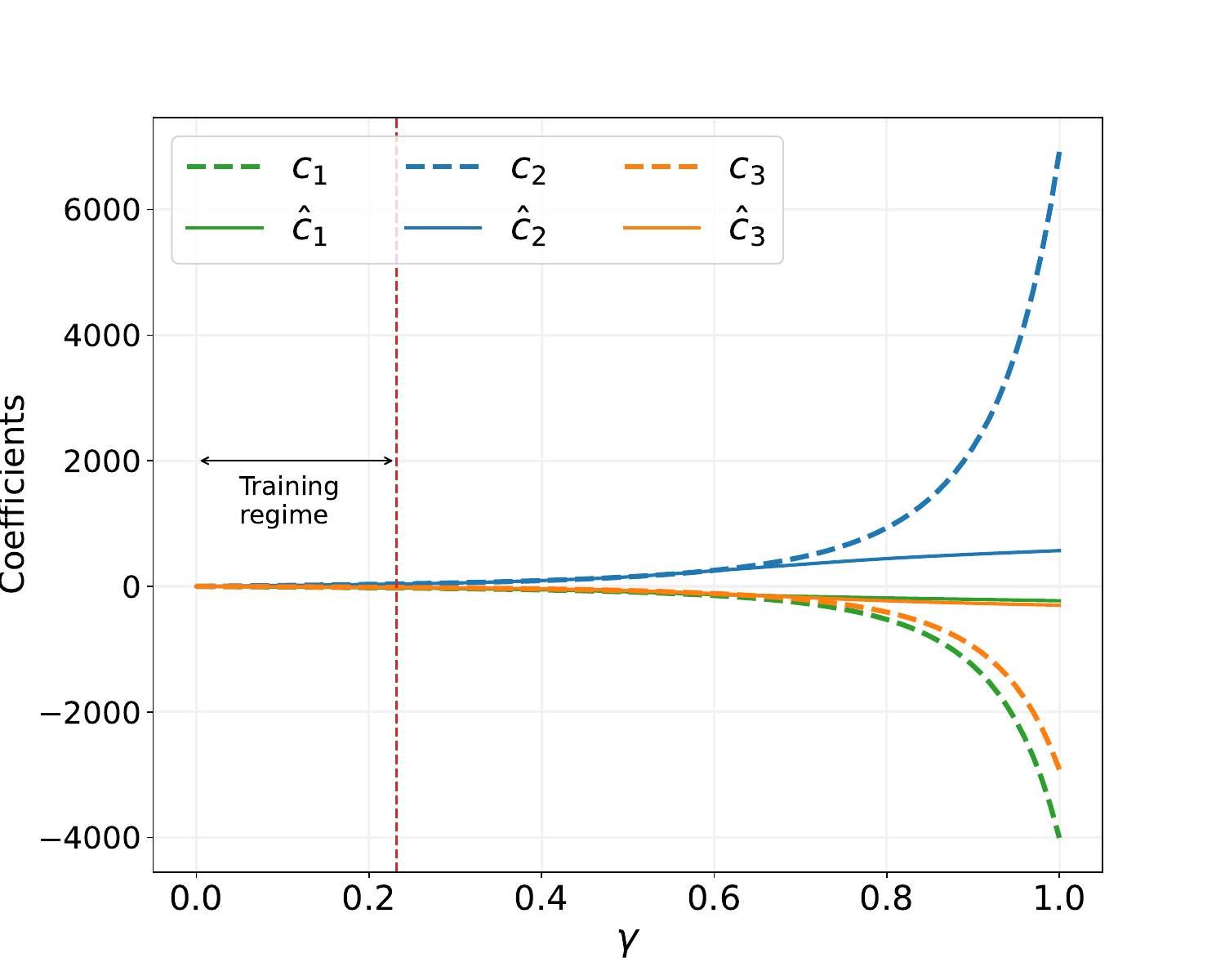}
    \caption{St.V-Coeff}
\end{subfigure}
    \caption{Selected Mooney-Rivlin coefficient curves over parameter $\gamma$. Dashed lines: extracted coefficients, solid lines: trained NN coefficients.} \label{fig:path_coefficients}
\end{figure}

\begin{figure}[h!]
\begin{subfigure}[b]{0.48\linewidth}
        \centering
\includegraphics[scale=0.3]{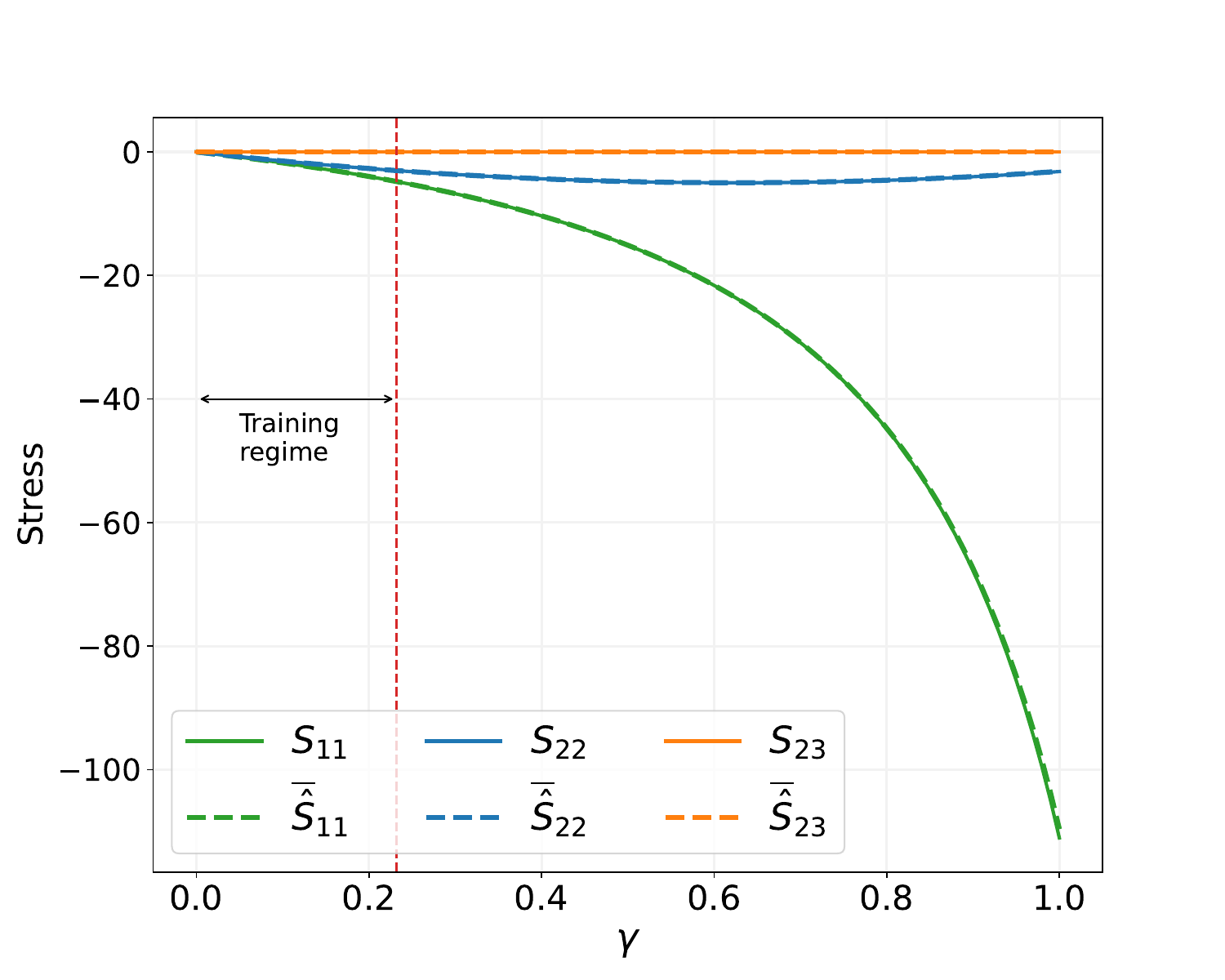}
    \caption{Convex-Pot}
\end{subfigure}
\begin{subfigure}[b]{0.48\linewidth}
        \centering
    \includegraphics[scale=0.3]{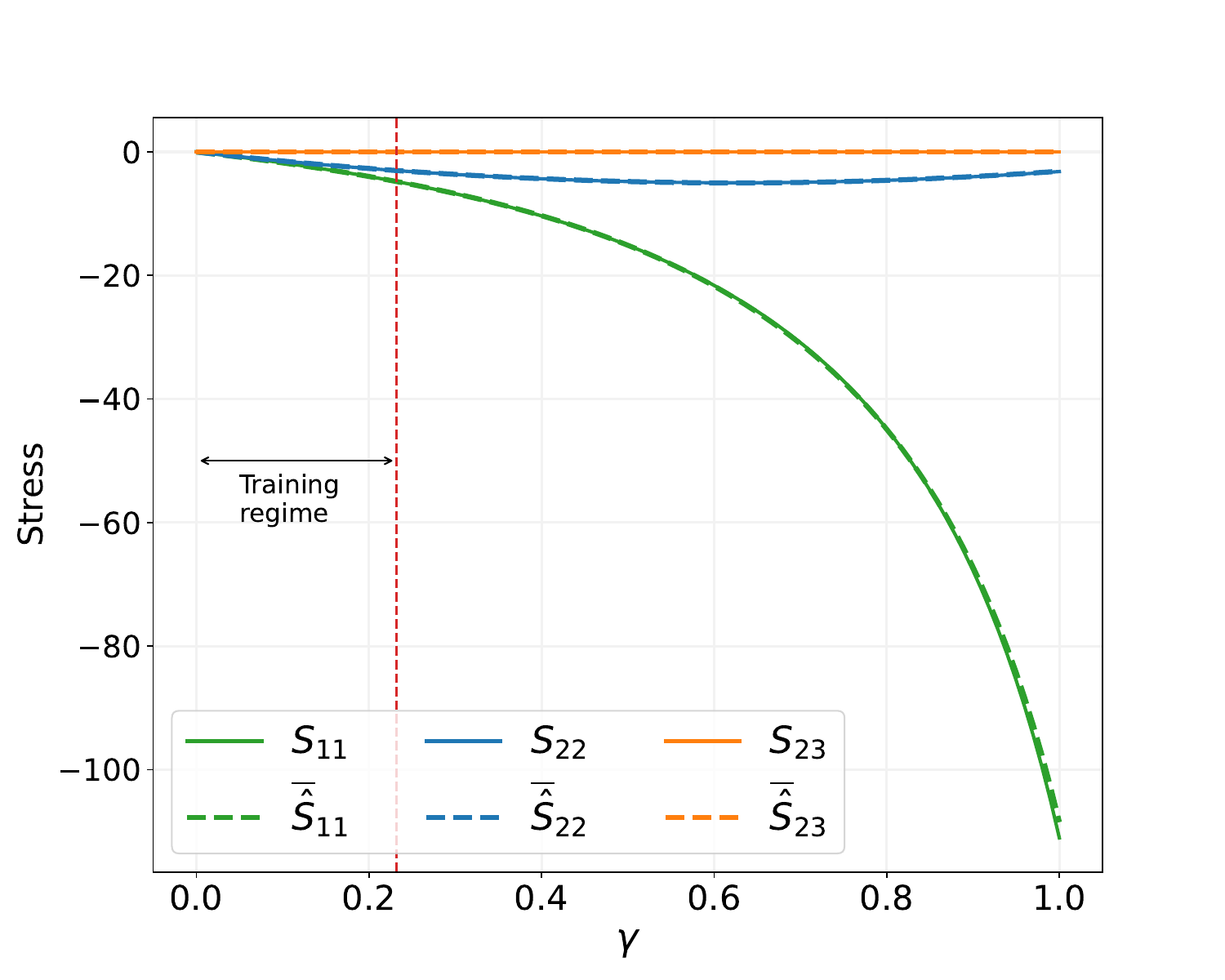}
    \subcaption{Mono-Coeff}
\end{subfigure}
\begin{subfigure}[b]{0.48\linewidth}
        \centering
    \includegraphics[scale=0.3]{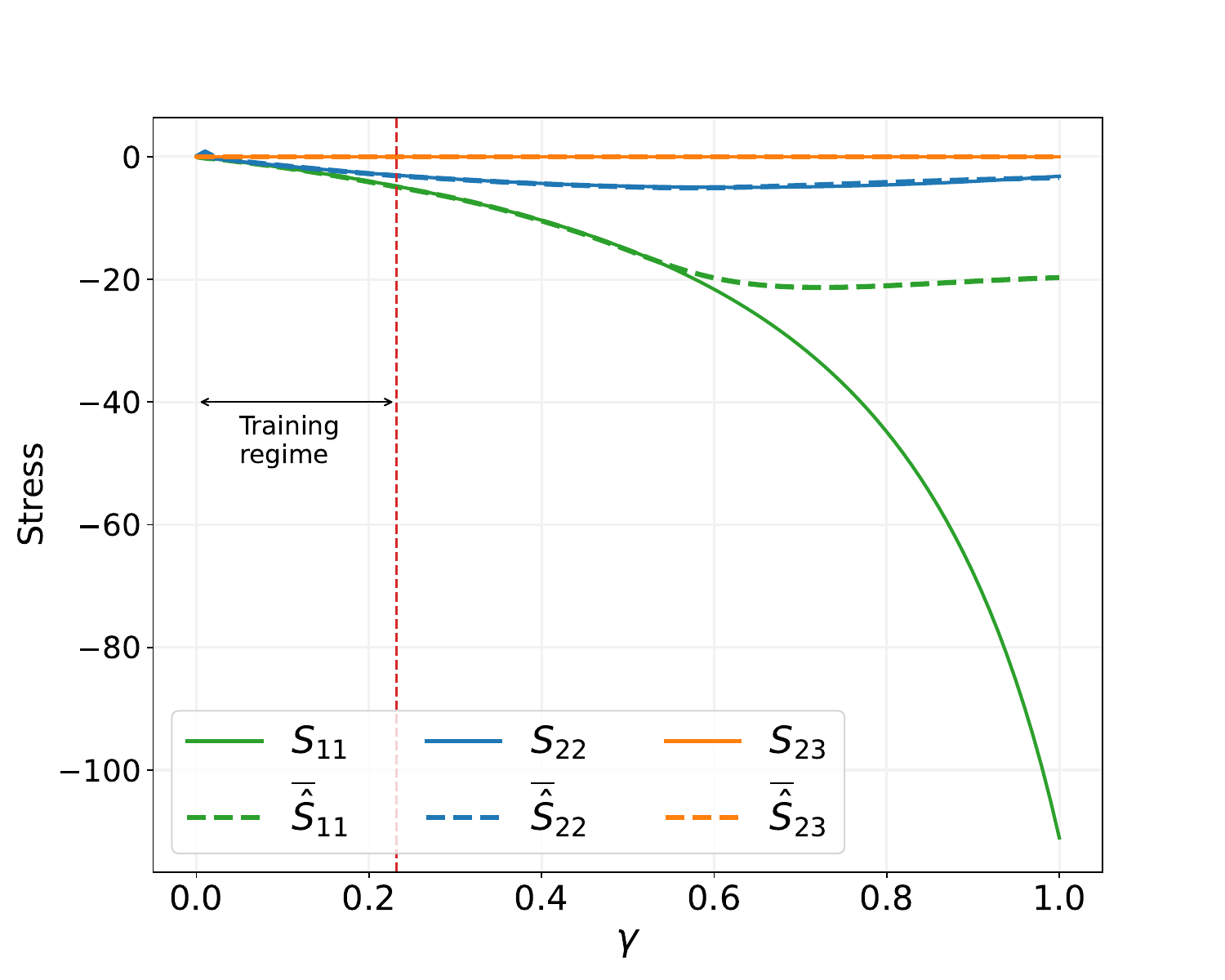}
    \subcaption{Crisc-Pot}
\end{subfigure}
\begin{subfigure}[b]{0.48\linewidth}
        \centering
\includegraphics[scale=0.3]{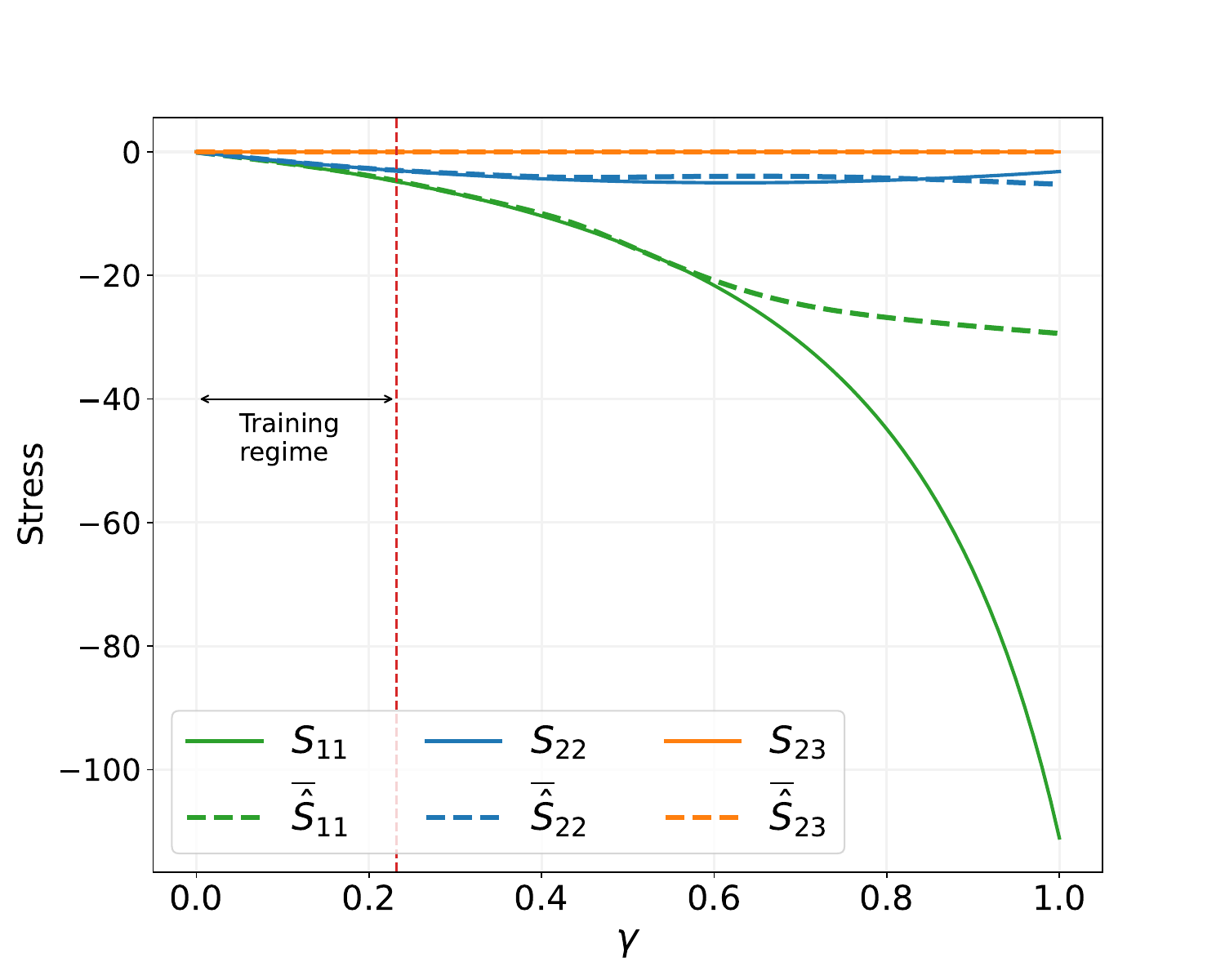}
    \caption{Ortho-Coeff}
\end{subfigure}
\begin{subfigure}[b]{1.0\linewidth}
        \centering
\includegraphics[scale=0.3]{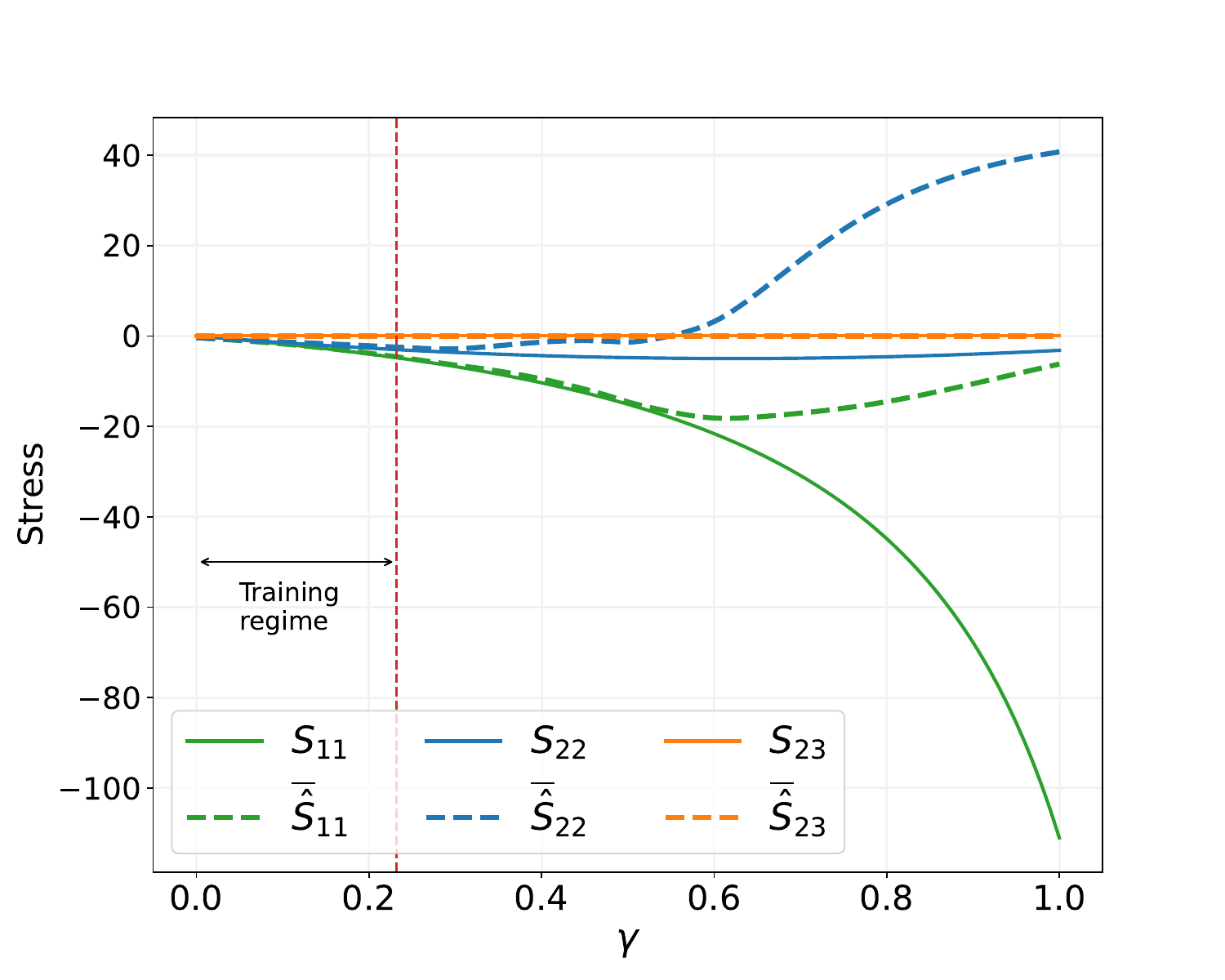}
    \caption{St.V-Coeff}
\end{subfigure}
    \caption{Selected Mooney-Rivlin stress-strain curves over parameter $\gamma$. Dashed lines: data, solid lines: trained NN.} \label{fig:path_stress}
\end{figure}

%.......................................................................
To highlight the predictive quality of selected representations consider a loading path given by the homogenous deformation \cite{currie2004attainable}
\begin{equation}
\begin{aligned}
        I_{1} &= 3 - 1.6\, \gamma  + \gamma^{2} \\
        I_{2} &= 3 - 3.2 \gamma  + 1.64\, \gamma^{2} \\
        I_{3} &= (1-0.8 \, \gamma)^{2} 
\end{aligned}
\end{equation}
where $\gamma\in [0,1]$. The loading path projected into invariant space is shown in
\Fref{fig:path}. We can see that it starts at the undeformed configuration and goes beyond even the range of the test data. The deformation leaves the hull of the training data points at $\gamma \approx 0.25$. We highlight that this path is not explicitly part of the training data set. 
For five of the representations, \Fref{fig:path_coefficients} shows the expected coefficients and trained coefficients respectively. The range of the training regime is highlighted. 
Surprisingly, all the selected models fit the reference coefficients sufficiently well, even St.V-Coeff which was by far the worst-performing approach in terms of generalization error. 
Comparing the fits of the models to the conclusions drawn from \Fref{fig:test_error_comparison}, it becomes evident that the magnitude-wise largest coefficient predictions of the poorly performing representations tend to stagnate earlier.
Similar reasoning can be followed when examining \Fref{fig:path_stress} which plots three components of the expected stress and their predicted counterparts from the neural network models, where we observe that the representations which have only one significant coefficient on the path are also the best performing.

%%%%%%%%%%%%%%%%%%%%%%%%%%%%%%%%%%%%%%%%%%%%%%%%%%%%%
\section{Conclusion} \label{sec:conclusion}
%%%%%%%%%%%%%%%%%%%%%%%%%%%%%%%%%%%%%%%%%%%%%%%%%%%%%
In this work, we investigated the a wide variety of tensor basis neural network models, including previously unexplored alternatives to the classical Finger-Rivlin-Ericksen stress representation, such as TBNNs with an orthogonal basis and others with a St. Venant basis. 
We compared coefficient-based TBNNs against potential-based models and discussed and summarized techniques to obtain reference coefficient values from stress-strain data pairs.

In our cases studies involving six test datasets for three materials and two noise levels, the representations derived from the classical Finger-Rivlin-Ericksen formulation yield the best generalization performance.
This was surprising to us, because we had initially believed that in particular, the potential advantages of orthogonal bases, e.g. continuity of the coefficients and linear independence, would also translate into better extrapolations. 
We found that the generalization capabilities of the stress representations is largely dependent on the simplicity and lower complexity of the coefficient functions in the sense that the coefficient functions are smooth and monotonic, like low-order polynomials.
This is the case when the stress generators (bases) already describe a majority of the stress-strain data complexities and hence the invariant-coefficient mappings are simpler.
The introduction of physics-constrained extensions to the TBNN in particular convexity~(potential) or monotonicity~(coefficients) appears to be the most beneficial for accurate extrapolation and generalization. 
We also observed that the assumption of the existence of a potential helps the performance, i.e. potential-based models generalize better than coefficient-based models for the same stress representation.

In future work the study will be extended to include anisotropic representations. We also aim to introduce monotonically increasing neural network formulations that eliminate the restrictiveness of current implementations, where a subset of the outputs is monotonically increasing with only a subset of the inputs, to potentially improve the performance of the monotonic Rivlin representation.

%%%%%%%%%%%%%%%%%%%%%%%%%%%%%%%%%%%%%%%%%%%%%%%%%%%%%
\section*{Acknowledgments}
%%%%%%%%%%%%%%%%%%%%%%%%%%%%%%%%%%%%%%%%%%%%%%%%%%%%%

REJ would like to thank Professor J.B. Estrada (University of Michigan) and NB would like to thank Professor K.T. Ramesh (Johns-Hopkins University) for independently pointing out Criscione's work on stress representations with an orthogonal basis.

JF and NB gratefully acknowledge support by the Air Force Office of Scientific Research under award number FA9550-22-1-0075.

Sandia National Laboratories is a multimission laboratory managed and operated by National Technology and Engineering Solutions of Sandia, LLC., a wholly owned subsidiary of Honeywell International, Inc., for the U.S. Department of Energy's National Nuclear Security Administration under contract DE-NA-0003525. This paper describes objective technical results and analysis. Any subjective views or opinions that might be expressed in the paper do not necessarily represent the views of the U.S. Department of Energy or the United States Government.

\clearpage
%\bibliography{hyperelasticity.bib}

\appendix
\renewcommand{\theequation}{\thesection.\arabic{equation}}
\numberwithin{equation}{section}
%%%%%%%%%%%%%%%%%%%%%%%%%%%%%%%%%%%%%%%%%
\section{Solutions for the coefficients} \label{app:multiplicity} \setcounter{equation}{0}
%%%%%%%%%%%%%%%%%%%%%%%%%%%%%%%%%%%%%%%%%

There are many variants to the solution of the coefficients at particular values of the stress and stretch/strain.

Using the collinearity of the eigenbasis for the stress and stretch tensors leads to a solution of the form:
\begin{equation} \label{eq:generic_coef_solve}
\Vs \cs = \mathsf{s}
\end{equation}
where the matrix $\mathsf{V}$ is a function of the eigenvalues of $\Cb$, $\mathsf{c}$ is the coefficient vector, and $\mathsf{s}$ is the vector of stress eigenvalues.
For instance, the solution for the power basis $\Bc = \{ \Cb^a, a=0,1,-1 \}$: 
\begin{equation} \label{eq:coef_solve1}
\begin{bmatrix}
c_1 \\
c_2 \\
c_3 
\end{bmatrix}
= 
\begin{bmatrix}
1 & \epsilon_1 & \epsilon_1^{-1}  \\
1 & \epsilon_2 & \epsilon_2^{-1}  \\
1 & \epsilon_3 & \epsilon_3^{-1}  
\end{bmatrix}^{-1}
\begin{bmatrix}
\sigma_1 \\
\sigma_2 \\
\sigma_3 
\end{bmatrix}
=
\begin{bmatrix}
 \frac{\epsilon_1 \left(\epsilon_2+\epsilon_3\right)}{\left(\epsilon_1-\epsilon_2\right) \left(\epsilon_3-\epsilon_1\right)} 
   & \frac{\epsilon_2 \left(\epsilon  _1+\epsilon_3\right)}{\left(\epsilon  _1-\epsilon_2\right) \left(\epsilon_2-\epsilon_3\right)} 
 & \frac{\epsilon_3 \left(\epsilon_1+\epsilon_2\right)}{\left(\epsilon_1-\epsilon_3\right) \left(\epsilon_3-\epsilon_2\right)} \\
 \frac{\epsilon_1}{\left(\epsilon_1-\epsilon_2\right) \left(\epsilon_1-\epsilon_3\right)} 
 & \frac{\epsilon_2}{\left(\epsilon_2-\epsilon_1\right) \left(\epsilon _2-\epsilon_3\right)} 
 & \frac{\epsilon_3}{\left(\epsilon_1-\epsilon_3\right) \left(\epsilon_2-\epsilon_3\right)} \\
 \frac{\epsilon_1 \epsilon_2 \epsilon_3}{\left(\epsilon_1-\epsilon_2\right) \left(\epsilon_1-\epsilon_3\right)} 
 & \frac{\epsilon_1 \epsilon_2 \epsilon_3}{\left(\epsilon_1-\epsilon_2\right) \left(\epsilon_3-\epsilon_2\right)} 
 & \frac{\epsilon_1 \epsilon_2 \epsilon_3}{\left(\epsilon_1-\epsilon_3\right) \left(\epsilon_2-\epsilon_3\right)} \\
\end{bmatrix}
\begin{bmatrix}
\sigma_1 \\
\sigma_2 \\
\sigma_3 
\end{bmatrix}
\end{equation}
Likewise for $\Bc = \{ \Cb^a, a=0,1,2 \}$
\begin{equation}  \label{eq:coef_solve2}
\begin{bmatrix}
c_1 \\
c_2 \\
c_3 
\end{bmatrix}
= 
\begin{bmatrix}
1 & \epsilon_1 & \epsilon_1^{2}  \\
1 & \epsilon_2 & \epsilon_2^{2}  \\
1 & \epsilon_3 & \epsilon_3^{2}  
\end{bmatrix}^{-1}
\begin{bmatrix}
\sigma_1 \\
\sigma_2 \\
\sigma_3 
\end{bmatrix}
= 
\begin{bmatrix}
 \frac{\epsilon _2 \epsilon _3}{\left(\epsilon _1-\epsilon _2\right) \left(\epsilon
   _1-\epsilon _3\right)} & \frac{\epsilon _1 \epsilon _3}{\left(\epsilon _1-\epsilon
   _2\right) \left(\epsilon _3-\epsilon _2\right)} & \frac{\epsilon _1 \epsilon
   _2}{\left(\epsilon _1-\epsilon _3\right) \left(\epsilon _2-\epsilon _3\right)} \\
 -\frac{\epsilon _2+\epsilon _3}{\left(\epsilon _1-\epsilon _2\right) \left(\epsilon
   _1-\epsilon _3\right)} & \frac{\epsilon _1+\epsilon _3}{\left(\epsilon _1-\epsilon
   _2\right) \left(\epsilon _2-\epsilon _3\right)} & \frac{\epsilon _1+\epsilon
   _2}{\left(\epsilon _1-\epsilon _3\right) \left(\epsilon _3-\epsilon _2\right)} \\
 \frac{1}{\left(\epsilon _1-\epsilon _2\right) \left(\epsilon _1-\epsilon _3\right)} &
   \frac{1}{\left(\epsilon _2-\epsilon _1\right) \left(\epsilon _2-\epsilon _3\right)} &
   \frac{1}{\left(\epsilon _1-\epsilon _3\right) \left(\epsilon _2-\epsilon _3\right)} \\
\end{bmatrix}
\begin{bmatrix}
\sigma_1 \\
\sigma_2 \\
\sigma_3 
\end{bmatrix}
\end{equation}
Note that the solution \eref{eq:coef_solve1} or \eref{eq:coef_solve2}  is permutational invariant to the ordering of the eigenvalues.
Also the symmetry of the least squares version of \eref{eq:generic_coef_solve}, $\Vs^T \Vs \cs = \Vs^T \mathsf{s}$ may aid the solution for the coefficient values.

An alternative solution in terms of invariants \cite{scheidler1996smoothness} (not eigenvalues) is simply the matrix of inner products:
\begin{equation}
\begin{bmatrix}
\Ib : \Ib & \Ib : \Cb & \Ib : \Cb^2 \\
\Cb : \Ib & \Cb : \Cb & \Cb : \Cb^2 \\
\Cb^2 : \Ib & \Cb^2 : \Cb & \Cb^2 : \Cb^2 \\
\end{bmatrix}
\begin{bmatrix}
c_0 \\
c_1 \\
c_2
\end{bmatrix}
\begin{bmatrix}
\Sb : \Ib \\
\Sb : \Cb \\
\Sb : \Cb^2 \\
\end{bmatrix}
\end{equation}
or more generally
\begin{equation}
[ \Bb_i : \Bb_j ] [ c_j ] = [ \Bb_i : \Sb ]
\end{equation}
From this form the connection between a linear least squares approach and projection is clear.
In fact, for an orthogonal basis the matrix $[ \Bb_i : \Bb_j ]$  is diagonal.
For the system in terms of eigenvalues
\begin{equation}
[ \ab_i \otimes \ab_i : \Bb_j ] [ c_j ] = [ \ab_i \otimes \ab_i : \Sb ]
\end{equation}

Conditioning of these systems of equations is an issue and depends on the stretch or strain measure.
The stretch $\Cb$ has $O(1)$ eigenvalues, while the strain $\Eb$ has $O(0)$ eigenvalues.
For basis with powers 0,1,2 of $\Cb$,
\begin{equation}
\det \Vs = (\epsilon_1 - \epsilon_2) (\epsilon_2 - \epsilon_3) (\epsilon_3 - \epsilon_1)
\end{equation}
while for powers 0,1,-1 of $\Cb$ 
\begin{equation}
\det \Vs = (\epsilon_1 - \epsilon_2) (\epsilon_2 - \epsilon_3) (\epsilon_3 - \epsilon_1) / (\epsilon_1 \epsilon_2 \epsilon_3)
\end{equation}
For inner product solve 
\begin{eqnarray}
\det \Vs &=& (\Bb_1 : \Bb_1) (\Bb_2 : \Bb_2) (\Bb_3 : \Bb_3)  + 2  (\Bb_1 : \Bb_2)  (\Bb_2 : \Bb_3)  (\Bb_3 : \Bb_1) \\
&-& (\Bb_1 : \Bb_1)^2  (\Bb_2 : \Bb_3)  - (\Bb_2 : \Bb_2)^2  (\Bb_3 : \Bb_1) - (\Bb_3 : \Bb_3)^2  (\Bb_1 : \Bb_2) \nonumber
\end{eqnarray}
Note the stretch eigenvalues are $O(1)$ but the strain eigenvalues are $O(0)$.
Vandermonde-like matrices are well-known to be ill-conditioned \cite{pan2016bad}.

None of these systems are solvable as is for repeated eigenvalues.
Gurtin \cite{gurtin1982introduction} provided a well-conditioned solution by way of solving a  reduced system.
For instance, in the case of $\Sb=\Sb(\Cb)$ and two of the eigenvalues of $\Cb$ are identical (as in uniaxial tension), $\Cb$ can be represented as $\Cb = \epsilon_1 \ab\otimes\ab + \epsilon_2 (\Ib - \ab\otimes\ab)$ where $\lambda_1$ is the unique eigenvalue and $\epsilon_2$ is the repeated one.
Any vector perpendicular to $\ab$ is an eigenvector.
Now instead of solving \eqref{eq:coef_solve1}
only
\begin{equation} \label{eq:system2}
\begin{bmatrix}
\sigma_1 \\ \sigma_2 
\end{bmatrix}
=
\begin{bmatrix}
1 & \epsilon_1 \\
1 & \epsilon_2 \\
\end{bmatrix}
\begin{bmatrix}
c_0 \\ c_1 
\end{bmatrix}.
\end{equation}
needs to be solved since the assumption is that $c_2 = 0$ in this case.
We proposed an alternative scheme in \cref{frankel2020tensor}.
In such a case, the Cayley-Hamilton theorem and continuity of the deformation to stress mapping require that the derivative of the stress eigenvalue with respect to the repeated eigenvalue must be zero, so
\begin{equation}
0 = c_0 + 2 \epsilon_0 c_1.
\end{equation}
should be substituted for one of the redundant equations.
If all the eigenvalues are equal, the second derivative must also be zero, giving the additional equation
\begin{equation}
c_1 = 0
\end{equation}
to replace another of the redundant equations in the system of equations, in which case the result is trivially $c_0 = \sigma_1$ and $c_1=c_2 = 0$.

Note these solutions for special cases do not ensure the continuity of the coefficient functions \cite{serrin1959derivation,man1995smoothness,scheidler1996smoothness,xiao2002basic}.
To allow for continuity the stresses must be twice differentiable with respect to $\Cb$.
Xiao \etal \cite{xiao2002basic} gives non-eigenvalue indicators of multiplicity
\begin{eqnarray}
\text{if} \ & \| \Bb_3 \| \neq 0 \ &\text{then} \ m=3 \\
\text{if}\  & \| \Bb_3 \| = 0 \ \text{and} \ \| \Bb_2\| \neq 0 \ & \text{then} \ m=2  \nonumber \\
\text{if}\ &  \| \Bb_3 \| = 0 \ \text{and} \ \| \Bb_2 \| = 0 \ &\text{then} \ m=1 \nonumber
\end{eqnarray}
where $\Bb_2 = \dev\Cb$ and $\Bb_3 = \| \dev \Cb \|^2 \dev(\Cb^2) - (\Cb^2 : \dev\Cb) \dev\Cb$.

%%%%%%%%%%%%%%%%%%%%%%%%%%%%%%%%%%%%%%%%%
\section{Orthogonal basis} \label{app:orthogonal} \setcounter{equation}{0}
%%%%%%%%%%%%%%%%%%%%%%%%%%%%%%%%%%%%%%%%%
Criscione \etal \cite{criscione2000invariant} and others \cite{xiao2002basic} have employed a stress representation with an orthogonal basis.
Criscione \etal was able to relate an orthogonal basis to derivatives of particular invariants.
Noting the form of the invariants is valid for invariants of the form applied to any tensor argument and we use a different scaling of the invariants. 

Derivatives of the first two invariants are straightforward:
\begin{equation}
\partialb_\Cb \tr \Cb =   \partialb_\Cb \Cb : \Ib = \Ib
\end{equation}
and
\begin{equation}
\partialb_\Cb \| \dev \Cb \| 
= 1/2 ( \dev \Cb : \dev \Cb )^{-1/2} \, 2 \dev \Cb : \Dbb
= \frac{\dev \Cb}{\| \dev \Cb \|}
\end{equation}
where 
\begin{equation}
    \Dbb = \Ibb - \frac{1}{3}\Ib \otimes \Ib
\end{equation} is such that $\dev \Cb = \Dbb \Cb$.
Note $\Dbb$ is in the form of a projector 
\begin{equation} \label{eq:projector}
\Pbb_{\Ab} = \Ibb - \frac{1}{( \Ab : \Ab )} \Ab \otimes \Ab 
\end{equation}
hence $\Dbb = \Pbb_\Ib$ and $\Dbb^n = \Dbb$.
Note $\Ibb = \delta_{il} \delta_{jl} \eb_i \otimes \eb_j \otimes \eb_k \otimes \eb_l$.

The derivative of the third invariant $K_3 = \det\left( {\dev \Cb}/{\| \dev \Cb \|} \right)$ 
\begin{eqnarray}
\partialb_\Cb K_3 
&=& \partialb_\Cb \det\left( \Ab\right) 
=\Ab^* \, : \partialb_\Cb \Ab 
= \Ab^* \, : \Pbb_{\dev{\Cb}} \Dbb   \\
&=& \Ab^* \, : \frac{1}{\| \dev \Cb \|}  \left[ \Ibb - \frac{1}{3} \Ib \otimes \Ib - \Ab \otimes \Ab \right]  \nonumber \\
&=& \det\left( \Ab\right) \Ab ^{-T} \, : \frac{1}{\| \dev \Cb \|} \left[ \Ibb - \frac{1}{3} \Ib \otimes \Ib - \Ab \otimes \Ab \right] \nonumber  \\
&=&  \frac{\det\left( \Ab\right)}{\| \dev \Cb \|} \left[ \Ab^{-1} - \frac{1}{3} \tr(\Ab^{-1}) \Ib - 3 \Ab \right] \nonumber 
\end{eqnarray}
results from $\partialb_{\Ab} \det\left( \Ab\right) = \Ab^* \equiv \det\left( \Ab\right) \Ab ^{-T}$ and the derivative of a unit tensor $\Ab$ is the projector $\Pbb_{\Ab}$
\begin{eqnarray}
\partialb_\Cb \Ab &=&
\partialb_\Cb \frac{\dev \Cb}{\| \dev \Cb \|} 
= \frac{1}{\| \dev \Cb \|} \left( \Ibb - \frac{1}{\| \dev \Cb \|^2} \dev \Cb \otimes \dev \Cb  \right) \Dbb \\
&=& \frac{1}{\| \dev \Cb \|} \Pbb_{\Ab} \Dbb  
= \frac{1}{\| \dev \Cb \|} \left[ \Ibb - \frac{1}{3} \Ib \otimes \Ib - \Ab \otimes \Ab \right] \nonumber
\end{eqnarray}
using the definition $\Ab \equiv {\dev \Cb}/{\| \dev \Cb \|}$ and $(\Ib \otimes \Ib) : (\Ab \otimes \Ab) = \mathbf{0}$.
Alternatively, Criscione \etal \cite{criscione2000invariant} take the route
\begin{eqnarray}
\partialb_\Cb K_3 &=& \partialb_{\Ab} \, \left( \frac{1}{3} \Ab^3 \right) \partialb_\Cb \Ab 
= \Ab^2 \, :  \Pbb_{\Ab} \Dbb \\
&=&  \Ab^2 : \frac{1}{\| \dev \Cb \|} \left[ \Ibb - \frac{1}{3} \Ib \otimes \Ib - \Ab \otimes \Ab \right]  \nonumber \\
&=&   \frac{1}{\| \dev \Cb \|} \left[  \Ab^2  - \frac{1}{3} \Ib  - \tr\left(\Ab^3\right) \Ab
\right] \nonumber
\end{eqnarray}
from the identity 
\begin{equation}
\det \Ab = \frac{1}{3} \tr \Ab^3
\end{equation}
derived from  the Cayley-Hamilton theorem \eref{eq:cayley-hamilton} applied to $\Ab$:
\begin{equation}
\tr \Ab^3 - \tr(\Ab) \tr(\Ab^2) + I_2 \tr \Ab - 3 \det(\Ab)  = 0
\end{equation}
Note $\tr \Ab^2 = 1$ since $\| \Ab \| = 1$.
The derivative, $\partialb_\Cb K_3$ can be recognized as the third element of the Gram-Schmidt basis in \eref{eq:orthogonal_basis}.
Others, for example, Xiao \etal \cite{xiao2002basic}, start with the usual spherical-deviatoric split and then use the Gram-Schmidt procedure to obtain the third basis element
\begin{equation}
\{ \Ib, \dev \Cb, 
\| \dev \Cb \|^2 \dev(\Cb^2) - (\Cb^2 : \dev \Cb) \dev \Cb \}
\end{equation}
where $\Bb_3 = \| \dev \Cb \|^2 \partialb_{\Cb} K_3$.

We can verify the orthogonality of the basis
\begin{eqnarray}
\Ib : \Ab &=&  \frac{1}{\| \dev{\Ab} \|} \tr \dev(\Ab) = 0 \\
\Ib : \left[ 
\Ab^2 
- \frac{1}{3} \Ib 
- \tr\left(\Ab^3\right) \Ab
\right] 
&=& \tr \Ab^2 - 1 + 0 =  0 \\
\Ab : \left[ 
\Ab^2 
- \frac{1}{3} \Ib 
- \tr\left(\Ab^3\right) \Ab
\right] 
&=& \Ab : \Ab^2 - 0 - \tr(\Ab^3) = 0
\end{eqnarray}
using the identities $\tr \Ab = 0$, $\tr \Ab^2 = 1$,  and $\det \Ab = 1/3 \tr \Ab^3$.
These identities could also be verified using the eigendecomposition $\Cb = \sum_a \epsilon_a \ab_a \otimes \ab_a$.

Smoother alternatives to these invariants: 
\begin{equation}
K'_2 = \| \dev(\Cb) \|^2 = \tr (\dev \Cb)^2
\end{equation}
which is essentially the second principal invariant of $\dev \Cb$, and 
\begin{equation}
K'_3 = \det (\dev \Cb)
\end{equation}
which is the second principal invariant of $\dev \Cb$,
have the derivatives
\begin{equation}
\partial_\Cb K'_2 = \partial_\Cb \tr (\dev \Cb)^2
= 2 \dev \Cb
\end{equation}
and
\begin{equation}
\partial_\Cb K'_3 = \partial_\Cb \det (\dev \Cb)
= (\det (\dev \Cb)) \dev( (\dev(\Cb))^{-1} ) 
\end{equation}

%%%%%%%%%%%%%%%%%%%%%%%%%%%%%%%%%%%%%%%%%%%%%%%%%%%%%
\section{Potential-based neural network} \label{app:potentialNN} \setcounter{equation}{0}
%%%%%%%%%%%%%%%%%%%%%%%%%%%%%%%%%%%%%%%%%%%%%%%%%%%%%

Given a set of invariants $\Ic=\lbrace I_{1},I_{2},I_{3} \rbrace$ and a set of bases $\mathcal{B} = \lbrace \Bb_{1},\Bb_{2},\Bb_{3} \rbrace$ and let the stress $\Sb$ be a function of $\Ic$, $\mathcal{B}$ as well as of the derivatives of a potential $\Psi$ to its arguments $\Ib$, \ie $\Sb = \Sb(I_{1},I_{2},I_{3}, \Bb_{1},\Bb_{2},\Bb_{3},\partialb_{I_1}  \Psi,\partialb_{I_2}  \Psi,\partialb_{I_3}  \Psi)$.

Let $\mathcal{N}: \mathbb{R}^{3} \rightarrow \mathbb{R}$ be a feedforward neural network with $L$ hidden layers that takes the invariants as an input and outputs a potential-like scalar quantity.
The updating formula of the neural networks reads
\begin{equation}
    \begin{aligned}
            \bm{x}_{0}  &\in \mathbb{R}^{3} \\
            \bm{x}_{1} = \sigma_{1} \left( \bm{x}_{0} \bm{W}_{1}^{T} + \bm{b}_{1} \right) &\in \mathbb{R}^{n^{1}} \\
            \bm{x}_{l} = \sigma_{l} \left( \bm{x}_{l-1} \bm{W}_{l}^{T} + \bm{b}_{l} \right) &\in \mathbb{R}^{n^{l}}, \qquad l=1, \ldots, L-1 \\
            \hat{\phi} = \bm{x}_{L-1} \bm{W}_{L}^{T} + \bm{b}_{L}, &\in \mathbb{R}
    \end{aligned}
\end{equation}
with the trainable weights $\bm{W}_{i}$ and biases $\bm{b}_{i}$ and the elementwise activation functions $\sigma_{i}$. The output of the network is convex with regards to the inputs if all the weights are non-negative and the activation functions are convex and non-decreasing. This can be proven based on the fact that the composition of a convex and convex non-decreasing
function is convex as well and that non-negative sums of convex functions are also convex, see \cref{amos2017input}.
We can then set the predicted value of the potential $\hat{\Psi}$ as
\begin{equation}
    \hat{\Psi}(I_{1},I_{2},I_{3}) = \hat{\phi}(I_{1},I_{2},I_{3})  - \hat{\phi}^{S_{0}}(I_{1},I_{2},I_{3}) - \hat{\phi}(I_{1}|_{\Fb=\Ib},I_{2}|_{\Fb=\Ib},I_{3}|_{\Fb=\Ib})
\end{equation}
where $\hat{\phi}(I_{1}|_{\Fb=\Ib},I_{2}|_{\Fb=\Ib},I_{3}|_{\Fb=\Ib})$ ensures that the value of the potential is zero at the undeformed configuration $\Fb=\Ib$ while $\hat{\phi}^{S_{0}}(I_{1},I_{2},I_{3})$ is a linear function in its arguments and can be chosen so that the stress prediction $\hat{\Sb} = \hat{\Sb}(I_{1},I_{2},I_{3},\-\Bb_{1},\Bb_{2},\Bb_{3},\partialb_{I_1} \hat{\Psi},\partialb_{I_2}  \hat{\Psi},\partialb_{I_3}  \hat{\Psi})$ is also zero at the undeformed configuration.
For a practical example of the latter consider the set of invariants given by 
\begin{equation}
    I_1 = \tr(\Cb), \quad 
    I_2 = \tr(\Cb^{-1}) \det (\Cb), \quad 
    I_3 = \det(\Cb)
\end{equation}
and let the basis set read
\begin{equation}
    B_{1} = \Ib, \quad 
    B_{2} = \Cb, \quad 
    B_{3} = \Cb^{-1}.
\end{equation}
The function $\hat{\phi}^{S_{0}}(I_{1},I_{2},I_{3})$ can then for example be chosen to be 
\begin{equation}
    \hat{\psi}^{S_{0}}(I_{1},I_{2},I_{3}) = \left(  \partialb_{I_1}  \hat{\psi}|_{\Fb=\Ib} + 2 \partialb_{I_2}  \hat{\psi}|_{\Fb=\Ib} +  \partialb_{I_3}  \hat{\psi}|_{\Fb=\Ib} \right) (I_{3}-1)
\end{equation}

The explicit formula for the stress from the predicted potential value would then yield
\begin{equation} 
\begin{aligned}
        \hat{\Sb} &= 2 \left[ \partial_{I_1} \hat{\psi} +  I_{1}  \partial_{I_2} \hat{\psi} \right] B_{1}
    - 2 \partial_{I_2}\hat{\psi}\, B_{2}
    + 2 I_{3} \, \partial_{I_3} \hat{\psi} \, B_{3}
- 2 I_{3} \partial_{I_3} \hat{\psi}^{S_{0}} \, B_{3} \\
&= 2 \left[ \partial_{I_1} \hat{\psi} +  I_{1}  \partial_{I_2} \hat{\psi} \right] B_{1}
    - 2 \partial_{I_2}\hat{\psi}\, B_{2}
    + 2 I_{3} \, \partial_{I_3} \hat{\psi} \, B_{3}
- 2 I_{3} \left(  \partialb_{I_1}  \hat{\psi}|_{\Fb=\Ib} + 2 \partialb_{I_2}  \hat{\psi}|_{\Fb=\Ib} +  \partialb_{I_3}  \hat{\psi}|_{\Fb=\Ib} \right) \, B_{3}
\end{aligned}
\end{equation}
which in the undeformed configuration would mean 
\begin{equation}
    \hat{\Sb}|_{\Fb=\Ib}= 2 \left[ \partial_{I_1} \hat{\psi}|_{\Fb=\Ib} +  2  \partial_{I_2} \hat{\psi}|_{\Fb=\Ib} \right] \Ib
    + 2 \, \partial_{I_3} \hat{\psi}|_{\Fb=\Ib} \, \Ib
- 2 \left(  \partialb_{I_1}  \hat{\psi}|_{\Fb=\Ib} +  2 \partialb_{I_2}  \hat{\psi}|_{\Fb=\Ib} +  \partialb_{I_3}  \hat{\psi}|_{\Fb=\Ib} \right) \, \Ib = \bm{0}
\end{equation}

%%%%%%%%%%%%%%%%%%%%%%%%%%%%%%%%%%%%%%%%%%%%%%%%%%%%%
\section{Coefficient-based neural network} \label{app:coeffNN} \setcounter{equation}{0}
%%%%%%%%%%%%%%%%%%%%%%%%%%%%%%%%%%%%%%%%%%%%%%%%%%%%%

Given a set of invariants $\Ic=\lbrace I_{1},I_{2},I_{3} \rbrace$ and a set of bases $\mathcal{B} = \lbrace \Bb_{1},\Bb_{2},\Bb_{3} \rbrace$ and let the stress $\Sb$ be a function of $\Ib$, $\mathcal{B}$, \ie 
\begin{equation}
    \Sb = \sum_{a=1}^3 c_{a}(\Ic) \Bb_{a}
\end{equation}
where $c_{i}(\Ib)$ are stress coefficient functions.

A potential feedforward neural network $\mathcal{N}: \mathbb{R}^{3} \rightarrow \mathbb{R}^{3}$ to model this behavior could have $L$ hidden layers and takes the invariants inputs and has coefficient-like outputs.
The updating formula of the neural networks reads
\begin{equation}
    \begin{aligned}
            \bm{x}_{0} &\in \mathbb{R}^{3} \\
            \bm{x}_{1} = \sigma_{1} \left( \bm{x}_{0} \bm{W}_{1}^{T} + \bm{b}_{1} \right) &\in \mathbb{R}^{n^{1}} \\
            \bm{x}_{l} = \sigma_{l} \left( \bm{x}_{l-1} \bm{W}_{l}^{T} + \bm{b}_{l} \right) &\in \mathbb{R}^{n^{l}}, \qquad l=1, \ldots, L-1 \\
            \hat{\bm{c}} = \bm{x}_{L-1} \bm{W}_{L}^{T} + \bm{b}_{L}, &\in \mathbb{R}
    \end{aligned}
\end{equation}
with the trainable weights $\bm{W}_{i}$ and biases $\bm{b}_{i}$, the element-wise activation functions $\sigma_{i}$ and where $\hat{\bm{c}}$ are the predicted coefficients. 
The network output is monotonically non-decreasing with regards to the input if all the activation functions are monotonically non-decreasing and all the weights are positive. The proof follows from the fact that the sum of monotonically non-decreasing functions is monotonically non-decreasing and that the composition of  monotonically non-decreasing functions is also monotonically non-decreasing.

\end{document}